\documentclass[prapplied,twocolumn]{revtex4-2}
\usepackage[colorlinks, allcolors=blue]{hyperref}
\usepackage[utf8]{inputenc}
\usepackage[english]{babel}
\usepackage[T1]{fontenc}
\usepackage{amsmath}
\usepackage{bbm}
\usepackage{hyperref}
\usepackage{graphicx}
\usepackage{tikz}
\usepackage{lipsum}
\usepackage{braket}

\usepackage{hyperref}
\usepackage{subcaption}
\usepackage{amsmath}       
\usepackage{booktabs}     
\usepackage{array} 
\usepackage{pifont}
\usepackage{array}
\newcolumntype{P}[1]{>{\raggedright\arraybackslash}p{#1}}

\begin{document}

\title{Gate-Based Initialization and Fidelity in Correlated Open Quantum Systems}

\author{Sirui Chen}\thanks{Contact author: \href{mailto:schen931@gatech.edu}{schen931@gatech.edu}}
\author{Jiahao Chen}
\affiliation{Georgia Institute of Technology, Atlanta, Georgia, United States}
\author{Dragomir Davidović}\thanks{Contact author: \href{mailto:dragomir.davidovic@physics.gatech.edu}{dragomir.davidovic@physics.gatech.edu}}

\affiliation{Georgia Institute of Technology, Atlanta, Georgia, United States}

\begin{abstract}
We present a minimal and general framework for initializing open quantum systems via gate operations, treating system–bath correlations and control dynamics on equal footing. Our protocol simulates thermal equilibration followed by a gate pulse, modeled using a time-dependent Bloch–Redfield equation accurate to second order in coherence. After the gate, the system exhibits hybrid dynamics: populations evolve Markovianly, while coherences dephase as if the system were initially factorized. In fast, strongly dephasing regimes, initial system–bath correlations can revive coherence. Gate fidelities reveal spin-echo–like suppression of errors near $2\pi$-rotations, indicating an intrinsic mechanism for error cancellation. The approach is efficient and broadly applicable to both quantum devices and molecular complexes. Applications include fidelity optimization with respect to bath correlation times and coherence tracking in systems such as the Fenna–Matthews–Olson complex. Our results show that long-lived excitonic coherences can arise from strongly non-Markovian dynamics triggered by ultrafast pulses.
\end{abstract}

\maketitle
\section{Introduction}

Quantum coherence is essential for quantum computing. A fundamental operation—creating superposition via a $\pi/2$ rotation—is ideally unitary but degraded in practice by environmental coupling, which lowers gate fidelity.

A central challenge is characterizing the qubit state immediately after the gate. Continuous system–bath interaction during gating introduces ambiguity, especially when initial correlations exist. These correlations affect post-gate coherence and can induce non-Markovian behavior~\cite{rosario2008no,modi2012classical,paz2019dynamics}.

Standard models assume factorized system–bath states, enabling Kraus maps~\cite{hayashi2003kraus} or master equations~\cite{redfield1957theory,1976CMaPh..48..119L,gorini1976completely}. This simplifies analysis but neglects equilibrium correlations, which modify dephasing, especially at low temperatures~\cite{subasi2012equilibrium, Wu2008}.

Correlated initial states, such as mean-force Gibbs states, encode the environment’s dressing of the system and shift coherence dynamics. Pre-gate correlations persist through the gate, transferring information and modifying dephasing and relaxation~\cite{breuer2009measure,rivas2014quantum}.

\begin{figure}[h]
    \centering
    \includegraphics[width=1\columnwidth]{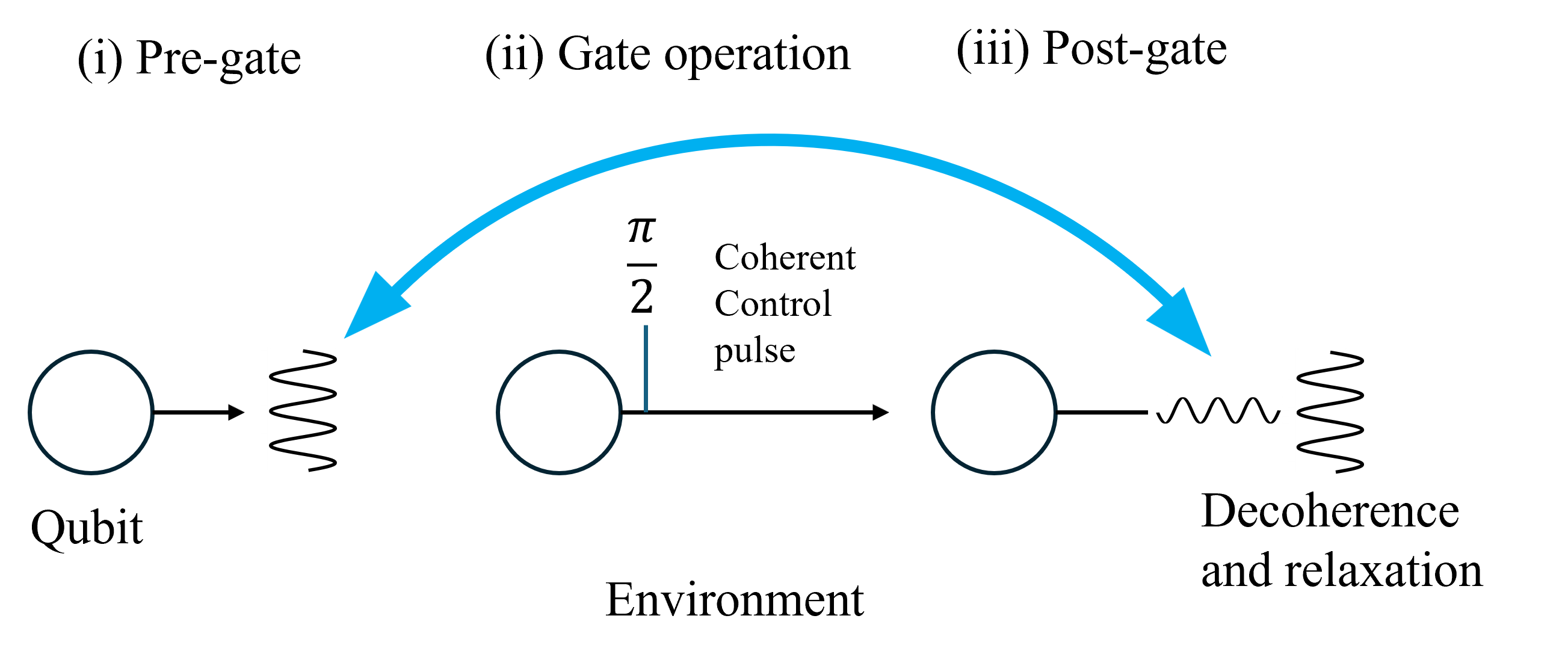}
    \captionsetup{justification=raggedright, singlelinecheck=false}
    \caption{
Schematic of a qubit’s evolution in an open quantum system. The process includes three stages: (i) pre-gate interaction with the environment, (ii) a gate operation (e.g., a $\pi/2$ pulse) applied while the qubit remains coupled, and (iii) post-gate relaxation and decoherence. Blue arrows indicate the bidirectional transfer of system–environment correlations through the gate, ensuring that the pre-gate correlations are conjugated by the gate's unitary operation. This setting mirrors superconducting qubit~\cite{martinis2003decoherence,bylander2011noise,yan2016flux} and 2DES experiments~\cite{engel2007evidence,turner2010coherent,westenhoff2012coherent}, where coherent dynamics unfold under continuous environmental coupling.
}
    \label{Fig:PrePostGate}
\end{figure}

We address:  
\textit{How do preexisting system–bath correlations shape dephasing and relaxation after a gate and gate fidelity?} Figure~\ref{Fig:PrePostGate} illustrates the process: from pre-gate correlations, through the gate applied under continuous coupling, to post-gate decoherence.

\subsection{Methodology}

We require a method that (i) captures correlated coherence, (ii) handles time-dependent control, and (iii) tracks post-gate dynamics under continuous coupling. To this end, we extend Bloch–Redfield master equations to account for correlation transfer across gates.

\paragraph*{(1) Accuracy in coherence.}  
We adopt the Bloch–Redfield master equation for its ability to resolve qubit coherences with high precision. While several master equations exist—including Gorini-Kossakowski-Sudarshan-Lindblad (GKSL)~\cite{gorini1976completely,1976CMaPh..48..119L}, Floquet~\cite{PhysRev.138.B979}, and adiabatic~\cite{spohn1980kinetic,albash2012quantum} forms—most rely on factorized initial states and neglect persistent system–bath correlations.

The Bloch–Redfield equation uniquely captures the asymptotic-state coherences of the mean-force Gibbs state (or ground state at $T=0$) with \emph{quadratic} accuracy in the coupling strength—beyond the reach of Lindblad models. Despite assuming an uncorrelated state at all times (Born approximation), it reproduces subtle coherence features due to entanglement~\cite{fleming2011accuracy,thingna2012generalized,hartmann2020accuracy,lee2022perturbative,crowder2024invalidation,tupkary2022fundamental}. Minor violations of complete positivity are tolerated in favor of improved coherence modeling.

\paragraph*{(2) Time-dependent control.}  
To address driven dynamics under environmental coupling, we employ a time-dependent Bloch–Redfield equation in the interaction picture~\cite{chen2022hamiltonian}. This avoids secular approximations that decouple coherences and populations~\cite{davies1974markovian} and handles strong or aperiodic drives more reliably than adiabatic~\cite{albash2012quantum} or Floquet-based schemes~\cite{yamaguchi2017markovian}. Although earlier work enforces complete positivity in time-dependent settings~\cite{mozgunov2020completely,nathan2020universal,davidovic2022geometric}, such approaches often lose accuracy in coherence or energy conservation~\cite{tupkary2022fundamental}.

Recent time-dependent Markovian master equations~\cite{di2024time} agree well with tensor network simulations under periodic driving~\cite{paeckel2019time,schollwock2011density,orus2014practical}, but remain limited by secular assumptions. Our method allows efficient simulation over long times without Floquet truncation or quasi-energy ambiguity.

\paragraph*{(3) Post-gate correlation tracking.}  
We introduce \textit{dynamical state preparation} to model system–environment evolution across gate operations. This reflects experimental conditions in superconducting circuits~\cite{mielke2021nuclear,blais2004cavity}, where qubits remain coupled to their environment during initialization and gating. Our method simulates coherent evolution pre-, during, and post-gate, offering a more accurate alternative to stagewise or factorized treatments.

This aligns with the general framework by Paz-Silva et al.~\cite{paz2019dynamics}, which decomposes initial states into bath-positive components to preserve complete positivity. However, we focus on a concrete regime: fast gate operations and their coherence consequences. Our goal is not formal generality, but physical insight—capturing hybrid post-gate states and phenomena such as spin-echo–like fidelity recovery. These features are increasingly relevant for quantum error correction and coherent gate engineering, where small residual correlations can impact system performance.

\subsection{Main Results}

Our findings reveal that quantum gate operations generally prepare the system in an initial state that is neither fully factorized nor fully equilibrated with the environment. This hybrid condition gives rise to distinct post-gate dynamical signatures: coherences decay rapidly, as if initialized from a factorized state, while populations evolve as if originating from a correlated, partially equilibrated one. As a result, both dephasing and relaxation proceed at their maximal rates—defining an operationally unfavorable regime for quantum information processing.

These hybrid dynamics closely resemble the framework proposed by Paz-Silva et al.~\cite{paz2019dynamics}, which expresses the evolution from initially correlated states as a mixture of inequivalent bath-positive maps. Our results provide a concrete instantiation of this framework, where gate operations mediate a nontrivial interplay between inherited correlations and  decoherence.

\subsubsection{Fidelity and Correlation Transfer}

To assess the performance of quantum gates under continuous environmental coupling, we examine the fidelity of the post-gate reduced state. We find that the fidelity loss contains a static component, such that the maximum fidelity is achieved \textit{near}, but not exactly at, the target state of the applied unitary. This offset on the Bloch sphere arises from latency and dephasing induced by the slowly relaxing bath. The maximum fidelity remains strictly less than one, reflecting an irreducible phase uncertainty (i.e., noise) introduced by the bath during and prior to the gate operation. This fidelity loss cannot be compensated by simply rotating or offsetting the Bloch vector—it represents a fundamental limitation imposed by the system–bath interaction.

We further show that the gate fidelity can be significantly enhanced through pulse shaping. In environments with strong memory (e.g., sub-Ohmic spectral densities), optimized pulse profiles leverage the transfer of pre-existing bath correlations across the gate, as well as the dissipative dynamics during the gate itself. This coordinated control leads to substantial fidelity gains, particularly in non-Markovian regimes where system–bath correlations persist across operational timescales.

\subsubsection{Spin-Echo–Like Behavior and Intrinsic Error Mitigation}

In the regime of long gate durations, we observe a fidelity dip near a rotation angle of $\pi$, reminiscent of the Spin-Echo effect. Just as Spin-Echo refocuses coherence lost to dephasing, the gate operation induces a similar rephasing phenomenon—suggesting a form of intrinsic error cancellation. This effect implies that partial elements of dynamical decoupling or error correction may arise naturally, even in imperfect gate implementations. Such behavior points to novel pathways for coherence preservation in noisy intermediate-scale quantum systems.

\subsubsection{Ultrashort Gates and Echo-Like Rephasing}

In the opposite regime—ultrashort gate pulses under strong dephasing—we observe transient coherence recovery following its initial decay. This echo-like behavior parallels rephasing phenomena seen in nonlinear spectroscopy, where ultrafast pulses initiate and probe non-Markovian system–bath dynamics. Our results suggest that gate operations in quantum processors may play an analogous role.

\subsubsection{Connections to Light-Harvesting Systems}

Interestingly, these results parallel phenomena reported in biological systems, particularly in photosynthetic light-harvesting complexes such as the Fenna--Matthews--Olson (FMO) complex. Two-dimensional electronic spectroscopy (2DES) studies have demonstrated long-lived quantum coherences at physiological temperatures~\cite{engel2007evidence,panitchayangkoon2010}.  Theoretical models attributed these effects to structured environmental couplings, vibronic modes, and correlated initial conditions~\cite{Ishizaki2009,Rebentrost2009}. In our explanation, long-lived excitonic coherence arises from the interference between pre-gate and post-gate dynamics in the presence of a slow bath. Post-gate evolution induces rapid dephasing, while residual pre-gate correlations generate rephasing through non-secular population–coherence transfer. This balance persists until the bath response fully decays, at which point the system relaxes into a conventional Markovian regime.
This provides a unified explanation for coherent energy transport in biological systems and highlights the broader relevance of our approach beyond quantum information platforms.

\subsection*{Paper Organization and Guide to the Reader}

This work is structured to separate formal developments, analytical results, numerical evidence, and physical applications. The reader is advised to proceed according to purpose.

\begin{itemize}
    \item \textbf{Section~\ref{Sec: derivBloch-Redfield}} contains the derivation of the Bloch--Redfield master equation for time-dependent Hamiltonians. It addresses the treatment of initial correlations and the formulation of reduced dynamics beyond the Markov approximation. This section provides the theoretical basis for the rest of the paper.

    \item \textbf{Section~\ref{Sec:dp}} presents analytic results in regimes where exact or approximate solutions are possible. Special attention is given to dynamically prepared states, instantaneous gates, and long-lived coherences. These subsections clarify the essential mechanisms of dephasing and correlation transfer.

    \item \textbf{Section~\ref{Sec:Finite-Duration}} extends the analysis to finite-duration gates. It shows how relaxation and dephasing evolve in dynamically prepared states, and characterizes fidelity loss due to interaction with the environment during gate operation.

    \item \textbf{Section~\ref{Sec:Applications}} contains applications. Section~\ref{Sec:application: optimization} discusses how pulse shapes may be optimized to improve gate fidelity in non-Markovian regimes. Section~\ref{Sec:FMO} applies the formalism to excitonic coherence in light-harvesting complexes. These results may be of interest beyond quantum computing, including to those working in chemical and biological physics.

    \item \textbf{The Conclusion} summarizes main findings and outlines possible extensions, including multi-qubit dynamics, structured baths, and implications for fault-tolerant architectures.
\end{itemize}

\noindent
Readers interested in theoretical consistency and mathematical derivation should begin with Section~\ref{Sec: derivBloch-Redfield}. Readers concerned with observable effects and practical behavior may focus on Sections~\ref{Sec:dp} and~\ref{Sec:Finite-Duration}. Applications, both technological and biological, are presented in Section~\ref{Sec:Applications}. Each section is logically independent, but intended to be read in sequence.

\section{Rederivation of the Bloch-Redfield Master Equation for Time-Dependent Hamiltonians}\label{Sec: derivBloch-Redfield}

To keep the presentation self-contained and accessible, we rederive the Bloch--Redfield equation for a time-dependent system Hamiltonian. This rederivation not only clarifies foundational aspects but also captures the emergence of non-Markovian dynamics induced by the applied gate. While the core formalism has appeared previously~\cite{chen2022hamiltonian,davidovic2022geometric}, our treatment extends it by showing how a gate operation disrupts otherwise indivisible dissipative quantum evolution.

We intentionally adopt a minimal yet exact mathematical formulation, prioritizing clarity without sacrificing rigor. This streamlined approach is not a simplification for its own sake; rather, it sharpens focus on the assumptions underpinning the derivation—assumptions that are often hidden within more formal or pedagogically layered presentations.

We consider a quantum system \( S \) weakly coupled to an environment \( B \), described by the total Hamiltonian in the Schr\"odinger picture:
\begin{equation}
    H(t) = H_S(t) + H_B + H_I,
\end{equation}
where \( H_S(t) \) is a time-dependent Hamiltonian acting exclusively on the system, \( H_B \) is the time-independent bath Hamiltonian, and \( H_I \) is the time-independent system-bath interaction Hamiltonian (assumed to couple linearly to a bath of harmonic oscillators).

To study the time evolution of the system, we introduce the unitary propagator corresponding to the time-dependent Hamiltonian \( H_S(t) \), defined as
\begin{equation}
    U_S(t,t_0) = T_{\leftarrow} \exp\left[-i \int_{t_0}^t d\tau\, H_S(\tau) \right],
\label{Eq:TimeOrder}
\end{equation}
where \( T_{\leftarrow} \) denotes the time-ordering operator. This propagator describes the unitary evolution of the isolated system, from an initial time \( t = t_0 \) to a later time \( t \); the choice of \( t = t_0 \) is arbitrary for now. 
In the interaction picture of the isolated system, a Schrödinger picture operator \( X_S(t) \) evolves as
\begin{equation}
\label{Eq:X_IS}
    X_I(t) = U_S^\dagger(t,t_0) X_S(t) U_S(t,t_0).
\end{equation}

The \textit{interaction picture} of the total system is defined with respect to the decoupled evolution of both the system and the environment. The corresponding unitary transformation is given by
\begin{equation}
    U_{\text{tot}}(t, t_0) = U_S(t, t_0) \otimes e^{-i H_B (t - t_0)}.
\end{equation}
As an example, the interaction Hamiltonian in the interaction picture is given by 
\begin{equation}
    H_I(t) = \left[ U_S^\dagger(t, t_0) \otimes e^{i H_B (t - t_0)} \right]
             H_I
             \left[ U_S(t, t_0) \otimes e^{-i H_B (t - t_0)} \right].
\end{equation}

The evolution of the total system in the interaction picture is governed by the von Neumann equation:
\begin{equation}
    \frac{d\varrho}{dt} = -i \left[ H_I(t), \varrho \right].
\end{equation}
Throughout this work, \( \varrho \) denotes the state of the system in the interaction picture, while \( \rho \) refers to the state in the Schrödinger picture.

We now assume that at time \( t = t_0 \), the total system-environment state is factorized, that is, \( \varrho(t_0) = \varrho_S(t_0) \otimes \varrho_B \). We further assume that the initial bath state commutes with the isolated bath Hamiltonian, i.e., \( [\varrho_B, H_B] = 0 \), which implies that \( \varrho_B \) is diagonal in the energy eigenbasis of the bath and, therefore, remains constant in time within the Heisenberg picture of the isolated bath. 

Next, 
starting from the formal solution of the von Neumann equation,
\begin{equation}
    \varrho(t)= T_{\leftarrow} \exp\left[-i\int_{t_0}^t d\tau\, [H_{I}(\tau),]\right] \varrho(t_0),
\label{Eq:Formal}
\end{equation}
we apply the Born approximation, where \( \varrho(t) \approx \varrho_S(t) \otimes \varrho_B \), and expand the formal solution perturbatively in \( H_I(t) \).
Tracing over the bath degrees of freedom, and assuming that the odd-order terms in \( H_I \) vanish under the partial trace (i.e., they are traceless with respect to the bath), and integrating, we obtain:
\begin{equation}
\begin{aligned}
\varrho_S(t)
&= \varrho_S(t_0)
  - \int_{t_0}^{t}\!dt_1 \int_{t_0}^{t_1}\!dt_2\,
\\
&\quad \mathrm{Tr}_B\!\big[\, H_I(t_1),\,[\,H_I(t_2),\,\varrho_S(t_0)\!\otimes\!\varrho_B\,] \big]
  + \mathcal{O}\!\big(H_I^4 t^2 \tau_c^2\big).
\end{aligned}
\label{Eq:exact}
\end{equation}

where $\tau_c$ is the range of bath-correlations in time.

Taking the time derivative of Eq.~(8), neglecting higher-order terms, and replacing $\varrho_S(t_0) \to \varrho_S(t)$, one arrives at the Bloch--Redfield master equation:
\begin{equation}
\frac{d\varrho_S(t)}{dt} = -\mathrm{Tr}_B \int_{t_0}^t ds\, [H_I(t), [H_I(s), \varrho_S(t) \otimes \varrho_B]].
\label{Eq:BRinitial}
\end{equation}
While the approximations involved—such as the replacement of $\varrho_S(t_0)$ with $\varrho_S(t)$—may initially appear uncontrolled, their consistency has been rigorously justified. In particular, van Kampen~\cite{vanKampen1974stochastic} showed that such procedures yield finite, stable dynamics over arbitrary timescales within the weak-coupling limit and the bath with short correlation time.

Two central issues arise in this formulation. First, although the derivation assumes a factorized system–environment state throughout (the Born approximation), the resulting dynamics nonetheless capture system–bath correlations. This apparent contradiction is resolved by noting that the master equation generates perturbative corrections to the uncorrelated ansatz. As in standard perturbation theory, the correlated state is not required \textit{a priori}; it is built order by order from the factorized unperturbed state.

In this case, perturbative evaluation of the asymptotic reduced-state coherences captures entanglement-induced corrections at the correct order~\cite{fleming2011accuracy,tupkary2022fundamental,crowder2024invalidation}. Specifically, Schrödinger--Rayleigh perturbation theory shows that the first-order correction to the product ground state yields entangled system--bath states, with amplitudes proportional to the interaction Hamiltonian $H_I$. Tracing over the bath then produces reduced system states whose coherences match those predicted by the Bloch--Redfield equation at the same perturbative order~\cite{fleming2011accuracy,tupkary2022fundamental,crowder2024invalidation}.

Although the Bloch--Redfield equation does not fully capture the structure of entanglement—particularly in the populations—it accurately reflects the system coherences induced by entanglement with the bath. Crucially, no master equation in Lindblad form achieves this fidelity, as the requirement of complete positivity imposes constraints that suppress such coherence corrections~\cite{tupkary2022fundamental}. In this sense, the Bloch--Redfield formalism stands out by uniquely preserving partial information about system--bath entanglement through the coherence structure of the reduced state.

The second issue concerns the long-time behavior of the expansion in Eq.~(\ref{Eq:exact}). The second-order correction to the reduced state grows as \( O(H_{\text{I}}^2 t\tau_c) \), i.e., linearly in time. Higher-order terms, omitted from the expansion, scale as \( O(H_{\text{I}}^4 t^2\tau_c^2) \) or faster, and eventually dominate. This divergence, known as \textit{secular growth}, limits the validity of time-dependent perturbation theory to short times. In harmonic baths, it arises from statistically independent bath correlations evaluated at different time arguments~\cite{lampert2025sixth}.

The master equation avoids secular growth by effectively resumming Eq.~(\ref{Eq:exact}) using time-ordered cumulants, enabling stable evolution at long times. Van Kampen~\cite{vanKampen1974stochastic} showed that the dominant secular terms cancel at \textit{all} perturbative orders—a nontrivial result that extends the conventional cumulant expansion to time-ordered cases, where the interaction Hamiltonian does not commute at different times. In the context of quantum gates, it is crucial that the transformation to time-ordered cumulants involves both forward and backward propagation in time, thereby breaking strict chronological ordering~\cite{vanKampen1974stochastic,breuer2002theory}. This time-symmetric exchange of correlations between system and bath across the gate is a defining feature of the dynamics illustrated in Fig.~\ref{Fig:PrePostGate}. As a result, gate operations induce a bidirectional flow of quantum information—post-gate to pre-gate and vice versa. Achieving high-fidelity control in this regime requires pulse shaping tailored to the structure of these correlations.

The cancellation of secular growth by time-ordered cumulant expansion is not universally guaranteed. Recent work has shown that in environments with algebraically decaying bath correlations—unlike the more commonly assumed exponential decay—secular growth can reemerge even within the master equation itself, a phenomenon we termed \textit{secular inflation}~\cite{lampert2025sixth}. For the parameter regimes considered here, the effect of secular inflation is negligible and thus omitted from our analysis. Nonetheless, its emergence in strongly correlated or structured environments underscores the need for caution when applying master equations beyond the weak-coupling or Markovian limits.

To make the structure of Eq.~(\ref{Eq:BRinitial}) more transparent—and to emphasize the environment’s finite memory—we change the integration variable from $s$ to $s' = t - s$, yielding
\begin{equation}
    \frac{d\varrho_S(t)}{dt} = -\mathrm{Tr}_B \int_0^{t - t_0} ds'\, [H_I(t), [H_I(t - s'), \varrho_S(t) \otimes \varrho_B]].
    \label{Eq:Rmaster equation}    
\end{equation}
This form of the Bloch--Redfield master equation, previously derived in Refs.~\cite{chen2022hamiltonian,davidovic2022geometric}, represents a natural extension of the standard formulation to systems governed by time-dependent Hamiltonians. Only recently have its solutions begun to receive detailed scrutiny~\cite{tripathi2024modeling,fernandez2024recovering}, revealing rich dynamical features beyond the reach of time-independent or Markovian approximations.

In the conventional Markov approximation, the integration limit is extended to $t_0 \to -\infty$, effectively erasing any memory of initial correlations. By contrast, our approach retains a finite $t_0$ throughout the derivation in order to capture the environment’s finite correlation time and systematically go beyond the Markovian regime. Only after incorporating the effects of time-dependent control fields do we take the $t_0 \to -\infty$ limit—removing residual dependence on the initially factorized state while preserving the non-Markovian features induced by the gate.

\subsubsection{Implementation}
To apply the master equation in Eq.~(\ref{Eq:Rmaster equation}) to concrete systems, we begin by specifying the form of the interaction Hamiltonian in the Schrödinger picture:
\begin{equation}
    H_I = \sum_\alpha A_\alpha \otimes B_\alpha,
\end{equation}
where \( A_\alpha \) and \( B_\alpha \) are time-independent Hermitian operators acting on the system and bath, respectively. In this formulation, the physical units are absorbed into the bath coupling operators \( B_\alpha \), rendering the system operators \( A_\alpha \) dimensionless.

To proceed, we express the system operators in the interaction picture. Let \( A_\alpha(t) = U_S^\dagger(t,t_0) A_\alpha U_S(t,t_0) \) denote the evolution of \( A_\alpha \) under the system Hamiltonian. Following Ref.~\cite{davidovic2022geometric}, we apply the full Fourier transform to \( A_\alpha(t) \), yielding its frequency decomposition:
\begin{equation}
\begin{aligned}
    \mathcal{A}_\alpha(\omega) &= \int_{-\infty}^{\infty} dt\, A_\alpha(t)\, e^{i\omega (t - t_0)}, \\
    A_\alpha(t) &= \frac{1}{2\pi} \int_{-\infty}^{\infty} d\omega\, \mathcal{A}_\alpha(\omega)\, e^{-i\omega (t - t_0)}.
\end{aligned}
\end{equation}

This frequency-domain representation is particularly useful for identifying resonant transitions and simplifying the dissipative structure of the master equation. Substituting the Fourier decomposition into Eq.~(\ref{Eq:Rmaster equation}), the master equation takes the form~\cite{chen2022hamiltonian,davidovic2022geometric}:
\begin{equation}
\frac{d\varrho_S}{dt} = \sum_{\alpha,\beta} \left\{[\Lambda_{\alpha,\beta}^I(t)\varrho_S, A_\alpha(t)] + [A_\alpha(t), \varrho_S\, \Lambda_{\alpha,\beta}^{I\dagger}(t)] \right\},
\label{Eq:lidarform}
\end{equation}
where the \textit{filtered coupling operator} (or \textit{dissipator}) is defined as
\begin{equation}
\Lambda_{\alpha,\beta}^I(t) = \int_{-\infty}^{\infty} \frac{d\omega}{2\pi}\, \Gamma_{\alpha,\beta}(\omega, t - t_0)\, \mathcal{A}_\beta(\omega)\, e^{-i\omega (t - t_0)}.
\label{Eq:lambda}
\end{equation}

In the Schrödinger picture, the dissipator transforms as
\begin{equation}
\Lambda_{\alpha,\beta}^S(t) = U_S(t,t_0)\, \Lambda_{\alpha,\beta}^I(t)\, U_S^\dagger(t, t_0),
\label{Eq:lambdaSP}
\end{equation}
and the corresponding Schrödinger-picture master equation becomes
\begin{equation}
\begin{aligned}
\frac{d\rho_S}{dt}
&= -i[H_S(t), \rho_S]
\\[-2pt]
&\quad + \sum_{\alpha,\beta} \left\{
   [\Lambda_{\alpha,\beta}^S(t)\rho_S, A_\alpha]
   + [A_\alpha, \rho_S\,\Lambda_{\alpha,\beta}^{S\dagger}(t)]
\right\}.
\end{aligned}
\label{Eq:lidarformSP}
\end{equation}

The dissipator encapsulates the effect of the environment’s spectral response on the system's dynamics, modulating the system–bath interaction in both frequency and time. The function \( \Gamma_{\alpha,\beta}(\omega, t - t_0) \) acts as a time-dependent spectral density, quantifying how the bath mediates dissipation and decoherence at different frequencies. It is derived from the bath correlation function (BCF) via
\begin{equation}
    \Gamma_{\alpha,\beta}(\omega, t - t_0) = \int_0^{t - t_0} d\tau\, e^{i\omega\tau}\, C_{\alpha,\beta}(\tau),
    \label{Eq:Gamma}
\end{equation}
with the BCF defined as
\begin{equation}
C_{\alpha,\beta}(\tau) = \text{Tr}_B[\rho_B\, B_\alpha(\tau)\, B_\beta(0)].
\label{Eq:BCFdef}
\end{equation}
This framework provides a natural language for describing how the environment's spectral properties shape the system's reduced dynamics, especially when memory effects are significant.

\subsection{Static Limit}

When the system Hamiltonian is time-independent, Eq.~(\ref{Eq:lidarform}) reduces to the standard Bloch--Redfield form~\cite{davidovic2022geometric}. In this limit, the continuous frequency integral in Eq.~(\ref{Eq:lambda}) simplifies to a sum over discrete Bohr frequencies, and the filtered coupling operator becomes
\begin{equation} 
\Lambda_{\alpha, \beta}^I(t) = \sum_\omega \mathcal{A}_{\omega,\beta}\, e^{-i\omega (t - t_0)}\, \Gamma_{\alpha,\beta}(\omega, t - t_0),
\label{Eq: discreteA}
\end{equation}
where the frequency components \( \mathcal{A}_{\omega,\beta} \) are defined as
\begin{equation}
\mathcal{A}_{\omega,\beta} = \sum_{E_m - E_n = \omega} \Pi_n A_\beta \Pi_m,
\end{equation}
and \( \Pi_n \) are projectors onto the eigenspaces of the system Hamiltonian associated with eigenenergy \( E_n \)~\cite{breuer2002theory}.

In the Schrödinger picture, the corresponding dissipator is obtained by undoing the interaction-picture transformation using the free evolution operator \( e^{-iH_0 (t-t_0)} \), where \( H_0 \) is the time-independent system Hamiltonian. This yields:
\begin{equation}
\begin{aligned}
\Lambda^S_{\alpha,\beta}(t) 
&= e^{-iH_0 \Delta t}\, \Lambda_{\alpha,\beta}(t)\, e^{iH_0 \Delta t} \\
&= \sum_{\omega}
    \bigl(e^{-iH_0 \Delta t}\,\mathcal{A}_{\omega,\beta}\,e^{iH_0 \Delta t}\bigr)
\\[-2pt]
&\qquad\qquad\times e^{-i\omega \Delta t}\,
   \Gamma_{\alpha,\beta}\!\left(\omega,\Delta t\right) \\
&= \sum_{\omega} \mathcal{A}_{\omega,\beta}\,
   \Gamma_{\alpha,\beta}\!\left(\omega,\Delta t\right),
\end{aligned}
\end{equation}
where \(\Delta t = t-t_0\). We used the fact that \( \mathcal{A}_{\omega,\beta} \) commutes with \( H_0 \) up to a phase, due to its definition via energy eigenprojectors. 
Alternatively, the same expression can be written in the time domain as
\begin{equation}
\Lambda^S_{\alpha,\beta}(t) = \int_0^{t-t_0} d\tau\, C_{\alpha,\beta}(\tau)\, e^{-iH_0 \tau} A_\beta e^{iH_0 \tau}.
\label{Eq:StaticNonMarkov}
\end{equation}

We shall refer to the Bloch-Redfield equation in this form—where the system Hamiltonian is time-independent and the spectral density retains an explicit time dependence through a finite upper limit in Eq.~(\ref{Eq:Gamma})—as the \textit{static-nonMarkovian} equation. It is non-Markovian in the sense that the dynamics are sensitive to the initial time at which the factorized system-bath state is prepared; this memory effect is embedded in the time-dependent spectral density \( \Gamma_{\alpha,\beta}(\omega, t-t_0) \).

In the limit \( t_0 \to -\infty \), the spectral density converges to a stationary form, eliminating explicit time dependence in Eq.~(\ref{Eq:Gamma}). Specifically, we replace
\begin{equation}
    \Gamma_{\alpha,\beta}(\omega, t - t_0) \mapsto \Gamma_{\alpha,\beta}(\omega) = \int_0^{\infty} d\tau\, e^{i\omega \tau} C_{\alpha,\beta}(\tau),
    \label{Eq: Gamma1b}
\end{equation}
which corresponds to the Fourier--Laplace transform of the bath correlation function. 

In the static case, this stationary spectral density can be substituted into the filtered coupling operator by extending the upper integration limit in Eq.~(\ref{Eq:StaticNonMarkov}) to infinity. The result is the \textit{static-Markovian} form:
\begin{equation}
\Lambda_{\alpha,\beta}^{\text{SM}} = \int_0^{\infty} d\tau\, C_{\alpha,\beta}(\tau)\, e^{-iH_0 \tau} A_\beta e^{iH_0 \tau},
    \label{Eq:StaticMarkov}
\end{equation}
where the bath is assumed to lose memory rapidly relative to system timescales. This separation of time scales underpins the conventional Born--Markov approximation used in the standard Bloch--Redfield framework.

\subsection{Erasing the Initial Condition for Time-Dependent Hamiltonians}

A similar simplification can be made in the case of time-dependent system Hamiltonians. Here, the Markovian approximation is applied by neglecting the explicit \( t_0 \) dependence in the spectral density appearing in Eq.~(\ref{Eq:lambda}), using instead the stationary form given in Eq.~(\ref{Eq: Gamma1b}). As a result, the filtered coupling operator becomes independent of the initial time:
\begin{equation}
\begin{aligned}
   \label{Eq:FFTd} 
   \Lambda_{\alpha,\beta}^I(t) 
   &= \int_{-\infty}^{\infty} \frac{d\omega}{2\pi}\, \Gamma_{\alpha,\beta}(\omega)\, \mathcal{A}_\beta(\omega) e^{-i\omega (t - t_0)} \\ 
   &= \int_{-\infty}^{\infty} \frac{d\omega}{2\pi}\, \Gamma_{\alpha,\beta}(\omega)\, e^{-i\omega (t - t_0)}
      \int_{-\infty}^{\infty} d\tau\, A_\beta(\tau) e^{i\omega(\tau - t_0)} \\
   &= \int_{-\infty}^{\infty} d\tau\, A_\beta(\tau)
      \int_{-\infty}^{\infty} \frac{d\omega}{2\pi}\, \Gamma_{\alpha,\beta}(\omega) e^{-i\omega(t - \tau)}.
\end{aligned}
\end{equation}

This final expression reveals that \( \Lambda_{\alpha,\beta}(t) \) is a convolution of the spectral density with the time-dependent system operator \( A_\beta(t) \). Importantly, its evaluation reduces to a standard Fourier transform, allowing for efficient numerical implementation using the Fast Fourier Transform (FFT) algorithm. That said, the analytical result presented in Sec.~\ref{Sec:dp} yields a master equation that does not involve any integrals or Fourier transforms, thereby offering improved computational efficiency

To make this structure more explicit, we now recast the spectral density in the time domain. The correlation function \( C_{\alpha,\beta}(t - \tau) \) serves as the time-domain counterpart to the spectral density \( \Gamma_{\alpha,\beta}(\omega) \), with causality enforced by the Heaviside step function \( \Theta(t - \tau) \). We evaluate:
\begin{equation}
\begin{aligned}
&\int_{-\infty}^{\infty} \frac{d\omega}{2\pi}\,\Gamma_{\alpha,\beta}(\omega)\,e^{-i\omega(t-\tau)}\\
&= \int_{-\infty}^{\infty} \frac{d\omega}{2\pi} \int_{0}^{\infty} dt_1\,
   C_{\alpha,\beta}(t_1)\, e^{i\omega(t_1 - t + \tau)}\\
&= \int_{0}^{\infty} dt_1\, C_{\alpha,\beta}(t_1)
   \int_{-\infty}^{\infty} \frac{d\omega}{2\pi}\, e^{i\omega(t_1 - t + \tau)}\\
&= \int_{0}^{\infty} dt_1\, C_{\alpha,\beta}(t_1)\, \delta(t_1 - t + \tau)\\
&= C_{\alpha,\beta}(t - \tau)\, \Theta(t - \tau).
\end{aligned}
\end{equation}

Substituting this result into the final line of Eq.~(\ref{Eq:FFTd}) yields a manifestly time-domain form for the dissipator:
\begin{equation}
    \Lambda_{\alpha,\beta}^I(t) = \int_{-\infty}^{t} d\tau\, C_{\alpha,\beta}(t - \tau)\, A_\beta(\tau),
    \label{Eq:asymptoticLambda}
\end{equation}
which clearly illustrates how environmental memory effects enter through a convolution of the bath correlation function with the system operator’s evolution.

To express this result in the Schrödinger picture, we rewrite the interaction-picture operator \( A_\beta(\tau) \) in terms of its time-evolved Schrödinger-picture form. This gives:
\begin{equation}
    \Lambda_{\alpha,\beta}^S(t) = \int_{-\infty}^{t} d\tau\, C_{\alpha,\beta}(t - \tau)\, U_S(t,\tau)\, A_\beta\, U_S(\tau,t),
    \label{Eq:asymptoticLambdaS}
\end{equation}
where \( A_\beta \) is time-independent and \( U_S(t, \tau) \) is the system propagator defined in Eq.~(\ref{Eq:TimeOrder}). This formulation retains both the full time-dependence of the system Hamiltonian and the memory kernel imposed by the bath, while respecting causality.

Beyond its compact structure, Eq.~(\ref{Eq:asymptoticLambdaS}) offers a clear physical interpretation of gate-driven dynamics in open quantum systems. The filtered coupling operator \( \Lambda_{\alpha,\beta}^S(t) \) encodes a time-symmetric process: it propagates a quantum state backward from time \( t \) to \( \tau \), applies the coupling operator \( A_\beta \), and then evolves the result forward to \( t \). This two-way propagation is convolved with the bath correlation function, which weights contributions according to system--environment correlations over the interval \( [\tau, t] \). As a result, the dissipator dynamically encodes non-Markovian effects that span both pre- and post-gate interactions.

Crucially, this interpretation is not merely illustrative—it underpins the formal structure of state initialization through gates. Together, Eqs.~(\ref{Eq:lidarformSP}) and~(\ref{Eq:asymptoticLambdaS}) define an effective non-Markovian master equation that captures memory effects induced by coherent control protocols, without relying on explicit knowledge of the system’s initial preparation.

\subsection{Bloch--Redfield Master Equation for a Qubit Coupled to a Single Bath}

To illustrate the above framework, we now apply it to a specific and widely studied system: the spin-boson model. Here, a single qubit with free Hamiltonian \( H_0 = -\frac{\Delta}{2} \sigma_z \) interacts with a bosonic environment through a single system--bath coupling channel. Accordingly, we set \( \alpha = \beta = 1 \) and suppress these indices in what follows. For notational convenience, we absorb the frequency dependence into the subscript and write
$\Gamma(\omega,t)\equiv \Gamma_{\omega}(t)$ henceforth.

When the qubit is subject to a time-dependent control Hamiltonian \( H_c(t) \), the total system Hamiltonian becomes
\begin{equation}
    H_S(t) = H_0 + H_c(t) = -\frac{\Delta}{2} \sigma_z + H_c(t),
\end{equation}
and we continue to assume, as before, that the system and environment begin in a factorized state at \( t = t_0 \).

The system propagator is expressed as
\begin{equation}
    U_S(t,t_0) = U_0(t - t_0)\, U_c(t,t_0),
    \label{Eq:US}
\end{equation}
where \( U_0(t - t_0) = e^{-i H_0 (t - t_0)} \) describes the free evolution, and \( U_c(t,t_0) \) captures the action of the time-dependent control field in the {\it rotating frame} defined by \( H_0 \). A Schrödinger-picture operator \( X_S(t) \) transforms into the  rotating frame as
\begin{equation}
    X_R(t) = U_0^\dagger(t - t_0)\, X_S(t)\, U_0(t - t_0),
    \label{Eq:transform}
\end{equation}
which allows the separation of fast free evolution from the control dynamics. The corresponding evolution equation for \( U_c(t,t_0) \) is
\begin{equation}
    i\frac{dU_c(t,t_0)}{dt} = H_{c,R}(t)\, U_c(t,t_0),
    \label{Eq:propagator}
\end{equation}
where \( H_{c,R}(t) \) is the control Hamiltonian in the rotating frame. To implement this control experimentally, the pulse sequence must be specified in the Schrödinger picture.

The system's interaction operator is taken to be
\begin{equation}
    A = \frac{1}{2} \left( \sigma_x \cos\phi + \sigma_y \sin\phi + \xi \sigma_z \right),
    \label{Eq:A0}
\end{equation}
where \( \xi \) determines the relative strength of longitudinal to transverse coupling. The corresponding interaction-picture operator is given by
\begin{equation}
    A(t) = U_S^\dagger(t, t_0)\, A\, U_S(t, t_0),
    \label{Eq:Ainteraction}
\end{equation}
as in Eq.~(\ref{Eq:US}).

\subsubsection{Spectral Density}

On the real frequency axis and at zero temperature, the time-dependent spectral density \( \Gamma(\omega, t) = J_\omega(t) + i S_\omega(t) \), derived from Eq.~(\ref{Eq:Gamma}), captures the environment’s dissipative response. This expression, earlier derived in Ref.~\cite{crowder2024invalidation}, is given by
\begin{equation}
\label{Eq:MarkovGamma}
\Gamma_\omega(t)=
\left\{
\begin{aligned}
&-2i\lambda^2\Gamma(s+1)\,\omega_c
\left(-\frac{\omega}{\omega_c}\right)^s e^{-\omega/\omega_c}
\\[-1pt]
&\qquad\times\Bigl[
\Gamma\!\left(-s,-\frac{\omega}{\omega_c}\right)
-\Gamma\!\left(-s,-\frac{\omega}{\omega_c}-i\omega t\right)
\Bigr]
\\[-1pt]
&\text{for }\omega\neq 0,\\[6pt]
&-2i\lambda^2\,\omega_c\,\Gamma(s)\,[\,1-(1+i\omega_c t)^{-s}\,]
\\[-1pt]
&\text{for }\omega=0.
\end{aligned}
\right.
\end{equation}

where \( \Gamma(-s, z) \) is the upper incomplete gamma function.\footnote{The spectral density is evaluated in the limit \( \omega + i0 \) to resolve the branch cut structure of the incomplete gamma function as implemented in \textsc{Matlab}.}

In the long-time limit \( t \to \infty \), the spectral density converges to its asymptotic (time-independent) form:
\[
\Gamma_\omega = J_\omega + i S_\omega, \quad \text{with} \quad
J_\omega = 2\pi\lambda^2 \frac{\omega^s}{\omega_c^{s-1}}\, \Theta(\omega)\, e^{-\omega/\omega_c},
\]
where \( \Theta(\omega) \) is the Heaviside step function. The BCF defined by Eq.~\ref{Eq:BCFdef}
is given by
\begin{equation}
    C(\tau)=\frac{1}{\pi}\int_0^\infty d\omega\,J_\omega[\coth\,\frac{\beta\omega}{2}\cos\,\omega\tau-i\sin\,\omega\tau],
\end{equation}
where $\beta=1/(k_BT)$ and $T$ is the temperature. At $\tau\gg 1/\omega_c$, $C(\tau)\sim t^{-1-s}.$

The exponent \( s \) classifies the spectral density as sub-Ohmic (\( s < 1 \)), Ohmic (\( s = 1 \)), or super-Ohmic (\( s > 1 \)). In the Ohmic case, \( \lambda^2 \) is known as the Kondo parameter. The Kosterlitz--Thouless-type localization transition occurs at zero temperature in the limit \( \omega_c \to \infty \), with \( s = 1 \) and critical coupling \( \lambda^2 = 1 \)~\cite{leggett1987dynamics}.

The late-time asymptotic behavior of the real part of the spectral density is given by
\begin{equation}
J_\omega(t) - J_\omega \sim 
\begin{cases}
t^{-1-s} e^{i\omega t}, & \text{if } \omega \neq 0, \\
t^{-s}, & \text{if } \omega = 0,
\end{cases}
\label{Eq:OhmicSDas}
\end{equation}
indicating that the decay at zero frequency—relevant for dephasing processes—is significantly slower than at nonzero frequencies. Consequently, the characteristic time scales governing relaxation and dephasing dynamics can differ substantially. A direct comparison of this time dependence at zero and nonzero frequencies will be presented in Fig.~\ref{Fig: relaxation_delay}(b).



In the asymptotic limit ($t_0 \to -\infty$) employed throughout this paper, we compute the dissipator directly in the time domain using Eq.~(\ref{Eq:asymptoticLambdaS}). In this regime, the system is assumed to be in thermal equilibrium prior to the gate, and all dependence on the initial state vanishes. The extension to finite temperatures is standard~\cite{breuer2002theory} and detailed in our previous work~\cite{lampert2025sixth}.

Eqs.~(\ref{Eq:lidarformSP}) and~(\ref{Eq:propagator}) are solved either analytically or, when necessary, numerically using a fourth-order Runge--Kutta method. All simulations are carried out in dimensionless units by setting \( \Delta = 1 \) and \( \hbar = 1 \).

\section{Analytical Results\label{Sec:dp}}

Starting from Eq.~\ref{Eq:lidarformSP}, the Bloch-Redfield master equation in the Schr\"odinger picture can be separated between the unitary component of the free system and the dissipative contribution:
\begin{equation}
    \frac{d\rho_S}{dt} = -i [ H_{S}(t), \rho_S] + \mathcal{D}_{BR}^S(t)\rho_S,  \label{Eq:generatordecompose}
\end{equation}
where  \( \mathcal{D}_{BR}^S(t) \) is a super-operator in the Schr\"odinger picture that we refer to as {\it dissipative generator}.
Using the tensor-product basis of Hilbert-Schmidt space in column-major order,
\begin{equation}
    \{\ket{0}\bra{0}, \ket{1}\bra{0}, \ket{0}\bra{1}, \ket{1}\bra{1}\},
\end{equation}
and the Eq.~\ref{Eq:lidarformSP}, the dissipative generator takes the form:
\begin{equation}
    \mathcal{D}_{BR}^S(t) = \Lambda^S(t)^* \otimes A + A^* \otimes \Lambda^S(t) - \mathbbm{I} \otimes A \Lambda^S(t) - A^* \Lambda^S(t)^* \otimes \mathbbm{I}.
    \label{Eq:DBloch-Redfield}
\end{equation}
where
$\Lambda^{S}(t)$ is 
the dissipator in the Schr\"odinger picture defined by Eqs.~(\ref{Eq:lambda}) and~(\ref{Eq:lambdaSP}), and $\star$ is the complex-conjugate.

To represent the dynamics in a real-valued basis, we apply a unitary transformation 
$U$, yielding the transformed generator \( \mathcal{D}_{BR}'(t) = U^\dagger \mathcal{D}_{BR}^S(t) U \), where:
\begin{equation}
    U = \frac{1}{\sqrt{2}} \begin{bmatrix}
        1 & 0 & 0 & 1\\
        0 & 1 & i & 0\\
        0 & 1 & -i & 0\\
        1 & 0 & 0 & -1
    \end{bmatrix}.
\end{equation}
In this basis, the density matrix \( \rho \) is vectorized as:
\begin{equation}
    \ket{\rho} = \frac{1}{2} [1, n_x, n_y, n_z]^\dagger,
\end{equation}
which corresponds to the standard Bloch sphere form:
\begin{equation}
    \rho = \frac{1}{2}(\mathbbm{I} + n_x\sigma_x + n_y\sigma_y + n_z\sigma_z).
\end{equation}
In this representation, the generator is real and the first row of generator is identically zero, reflecting the trace-preserving nature of the dynamics. The structural form of the Bloch-Redfield dissipative generator given in Eq.~\eqref{Eq:DBloch-Redfield} remains valid in the presence of a time-dependent system Hamiltonian, as discussed in earlier sections. 

\begin{figure}[ht]
    \centering
    \includegraphics[width=1\columnwidth]{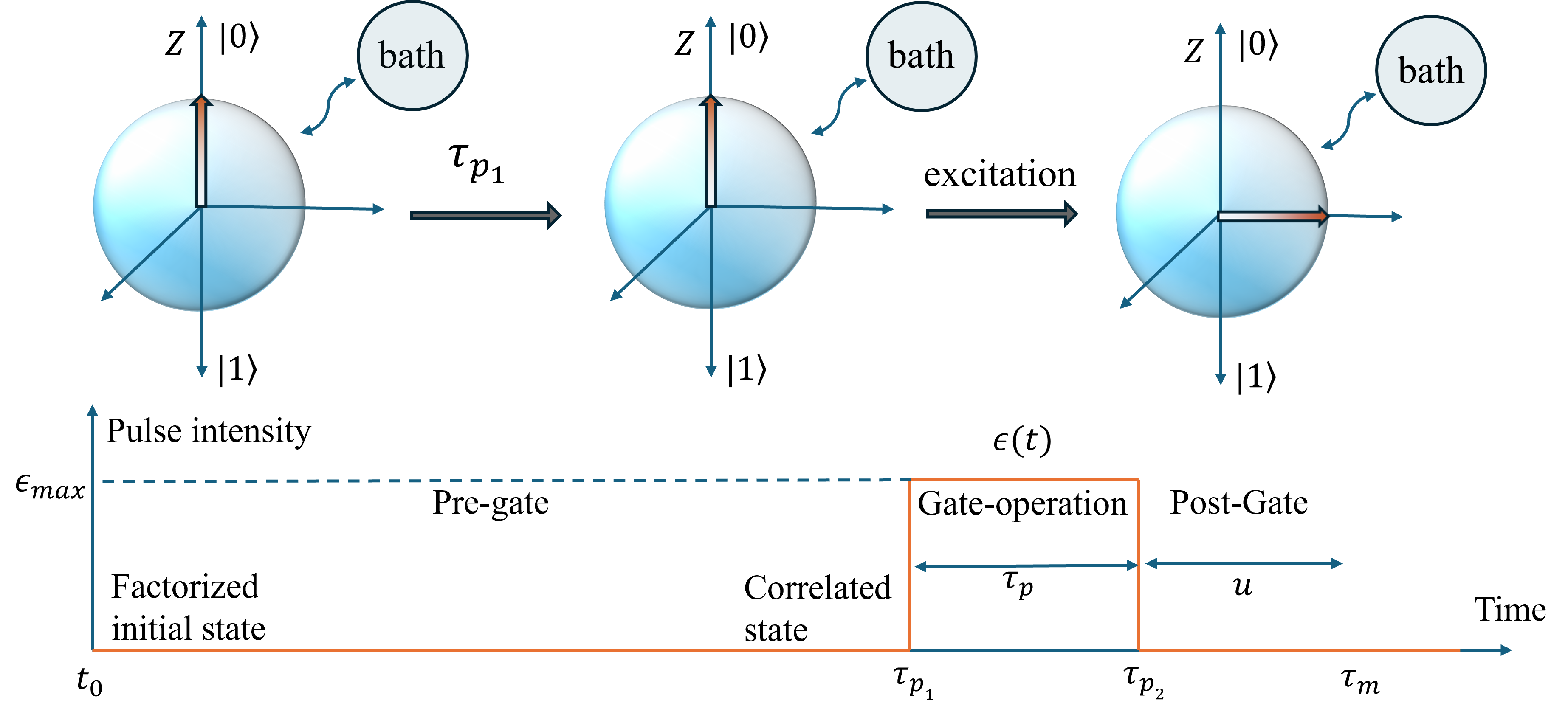}
    \captionsetup{justification=raggedright, singlelinecheck=false}
   \caption{Schematic of the time scales in both the dynamically prepared and static cases. \(\tau_{p_1}\) denotes the settling time for the system and bath. In the rotating frame, the pulse is rectangular with duration \(\tau_p\). In the static regime (Sec.~\ref{Sec:staticRegime}), where no pulse is applied, only a single time variable, \(\tau_m\), remains. In the strongly nonadiabatic regime (Sec.~\ref{Sec:Nonadiabatic}), \(\tau_{p_1} = \tau_{p_2}\) and \(\tau_p = 0\).}
    \label{Fig:pulse}
\end{figure}

In our specific case, the time dependence of the system Hamiltonian arises due to a control pulse applied to the qubit. In the rotating frame, this pulse appears as a rectangular modulation, as schematically illustrated in Fig.~\ref{Fig:pulse}.

Before turning to numerical simulations, we briefly examine two analytically tractable cases, presented in the next subsections.

\subsection{Static regime\label{Sec:staticRegime}}
To set the stage for post-gate dynamics, we first revisit dissipative dynamics under a time-independent Hamiltonian—the setting of standard Bloch-Redfield theory. 
The time-dependent dissipator is obtained using Eq.~(\ref{Eq:StaticNonMarkov}), e.g.,
\begin{equation}
\label{Eq:LambdaStaticMarkov}
\Lambda_{static}^S(t) = \int_0^{t-t_0} ds\, C(s)\, e^{-i H_0 s} A\, e^{i H_0 s}.
\end{equation}
Using the coupling operator defined in 
Eq.~(\ref{Eq:A0}), at $\phi=0$, and letting $t_0=0$, the Bloch-Redfield dissipative generator in the Schr\"odinger picture becomes, after some algebra:
\begin{equation}
\resizebox{\columnwidth}{!}{$
\displaystyle
\mathcal{D}_{BR}'(t)=\frac{1}{2}
\begin{bmatrix}
0 & 0 & 0 & 0\\
-[J_\Delta(t)-J_{-\Delta}(t)]\,\xi & -2J_0(t)\,\xi^2 & 0 & [J_\Delta(t)+J_{-\Delta}(t)]\,\xi\\
[\,S_\Delta(t)+S_{-\Delta}(t)-2S_0(t)\,]\,\xi & S_{-\Delta}(t)-S_\Delta(t) &
-2J_0(t)\,\xi^2 - J_\Delta(t) - J_{-\Delta}(t) &
[\,S_{-\Delta}(t)-S_\Delta(t)\,]\,\xi\\
J_\Delta(t)-J_{-\Delta}(t) & 2J_0(t)\,\xi & 0 & -J_\Delta(t)-J_{-\Delta}(t)
\end{bmatrix}
$}
\label{Eq:dcDBloch-RedfieldnonM}
\end{equation}

where \( J_\omega(t) = \text{Re}[\Gamma(\omega,t)] \) and \( S_\omega(t) = \text{Im}[\Gamma(\omega,t)] \). The second and third rows/columns correspond to transverse (coherence/dephasing) dynamics, while the fourth describes longitudinal (population/relaxation) dynamics.

Time-dependence in the spectral densities introduces memory effects and non-Markovianity. Even at zero temperature, $J_0(t)\neq 0$, indicating nonzero dephasing due to vacuum fluctuations. The relaxation rate $J_\Delta(t)+J_{-\Delta}(t)$ is not instantaneous, leading to initial slippage or delayed response. Nonsecular terms—such as \( \mathcal{D}_{42} \) - mediate coupling between coherence and population, converting off-diagonal elements into diagonal ones.

The well-known results for pure dephasing~\cite{breuer2002theory} are recovered by keeping only the leading-order (quadratic) terms in 
$\xi$, giving:
\begin{equation}
    \mathcal{D}_{BR,pd}'(t) = \begin{bmatrix}
        0 & 0 & 0 & 0\\ 
       0 & -J_0(t) \xi^2 & 0 & 0\\
       0 & 0 & -J_0(t) \xi^2 & 0\\
        0 & 0 & 0 & 0
    \end{bmatrix}.
    \label{Eq:dcDBloch-RedfieldnonMSD}
\end{equation}
To model dynamics, we include the unitary generator of free evolution:
\begin{equation}   
\mathcal{D}_f'=
    \begin{bmatrix}
        0 & 0 & 0 & 0\\ 
        0 & 0 & \Delta & 0\\
        0 & -\Delta & 0 & 0\\
        0 & 0 & 0 & 0
    \end{bmatrix}.
    \label{Eq:freeEvolution}
\end{equation}
Solving the resulting master equation yields transverse magnetization decaying as:
\begin{equation}
e^{-\xi^2\int_0^td\tau\,J_0(\tau)}.
\end{equation}
For an Ohmic bath using Eq.~\ref{Eq:OhmicSDas}, this becomes:
\begin{equation}(1+\omega_c^2t^2)^{-\xi^2\lambda^2},
\end{equation}
which predicts algebraic decay of coherence to zero. (For sub-Ohmic baths, the decay will be exponential.) Thus, dephasing persists even at zero temperature as a consequence of the factorized initial condition and vacuum fluctuations, which gradually generate system–bath correlations over long timescales.

{\it Markovian limit:} In the Markovian regime, we let $t_0\to -\infty$, and replace the time-dependent spectral functions by their asymptotic limits:
\[
J_\omega = \lim_{t \to \infty} J_\omega(t), \quad S_\omega = \lim_{t \to \infty} S_\omega(t).
\] At zero temperature, we have \( J_0 = J_{-\Delta} = 0 \), and the generator simplifies to:
\begingroup
\setlength{\arraycolsep}{2.5pt}      
\renewcommand{\arraystretch}{0.94}   
\begin{equation}
\mathcal{D}_{BR}'=\frac{1}{2}
\begin{bmatrix}
0 & 0 & 0 & 0\\
- J_\Delta\,\xi & 0 & 0 & J_\Delta\,\xi\\
\begin{aligned}[c]
&(S_\Delta + S_{-\Delta}\\[-1pt]
&\quad - 2S_0)\,\xi
\end{aligned}
& S_{-\Delta} - S_\Delta
& -J_\Delta
& (S_{-\Delta} - S_\Delta)\,\xi\\
J_\Delta & 0 & 0 & -J_\Delta
\end{bmatrix}
\label{Eq:dcDBloch-Redfield}
\end{equation}
\endgroup

In this limit, longitudinal (z-axis) relaxation decouples from coherence dynamics. The relaxation rate is:
\begin{equation}
T_1^{-1} = \frac{J_\Delta}{2}.
\end{equation}

The dephasing rate \( T_2^{-1} \) is found by diagonalizing the generator, which includes the free evolution term~\ref{Eq:freeEvolution}. The real parts of the eigenvalues yield \( T_2^{-1} = T_1^{-1}/2 \), indicating the absence of intrinsic dephasing in the Markovian zero-temperature case. The oscillation frequencies are slightly shifted due to damping effects:
\begin{equation}
\omega_{\pm} = \pm \Delta \sqrt{1 - \frac{S_{-\Delta} - S_\Delta}{4\Delta(1+\xi^2)} - \frac{J_\Delta^4}{64\Delta^2(1+\xi^2)^2}}.
\end{equation}

These analytic results provide a useful baseline for interpreting the role of non-Markovianity in gate-driven dynamics, which we now explore through explicit time-domain simulations.

\subsection{Dynamical State Preparation\label{Sec:Nonadiabatic}}

To investigate gate-induced non-Markovian effects, we now examine how dissipative dynamics are altered when the system is initialized via dynamical state preparation, as illustrated in Fig.~\ref{Fig:pulse}. This protocol is analyzed using the time-domain formulation in the asymptotic limit \( t_0 \to -\infty \), which eliminates unphysical memory of initial factorized conditions and captures bath correlations built up prior to the gate.

This approach avoids the sudden-onset approximation implicit in perturbative ``dressing'' arguments, offering a more accurate description of post-gate relaxation and dephasing. In particular, it may suppress the sharp transient (or jolt) commonly observed in physical observables when the system–bath interaction is switched on abruptly.

A related strategy was proposed in Ref.~\cite{fleming2011initial}, where the system–bath coupling is gradually turned on to reduce initial discontinuities. In contrast, our protocol allows the qubit to reach a near-ground state through equilibration with the bath before the gate is applied. A control pulse of duration \( \tau_p \) then excites the qubit into the superposition state \( \ket{\psi} = \frac{1}{\sqrt{2}}(\ket{0} - i\ket{1}) \), and the system's evolution is monitored at later times \( t \).

This preparation ensures minimal memory of the initial state at the onset of the gate, meaning that any non-Markovian behavior observed in the dynamics is genuinely induced by the gate operation itself. As a result, the protocol closely reflects experimental conditions and provides a robust framework for characterizing open-system dynamics under realistic control.

Starting from Eq.~(\ref{Eq:asymptoticLambdaS}), which assumes a factorized system–bath state in the infinite past, the dynamically prepared filtered coupling operator at time \( t > \tau_{p_2} \) can be naturally decomposed into three contributions corresponding to distinct phases of evolution:
\begin{equation}
\begin{aligned}
\Lambda_{dp}(t) &= \int_{-\infty}^{\tau_{p_1}} d\tau\, C(t - \tau)\, U_S(t,\tau)\, A\, U_S(\tau,t) \\
&\quad + \int_{\tau_{p_1}}^{\tau_{p_2}} d\tau\, C(t - \tau)\, U_S(t,\tau)\, A\, U_S(\tau,t) \\
&\quad + \int_{\tau_{p_2}}^{t} d\tau\, C(t - \tau)\, U_S(t,\tau)\, A\, U_S(\tau,t),
\end{aligned}
\label{Eq:PreOperPostGate}
\end{equation}
where \( \tau_{p_1} \) and \( \tau_{p_2} \) define the beginning and end of the gate pulse, respectively.

At times \( t > \tau_{p_2} \), the system propagator \( U_S(t, \tau) \) can be factorized based on the location of \( \tau \) relative to the gate window:
\begin{equation}
\label{Eq:splitU}
U_S(t,\tau) =
\begin{cases}
    U_S(t,\tau_{p_2})\, U_S(\tau_{p_2},\tau_{p_1})\, U_S(\tau_{p_1},\tau), & \tau < \tau_{p_1} \\
    U_S(t,\tau_{p_2})\, U_S(\tau_{p_2},\tau), & \tau \in (\tau_{p_1}, \tau_{p_2}) \\
    U_S(t,\tau), & \tau > \tau_{p_2}
\end{cases}
\end{equation}

For the pre-gate segment, this factorization yields
\begin{equation}
\label{Eq:firstInt}
\begin{aligned}
U_S(t,\tau) &=
e^{-iH_0(t-\tau_{p_2})}\, e^{-iH_0(\tau_{p_2}-\tau_{p_1})}
\\[-2pt]
&\quad \times U_c(\tau_{p_2},\tau_{p_1})\, e^{-iH_0(\tau_{p_1}-\tau)}
\\
&= e^{-iH_0(t-\tau_{p_1})}\, U_c\, e^{-iH_0(\tau_{p_1}-\tau)}.
\end{aligned}
\end{equation}

where \( H_0 \) is the free system Hamiltonian and \( U_c\equiv U_c(\tau_{p_2},\tau_{p_1}) \) represents the net effect of the gate in the rotating frame.

Substituting Eq.~(\ref{Eq:firstInt}) into the first integral in Eq.~(\ref{Eq:PreOperPostGate}) gives
\begin{equation}
\label{Eq:HistoryA1}
\begin{aligned}
&\int_{-\infty}^{\tau_{p_1}}\! d\tau\, C(t-\tau)\, U_S(t,\tau)\, A\, U_S(\tau,t)
\\
&= U_c(\tau_{p_1}-t)\!
\left[ \int_{-\infty}^{\tau_{p_1}}\! d\tau\, C(t-\tau)\, e^{-iH_0(t-\tau)} A\, e^{iH_0(t-\tau)} \right]\times
\\[-2pt]
&\quad U_c^\dagger(\tau_{p_1}-t).
\end{aligned}
\end{equation}

where we define the interaction-picture net rotation as
\begin{equation}
U_c(x) \equiv e^{iH_0 x} \, U_c \, e^{-iH_0 x}.
\end{equation}

Rewriting the integration range as
\[
\int_{-\infty}^{\tau_{p_1}} d\tau = \int_{-\infty}^{t} d\tau - \int_{\tau_{p_1}}^{t} d\tau,
\]
and applying Eqs.~(\ref{Eq:StaticMarkov}) and~(\ref{Eq:StaticNonMarkov}) allows us to simply express this contribution as:
\begin{equation}
\label{Eq:HistoryA2}
\begin{aligned}
&\int_{-\infty}^{\tau_{p_1}}\! d\tau\, C(t-\tau)\, U_S(t,\tau)\, A\, U_S(\tau,t)\\
&= U_c(\tau_{p_1}-t)\,\left[ \Lambda^{SM} - \Lambda_{\text{static}}(t-\tau_{p_1}) \right]\times\\[-2pt]
&\quad U_c^\dagger(\tau_{p_1}-t).
\end{aligned}
\end{equation}

Similarly, the third integral in Eq.~(\ref{Eq:PreOperPostGate}) corresponds to the post-gate regime and yields the static dissipator \( \Lambda_{\text{static}}^S(t - \tau_{p_2}) \) as in Eq.~(\ref{Eq:StaticNonMarkov}).

Substituting these results into Eq.~(\ref{Eq:PreOperPostGate}) gives the full time-dependent dissipator:
\begin{equation}
\label{Eq:PGP}
\begin{aligned}
\Lambda_{dp}(t) &= \Lambda_{\text{static}}(t - \tau_{p_2})  \\
&\quad + \int_{\tau_{p_1}}^{\tau_{p_2}} d\tau\, C(t - \tau)\, U_S(t,\tau)\, A\, U_S(\tau,t)  \\
&\quad + U_c( \tau_{p_1}-t)\left[\Lambda^{SM} - \Lambda_{\text{static}}(t - \tau_{p_1})\right] U_c^\dagger(\tau_{p_1}-t).
\end{aligned}
\end{equation}

This decomposition cleanly separates memory contributions into post-gate, intra-gate, and pre-gate intervals, illustrating how the gate pulse modulates dissipative dynamics by reshaping system–bath correlations. Specifically, the first term reflects the dissipator one would obtain if the system and bath were factorized at the end of the gate; the second term accounts for the correlations generated during the pulse; and the third term captures how preexisting environmental memory is rotated by the gate and carried forward into the subsequent evolution.

\subsubsection{Instantaneous Gate Approximation\label{Sec:InstaGate}}

To clarify the interplay between pre- and post-gate reduced dynamics in the simplest setting, we consider an idealized regime where the gate pulse is infinitely strong and instantaneous. In this limit, $\tau_{p_1} = \tau_{p_2}$, and the system–bath interaction during the gate can be neglected entirely, yielding our second analytic solution.

The gate is now modeled as an instantaneous unitary transformation \( U_c \), introducing a delta-function-like discontinuity into the otherwise continuous, non-Markovian evolution. This sharp unitary action partitions the dynamics at a point we refer to as a \textit{dissipative boundary}, across which quantum coherence is perfectly preserved, while environmental memory is momentarily frozen. By solving the corresponding master equation, we explore how non-Markovian dynamics responds to such coherent disruption—probing whether system–bath correlations are transferred, reshaped, or erased across the boundary. This analysis reveals fundamental structural features of open-system evolution under strong, localized control.

Importantly, this regime is not merely theoretical. It is realized experimentally in ultrafast optical spectroscopy, where femtosecond pulses excite excitonic systems such as the Fenna–Matthews–Olson (FMO) complex~\cite{engel2007evidence}, light-harvesting complexes~\cite{romero2014quantum}, and chlorosomes in green sulfur bacteria~\cite{cheng2023quantum}. In these systems, the pulse duration is much shorter than both the Rabi period set by exciton splittings and the bath correlation time, making the instantaneous gate approximation physically meaningful.

Neglecting the gate-operation term in Eq.~(\ref{Eq:PGP}) and setting the time origin at \( \tau_{p_{2}} \), we arrive at a simple dissipator in the Schr\"odinger picture:
\begin{equation}
\Lambda_{dp}^S(t) = U_c(-t)\left[\Lambda^{SM} - \Lambda_{\text{static}}(t)\right] U_c^\dagger(-t) + \Lambda_{\text{static}}(t).
\label{Eq:finalLambdaDP}
\end{equation}
The unitary gate operation first acts in an \textit{antichronological} (reverse-time) fashion via \( U_c^\dagger(-t) \), followed by the residual pre-gate dissipator  \( \Lambda^{SM} - \Lambda_{\text{static}}(t) \), and finally a \textit{chronological} (forward-time) application of \( U_c(-t) \).

The last equation refines the naive decomposition of the Markovian dissipator in the Schrödinger picture,
\begin{equation}
\Lambda^{SM} = \left[\Lambda^{SM} - \Lambda_{\text{static}}^S(t)\right] + \Lambda_{\text{static}}^S(t),
\label{Eq:decompSimplistic}
\end{equation}
by incorporating the effect of the instantaneous gate unitary \( U_c \) into the pre-gate component. This pre-gate term is ``dressed'' by the gate through a time-symmetric application of the unitary: it is conjugated by \( U_c(-t) \) and \( U_c^\dagger(-t) \), reflecting forward and backward evolution, respectively. Meanwhile, the post-gate dissipative term is modeled as a static contribution with factorized initial conditions, remaining unaffected by the unitary operation.

Both the dressed pre-gate and the static post-gate components evolve on a characteristic time scale determined solely by the bath properties, independent of the system-bath coupling strength. It is the timescale set by 
$1/\omega_c$
  that governs the temporal window following the gate application during which nontrivial non-Markovian effects can emerge.

This structure is consistent with non-Markovian or memory-aware frameworks such as the \textit{Keldysh non-equilibrium Green’s function formalism}~\cite{nazarov2009}, and the \textit{process tensor approach}~\cite{pollock2018, schmid2018}, both of which permit dissipation to reflect temporal influence in both forward and backward directions. However, our approach is considerably simpler and, in this minimal setting of an intervening gate, offers a more direct and illuminating perspective on the time-symmetric disruption of correlated system-bath dynamics. The system's post-gate past interacts bidirectionally with its pre-gate past; however, the overall dynamics remains causal in the sense that the control pulse influences the system only from the moment it is applied onward, without retroactively affecting earlier times.

Importantly, \( U_c(-t) \) is defined in the interaction picture at time \( -t \), while the dissipator in Eq.~\ref{Eq:finalLambdaDP} is evaluated in the Schr\"odinger picture. As a result, the Schr\"odinger picture dissipator exhibits oscillations at the system Bohr frequencies---similar to, though not identical to, those in the interaction picture.

Because of this structure, the usual strategy of averaging out free oscillations in the interaction picture, as in Davies’ approach~\cite{davies1974markovian}, does not lead to the decoupling between population and coherence dynamics. 
The emergence of nonsecular dissipative evolution in the reduced state signifies that coherences between energy eigenstates persist and actively influence the system's dynamics, leading to richer and more intricate post-gate evolution patterns, as will be shown next.

\subsubsection{Coherence Dynamics and Recovery}\label{Section: Coherence Dynamics and Recovery}

To analyze the system's evolution after the gate operation, we solve the master equation [Eq.~(\ref{Eq:lidarformSP})] for times \( t > 0 \). During this post-gate regime, the system Hamiltonian remains static, \( H_S(t) = H_0 \), while the influence of both prior correlations and the pulse is fully encoded in the time-dependent operator \( \Lambda_{dp}^S(t) \) in Eq.~(\ref{Eq:finalLambdaDP}). The initial condition is given by \( \rho(0) = U_c \rho_{\text{as}} U_c^\dagger \), where \( \rho_{\text{as}} \) denotes the asymptotic (pre-gate equilibrium) state.

\begin{figure}[ht]
    \centering
    \includegraphics[width=1\columnwidth]{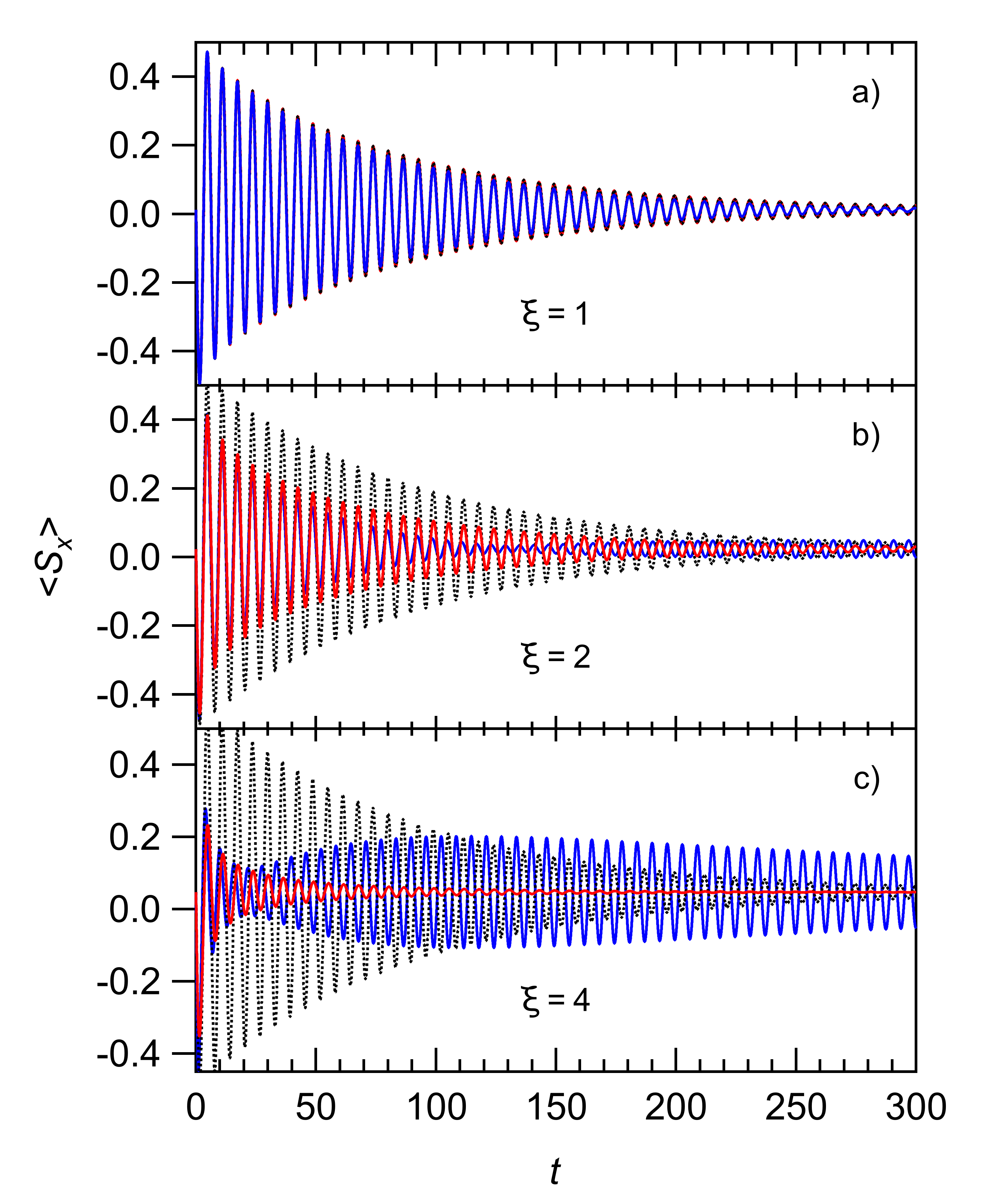}
    \captionsetup{justification=raggedright, singlelinecheck=false}
    \caption{
    X-component of the magnetization as a function of time for Markovian and non-Markovian dynamics. The initial state has the magnetization in the y-direction. The dephasing strength, characterized by $\xi$, increases from (a) to (c), while all other parameters remain fixed. Dotted black line: Markovian dynamics with  decoherence rate equal to half the relaxation rate. Red solid line: non-Markovian dynamics with a factorized initial state. Blue solid line: non-Markovian dynamics with a dynamically prepared initial state under the instantaneous gate approximation. Parameters: $\lambda^2 = 0.02$, $\phi = 0$, $\Delta = 1$, and $\omega_c = 1$.}
    \label{Fig:CoherenceInstant}
\end{figure}

In the following example, we analyze the dynamics of the transverse magnetization component, \(\langle s_x \rangle = \text{Tr}[\rho(t)s_x] = \text{Re}\,\rho_{12}(t)\), at zero temperature, $U_c=\exp(-i\pi\sigma_x/4)$ (rotation by $\pi/2$ around x-axis), with results shown in Fig.~\ref{Fig:CoherenceInstant}. As the coupling parameter \(\xi\) increases across the panels, the effective system--bath interaction strength, \(\lambda^2\xi^2\), takes on the values 0.02, 0.08, and 0.32, progressively pushing the system toward the strong-dephasing regime. However, in the Markovian regime (\(t \to \infty\)), the decoherence rate becomes independent of \(\xi\), as pure dephasing is absent at zero temperature. Consequently, the Markovian coherence dynamics—represented by the black dashed line—remains unaffected by variations in \(\xi\).

At weak coupling [Fig.~\ref{Fig:CoherenceInstant}(a)], coherence dynamics exhibit only minor differences across the three initialization regimes. At intermediate coupling [Fig.~\ref{Fig:CoherenceInstant}(b)], additional dephasing becomes significant for both factorized and dynamically prepared initializations, as evidenced by the pronounced contrast between the Markovian (dashed black) and non-Markovian (red and blue) cases. Notably, the initial loss of phase coherence is nearly identical for the factorized and dynamically prepared states, suggesting that vacuum fluctuations contribute similarly to early-time dephasing. However, their long-time behaviors diverge markedly, reflecting distinct memory effects and bath correlations retained in the non-Markovian dynamics.

At the strongest coupling, the qubit exhibits rapid decoherence in both non-Markovian regimes. However, the dynamically prepared state exhibits a subsequent coherence recovery, as illustrated by the blue curve in Fig.~\ref{Fig:CoherenceInstant}(c). Following this recovery, the transverse magnetization decays to zero (not shown), but at a significantly reduced rate. This long-lived coherence represents the first key result of the paper.

Such behavior contrasts with the expectations for open quantum systems, where coherence typically decays on timescales set by characteristic decoherence rates. For relaxation-limited systems, this timescale is generally bounded by $2T_1$. In dynamically prepared states, however, coherence can persist well beyond this limit due to a nontrivial interplay between pre-gate system--environment correlations and post-gate quantum dynamics.

The central mechanism enabling this effect is the \emph{unitary dressing of the pre-gate component of the post-gate dissipator}.
 When a coherent control operation $U_c(t)$ is applied—particularly in a near-instantaneous fashion—it effectively rotates the basis in which the system experiences environmental decoherence. This modifies the dissipative dynamics according to
\begin{equation}
    \Lambda^{SM} - \Lambda_{\text{static}}^S(t) \longmapsto U_c(-t) \left[ \Lambda^{SM} - \Lambda_{\text{static}}^S(t) \right] U_c^\dagger(-t),
    \label{Eq:UnitaryDressing}
\end{equation}
 This unitary transformation redistributes environmental noise across different quadratures of the Bloch sphere, transiently mitigating the effects of dephasing and enabling a partial recovery or persistence of coherence.

Such coherence revival is inherently dynamical and cannot be explained within the traditional secular approximation, which assumes decoupled evolution of populations and coherences~\cite{davies1974markovian}. Here, \emph{non-secular transitions}—specifically, transfers from population degrees of freedom to off-diagonal coherence—play a critical role. These transfers are driven by time-dependent control and pre-existing bath correlations, and are absent in static dissipation models. Capturing this requires analyzing post-gate evolution from dynamically prepared states, where system–bath correlations exist at the gate time.

\subsubsection{Coarse-graining the post-gate dynamics}

Recalling Eq.~(\ref{Eq:finalLambdaDP}), the interaction picture dissipator reads
\begin{equation}
\Lambda_{dp}^I(t) = U_c\left[\Lambda^{IM}(t) - \Lambda_{\text{static}}^I(t)\right] U_c^\dagger + \Lambda_{\text{static}}^I(t).
\label{Eq:LdpIP}
\end{equation}
From here, we compute the dissipative generator in the interaction picture. Specifically, we employ Eq.~\ref{Eq:DBloch-Redfield}, where the operator \( A \) is replaced by its interaction picture counterpart \( A(t) = e^{i H_0 t} A e^{-i H_0 t} \), and the dissipator is given by Eq.~\ref{Eq:LdpIP}. The post-gate generator naturally decomposes into a sum of pre- and post-gate contributions, corresponding to the structure of the splitting in Eq.~\ref{Eq:LdpIP}.

Finally, we apply coarse-graining to the post-gate generator in the spirit of Davies~\cite{davies1974markovian}, discarding rapidly oscillating terms \( \exp(\pm i\Delta t) \) in the interaction picture.

The resulting static component of the generator is then given by
\begin{equation}
\label{Eq:standardDSP}
\resizebox{\columnwidth}{!}{$\displaystyle
\mathcal{D}_{\text{static}}^{I,\text{cg}}(t)
= \frac{1}{2}
\begin{pmatrix}
0 & 0 & 0 & 0 \\
0 & -2\xi^2 J_0(t) - \tfrac{1}{2}\bigl(J_{\Delta}(t)+J_{-\Delta}(t)\bigr) & \tfrac{1}{2}\bigl(S_{\Delta}(t)-S_{-\Delta}(t)\bigr) & 0 \\
0 & \tfrac{1}{2}\bigl(S_{-\Delta}(t)-S_{\Delta}(t)\bigr) & -2\xi^2 J_0(t) - \tfrac{1}{2}\bigl(J_{\Delta}(t)+J_{-\Delta}(t)\bigr) & 0 \\
- J_{-\Delta}(t) + J_{\Delta}(t) & 0 & 0 & - J_{-\Delta}(t) - J_{\Delta}(t)
\end{pmatrix}
$}.
\end{equation}

This expression coincides with the dissipative generator of the Davies master equation~\cite{davies1974markovian}, except that we retain the full time dependence of the spectral density rather than replacing it with its asymptotic limit, as we are specifically interested in effects arising from slow baths.

In contrast, the unitary 
dressing in Eq.~(\ref{Eq:UnitaryDressing})
  mixes populations and coherences within the pre-gate contribution to the post-gate dissipator. Thus, the removal of oscillating terms in the interaction picture yields a generator that retains non-secular matrix elements, in contrast to the Davies generator. For a \(\pi/2\) \(X\)-rotation, the dissipative generator associated with the pre-gate term is:

\begin{equation}
\label{Eq:Nonsecular}
\resizebox{\columnwidth}{!}{$
\displaystyle
\mathcal{D}_{\text{pre-gate}}^{I,\text{cg}}(t)=\frac{1}{2}
\begin{pmatrix}
0 & 0 & 0 & 0 \\
2\xi^2 \delta S_0(t) + \frac{1}{2}(\delta S_{-\Delta}(t)+ \delta S_\Delta(t)) & 0 & \frac{1}{2}(\delta S_{-\Delta}(t)+ \delta S_\Delta(t)) & 0 \\
\frac{1}{2}(\delta J_{-\Delta}(t)- \delta J_\Delta(t)) & 0 &  -\frac{1}{2}(\delta J_{-\Delta}(t)+\delta J_\Delta(t)) & -2\xi^2\delta J_0(t) \\
-\frac{1}{2}(\delta J_{-\Delta}(t)- \delta J_{\Delta}(t)) & 0 & 0 & -\frac{1}{2}(\delta J_{-\Delta}(t)+\delta J_{\Delta}(t))
\end{pmatrix}
$}
\end{equation}

where \(\delta J_\omega(t)\) and \(\delta S_\omega(t)\) are defined as:

\begin{equation}
\begin{aligned}
    &\delta J_\omega(t) = J_\omega - J_\omega(t), \\
    & \delta S_\omega(t) = S_\omega - S_\omega(t).
\end{aligned}
\end{equation}

Additional data is provided in Appendix~\ref{appendix:cg}, where we demonstrate that the coherence dynamics computed using Eqs.~(\ref{Eq:standardDSP}) and~(\ref{Eq:Nonsecular}) closely match those shown in Fig.~\ref{Fig:CoherenceInstant}. Since the generator~(\ref{Eq:standardDSP}) decouples population and coherence dynamics, preventing any possibility of long-lived coherence, this indicates that the root cause of this effect lies in mechanisms captured by Eq.~(\ref{Eq:Nonsecular}).

In strongly dephasing systems, where \( \xi \gg 1 \), the dephasing rate given by \( \xi^2 \delta J_0(t) \) dominates over the remaining dissipative rates \( J_{\pm \Delta}(t) \). In the post-gate contribution described by Eq.~(\ref{Eq:standardDSP}), the dephasing is confined to the coherence-to-coherence block, leading to the expected rapid loss of phase coherence.

In contrast, the pre-gate contribution~(\ref{Eq:Nonsecular}) undergoes a unitary transformation that eliminates the dominant dephasing from the coherence-to-coherence block and relocates it to the population-to-coherence block of the generator, specifically in matrix element \( (3,4) \).
 As a result, while the post-gate component induces dephasing, the pre-gate component actively transfers population into coherence at a comparable rate. This dynamic balance stabilizes coherence at a relatively high level, as shown in Fig.\ref{Fig:CoherenceInstant}(c). Further details are provided in Appendix~\ref{appendix:cg}.

Moreover, at zero temperature, the dephasing rate decays very slowly. It is crucial to note that \( J_0(t) \) vanishes much more gradually than the saturation of \( J_\Delta(t) \), as shown in Eq.~\ref{Eq:OhmicSDas} and illustrated in Fig.~\ref{Fig: relaxation_delay}(b). When the coupling strength \( (\lambda \xi)^2 \) approaches unity, the dephasing matrix element \( 2\xi^2 J_0(t) \) continues to pump coherence from the population even after the system has nearly relaxed to its ground state. This mechanism enables prolonged population-to-coherence transfer and supports long-lived coherence. In the large-\( \xi \) regime, coherence persists well beyond the standard decoherence timescale, provided the bath relaxes slowly.

While the present analysis captures the essential structure of coherence dynamics under ultrafast control, a complete treatment of population dynamics requires a fourth-order master equation. This extension is necessary to properly account for the support that population dynamics provide to coherence—specifically, to ensure compliance with the Cauchy–Schwarz inequality relating coherence to population, thereby preserving the positivity of the density matrix—which lies beyond the capabilities of the standard Bloch–Redfield approach. A detailed exploration of these higher-order master equations is beyond the scope of this work and will be pursued in future studies. Our prior research provides a promising foundation for this direction~\cite{crowder2024invalidation}.

\subsubsection{Temperature dependence of long-lived coherences}

\begin{figure}[htbp]
    \centering
    \includegraphics[width=1\columnwidth]{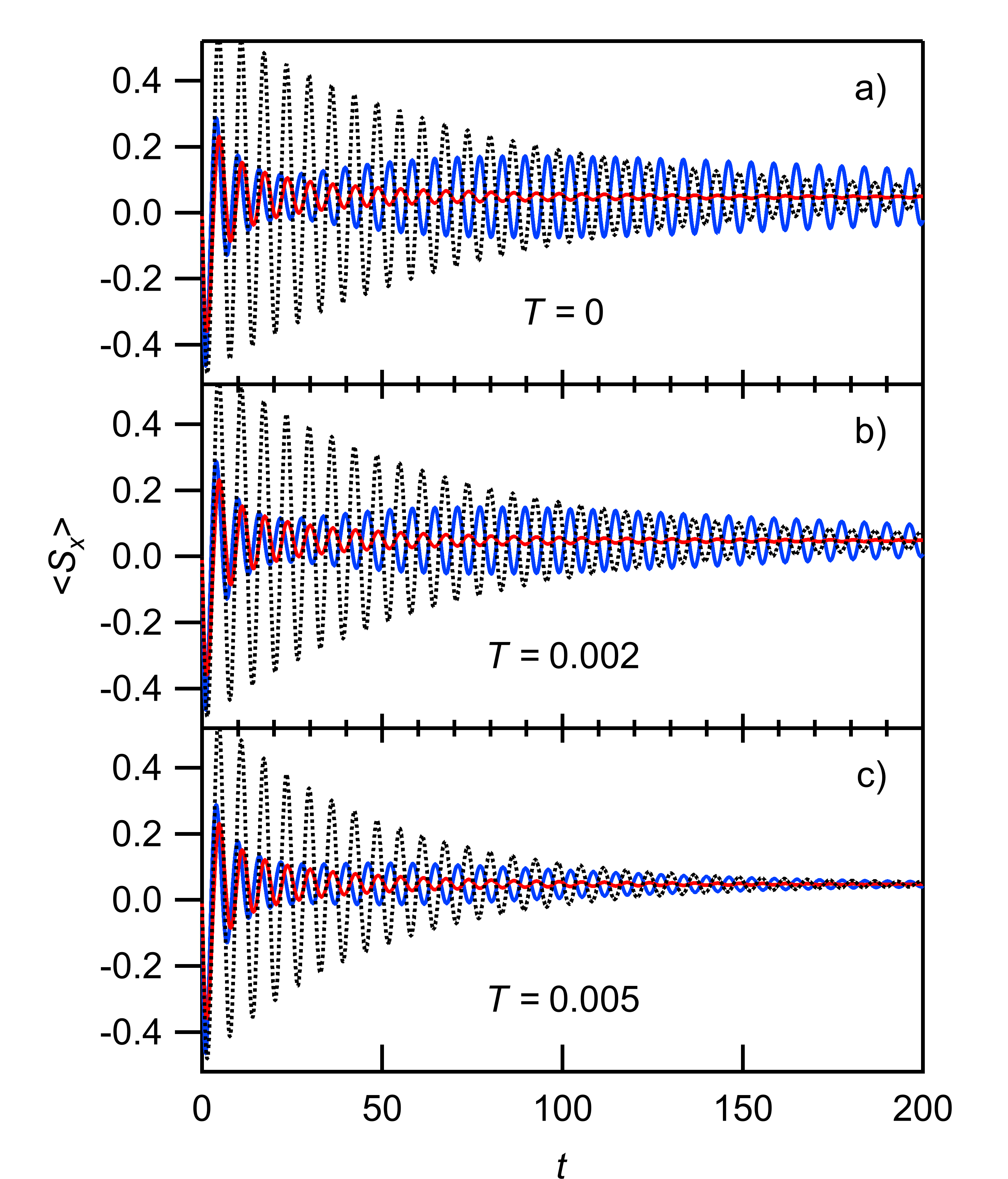}
    \captionsetup{justification=raggedright, singlelinecheck=false}
    \caption{X-component of magnetization as a function of time for Markovian and non-Markovian dynamics under the instantaneous gate approximation. Initial magnetization is along the \(y\)-axis. Temperature \(T\) increases from (a) to (c). Dotted black: Markovian dynamics, with decoherence rate set to half the relaxation rate. Red: non-Markovian dynamics with factorized initial state. Blue: non-Markovian dynamics with dynamically prepared initial state. Recovery disappears with increasing \(T\). Parameters: \(\lambda^2 = 0.02\), \(\phi = 0\), \(\Delta = 1\), \(\xi = 4\), \(\omega_c = 1\).}
    \label{Fig: Temperature}
\end{figure}

At finite temperature, these effects are suppressed. For the Ohmic bath, the coherence recovery disappears rapidly with increasing temperature, as shown in Fig.~\ref{Fig: Temperature}. Therefore, the mechanism cannot account for the observed coherence in FMO if the environment is Ohmic. The case of sub-Ohmic baths, where the recovery may survive at higher temperature, is discussed in Sec.~\ref{Sec:FMO}.

\noindent

In conclusion, dynamically controlled open quantum systems can exhibit coherence effects that exceed the limitations of static noise models. This opens new possibilities for coherence preservation and engineered noise control in near-term quantum devices. 

\section{Finite-Duration Quantum Gates\label{Sec:Finite-Duration}}

We now turn to the analysis of quantum gates with finite pulse duration. In contrast to the idealized instantaneous gate considered earlier, realistic gate operations necessarily involve pulses of finite length. Understanding the post-gate dynamics in this regime is essential for the design and optimization of coherent control protocols in practical quantum systems.

 In order to investigate the post-gate dynamics for finite pulse lengths, it is crucial to evaluate the second integral in Eq.~(\ref{Eq:PreOperPostGate}):
\begin{equation}
    \Lambda_{\text{dp-pulse}}(t) = \int_{\tau_{p_1}}^{\tau_{p_2}} d\tau \, C(t-\tau)\, U_S(t,\tau)\, A \, U_S(\tau,t).
\end{equation}

To proceed, it is essential to examine the system rotation occurring during the pulse. The time propagator for \( \tau_{p_1} < t < \tau_{p_2} \) is characterized as follows:

\begin{equation}
\begin{aligned}
    U_S(t,\tau_{p_1}) &= e^{-iH_0(t-\tau_{p_1})}\,e^{-i\frac{t-\tau_{p_1}}{\tau_{p_2}-\tau_{p_1}}\frac{\theta}{2}\sigma_x}
    \\ & = e^{-iH_0(t-\tau_{p_1})}\,e^{-i(t-\tau_{p_1})\frac{\omega_p}{2}\sigma_x}
    \label{Eq:SpRotation}
\end{aligned}
\end{equation}
where we define the {\it gate operating frequency:}

\begin{equation}
    \omega_p = \frac{\theta}{\tau_{p_2}-\tau_{p_1}}.
\end{equation}
From this point onward, $\omega_p$ is assumed constant over time. The instantaneous and the adiabatic limit correspond to  $\omega_p \rightarrow \infty$ and $\omega_p \rightarrow 0$, respectively. 

Within this framework, we partition the qubit dynamics into two distinct regimes: the dynamics during the gate operation, and the evolution following the gate. The evolution prior to the pulse is treated as the steady-state behavior of the open system under the influence of its environment, ensuring that any history effects observed after the gate are entirely induced by the control operation itself.

\subsubsection{Dissipator Within Gate}
Similar with Eq.~(\ref{Eq:PreOperPostGate}), when \(\tau_{p_1}<t<\tau_{p_2}\) the dissipator \(\Lambda_{\text{gate}}\) is:
\begin{equation}
\begin{aligned}
     \Lambda_{\text{gate}}(t)&=\int_{-\infty}^{\tau_{p_1}} d\tau\, C(t - \tau)\, U_S(t,\tau)\, A\, U_S(\tau,t) \\
&\quad + \int_{\tau_{p_1}}^{t} d\tau\, C(t - \tau)\, U_S(t,\tau)\, A\, U_S(\tau,t)
\end{aligned} 
\end{equation}
Recalling the pre-gate dissipator in Eq.~(\ref{Eq:HistoryA2}), the dissipator within the gate can be expressed as:
\begin{equation}
\begin{aligned}
     \Lambda_{\text{gate}}(t)&=U_r(t-\tau_{p_1})\left[\Lambda^{\text{SM}} - \Lambda_{\text{static}}(t - \tau_{p_1})\right] U_r^\dagger(t-\tau_{p_1})\\
&\quad + \int_{\tau_{p_1}}^{t} d\tau\, C(t - \tau)\, U_S(t,\tau)\, A\, U_S(\tau,t).
\label{Eq: dissipator in gate}
\end{aligned} 
\end{equation}
where 
\begin{equation}
    U_r(t) = e^{-iH_0t}\,e^{-i\frac{\omega_p}{2}\sigma_x t}\,e^{iH_0t} .
\end{equation}

The second integral involves the propagator \(U_S(t,\tau)\) for \(\tau_{p_1}<\tau<t<\tau_{p_2}\), which can be evaluated via:

\begin{equation}
\label{Eq:US_factor}
\begin{aligned}
&U_S(t,\tau) = U_S(t,\tau_{p_1})\,U_S(\tau_{p_1},\tau)\\[-2pt]
&= e^{-iH_0(t-\tau_{p_1})} e^{-i\frac{\omega_p}{2}\sigma_x (t-\tau)} e^{iH_0(\tau-\tau_{p_1})}.
\end{aligned}
\end{equation}

where we set the time origin to be \(\tau_{p_1}\). Assuming $\phi = 0$ for simplicity and after some algebra, we arrive at an analytic result expressed in terms of the time-dependent spectral densities evaluated at nine distinct frequencies: $-\Delta - \omega_p$, $-\Delta$, $-\Delta + \omega_p$, $-\omega_p$, $0$, $\omega_p$, $\Delta - \omega_p$, $\Delta$, and $\Delta + \omega_p$:
\begin{equation}
\label{Eq: full_dissipator_in_gate}
\begin{alignedat}[t]{1}
&\Lambda_{\text{gate}}(t)
= U_r(t)\,[\Lambda^{SM}-\Lambda_{\text{static}}(t)]\,U_r^\dagger(t)
\\[-2pt]
&\quad + \tfrac{1}{4}\sigma_z\biggl(
   \xi\bigl(\Gamma_{\omega_p}(t)+\Gamma_{-\omega_p}(t)\bigr)
\\[-2pt]
&\qquad\qquad\qquad
   - \tfrac{1}{2}\biggl(
      e^{i\Delta t}\bigl(\Gamma_{-\Delta+\omega_p}(t)-\Gamma_{-\Delta-\omega_p}(t)\bigr)
\\[-2pt]
&\qquad\qquad\qquad\qquad\qquad
      - e^{-i\Delta t}\bigl(\Gamma_{\Delta+\omega_p}(t)-\Gamma_{\Delta-\omega_p}(t)\bigr)
   \biggr)\biggr)
\\[-2pt]
&\quad + \tfrac{1}{4}\sigma_+\biggl(
    \Gamma_\Delta(t)
    + \tfrac{1}{2}\bigl(\Gamma_{\Delta+\omega_p}(t)+\Gamma_{\Delta-\omega_p}(t)\bigr)
\\[-2pt]
&\qquad\qquad\qquad
    + e^{2i\Delta t}\bigl(
       \Gamma_{-\Delta}(t)
       - \tfrac{1}{2}\bigl(\Gamma_{-\Delta+\omega_p}(t)+\Gamma_{-\Delta-\omega_p}(t)\bigr)
      \bigr)
\\[-2pt]
&\qquad\qquad\qquad
    + e^{i\Delta t}\bigl(\Gamma_{\omega_p}(t)-\Gamma_{-\omega_p}(t)\bigr)
  \biggr)
\\[-2pt]
&\quad + \tfrac{1}{4}\sigma_-\biggl(
    \Gamma_{-\Delta}(t)
    + \tfrac{1}{2}\bigl(\Gamma_{-\Delta+\omega_p}(t)+\Gamma_{-\Delta-\omega_p}(t)\bigr)
\\[-2pt]
&\qquad\qquad\qquad
    + e^{-2i\Delta t}\bigl(
       \Gamma_{\Delta}(t)
       - \tfrac{1}{2}\bigl(\Gamma_{\Delta+\omega_p}(t)+\Gamma_{\Delta-\omega_p}(t)\bigr)
      \bigr)
\\[-2pt]
&\qquad\qquad\qquad
    - e^{-i\Delta t}\bigl(\Gamma_{\omega_p}(t)-\Gamma_{-\omega_p}(t)\bigr)
  \biggr).
\end{alignedat}
\end{equation}

where $\Gamma_{\omega}(t)=\int_0^{t} e^{-i\omega s} C(s)ds $. It can be shown that this reproduces the result in Eq.~\ref{Eq:finalLambdaDP} in the instantaneous gate limit, $\omega_p \to \infty$, and reduces to the adiabatic regime in the limit $\omega_p \to 0$.

\subsubsection{Post Gate}
In case of \(\tau_{p_1} < \tau < \tau_{p_2} <t\), the chronological propagator \(U_S(t,\tau)\) is:

\begin{equation}
    \begin{aligned}
    U_S(t,\tau) & = U_S(t,\tau_{p_2})\,U_S(\tau_{p_2},\tau)\\
    & =  e^{-iH_0(t-\tau_{p_1})}\,e^{-i\frac{\omega_p}{2}\sigma_x (\tau_{p_2}-\tau)}\,e^{iH_0(\tau-\tau_{p_1})}
    \end{aligned}
\end{equation}

The second integral in Eq.~(\ref{Eq:PreOperPostGate}) is:
\begin{equation}
    \Lambda_{dp-pulse}(t) = \int_{\tau_{p_1}}^{\tau_{p_2}}d \tau \, C(t-\tau)\,U_S(t,\tau)\, A \, U_S(\tau,t).
\end{equation}

With some algebra:

\begin{equation}
    \Lambda_{dp-pulse}(t) = C_z \sigma_z + C_+\sigma_++C_-\sigma_-
    \label{Eq:pulse integral}
\end{equation}
Where:
\begin{equation}
\begin{alignedat}{1}
&C_z
= \frac{\xi}{4}\Bigl(
     e^{i\Phi_t}\,\delta\Gamma_{\omega_p}
\\[-2pt]
&\qquad\qquad\quad
   + e^{-i\Phi_t}\,\delta\Gamma_{-\omega_p}
  \Bigr)
\\[-2pt]
&\quad - \frac{1}{8}\Bigl(
     e^{i(\Phi_t+\Delta(t-\tau_{p_1}))}\,\delta\Gamma_{\Delta+\omega_p}
\\[-2pt]
&\qquad\qquad\quad
   + e^{-i(\Phi_t+\Delta(t-\tau_{p_1}))}\,\delta\Gamma_{-\Delta-\omega_p}
  \Bigr)
\\[-2pt]
&\quad + \frac{1}{8}\Bigl(
     e^{i(\Phi_t-\Delta(t-\tau_{p_1}))}\,\delta\Gamma_{-\Delta+\omega_p}
\\[-2pt]
&\qquad\qquad\quad
   + e^{-i(\Phi_t-\Delta(t-\tau_{p_1}))}\,\delta\Gamma_{\Delta-\omega_p}
  \Bigr).
\end{alignedat}
\end{equation}

\begin{equation}
\begin{aligned}
    C_+ & = \frac{1}{4}(e^{2i\Delta (t-\tau_{p_1})} \delta\Gamma_{-\Delta}+\delta\Gamma_{\Delta})\\
    & \quad -\frac{e^{2i\Delta(t-\tau_{p_1})}}{8}(e^{i\Phi_t}\delta\Gamma_{-\Delta+\omega_p}+e^{-i\Phi_t}\delta\Gamma_{-\Delta-\omega_p})\\&
    \quad +\frac{1}{8}(e^{i\Phi_t}\delta\Gamma_{\Delta+\omega_p}+e^{-i\Phi_t}\delta\Gamma_{\Delta-\omega_p})\\&
    \quad+\frac{\xi}{4} e^{i\Delta(t-\tau_{p_1})}(e^{i\Phi_t} \delta\Gamma_{\omega_p}-e^{-i\Phi_t} \delta\Gamma_{-\omega_p})
\end{aligned}
\end{equation}

\begin{equation}
\begin{aligned}
    C_- & = \frac{1}{4}(e^{-2i\Delta (t-\tau_{p_1})} \delta\Gamma_{\Delta}+\delta\Gamma_{-\Delta})\\
    & \quad -\frac{e^{-2i\Delta(t-\tau_{p_1})}}{8}(e^{i\Phi_t}\delta\Gamma_{\Delta+\omega_p}+e^{-i\Phi_t}\delta\Gamma_{\Delta-\omega_p})\\&
    \quad +\frac{1}{8}(e^{i\Phi_t}\delta\Gamma_{-\Delta+\omega_p}+e^{-i\Phi_t}\delta\Gamma_{-\Delta-\omega_p})\\&
    \quad+\frac{\xi}{4} e^{-i\Delta(t-\tau_{p_1})}(e^{-i\Phi_t} \delta\Gamma_{-\omega_p}-e^{i\Phi_t} \delta\Gamma_{\omega_p})
\end{aligned}
\end{equation}
\begin{equation}
    \begin{aligned}
        &\Phi_t = - \omega_p(t-\tau_{p_2})\\
        &\delta \Gamma_\omega = \Gamma_\omega(t-\tau_{p_1})-\Gamma_\omega(t-\tau_{p_2})
    \end{aligned}
\end{equation}

Since the derivation assumes the limit \( t_0 \to -\infty \), where the system and environment are factorized, we set the time origin at the end of the pulse, \( \tau_{p_2} = 0 \). The exact choice of \( t_0 \) is immaterial, as the system is assumed to have equilibrated long before the pulse is applied.

\subsection{Dichotomous Relaxation and Dephasing Dynamics in Dynamically Prepared States}

Building on the first major result of this work---the induction of nonsecular dynamics and long-lived coherences following gate operations, as described in the previous section---we now turn to a second key finding concerning the interplay between relaxation and dephasing in initially correlated system--environment states.

The dynamics of an open quantum system that is initially correlated with its environment can be described by a set of \( N^2 \) (or fewer) completely positive maps, where \( N \) is the dimension of the system~\cite{paz2019dynamics}. In the context of our system, as the pulse duration increases, the dynamics transitions from the regime of an instantaneous gate toward the adiabatic limit, where both relaxation and dephasing processes exhibit behavior that is effectively Markovian.

However, this convergence is not uniform: the dephasing dynamics approach the Markovian limit much more slowly than the relaxation dynamics as the pulse length increases. 
As a result, for a finite but short pulse, the system exhibits \textit{hybrid dynamics}: while relaxation becomes effectively Markovian, dephasing retains non-Markovian characteristics consistent with evolution from a uncorrelated initial state.
Such a dichotomy is permitted precisely because the system begins in a correlated state with its environment~\cite{paz2019dynamics}. The identification of this distinct behavior---where relaxation and dephasing processes follow qualitatively different trajectories toward Markovianity---constitutes the second major result of this paper.

To further elucidate this hybrid behavior, we now analyze the dephasing and relaxation dynamics in detail, beginning with the dephasing processes that exhibit the most pronounced non-Markovian features.
 To this end, we solve the Bloch-Redfield master equation for a representative case with \( \theta = \pi/4 \) and \( \phi = 0 \), using the in-gate and post-gate dissipators derived in the previous two subsections. More precisely, the unitary component of the dynamics corresponds to a rotation of the qubit magnetization from the \( z \)-direction into the \( xy \)-plane.

\begin{figure}[htbp]
    \centering
    \includegraphics[width=1\columnwidth]{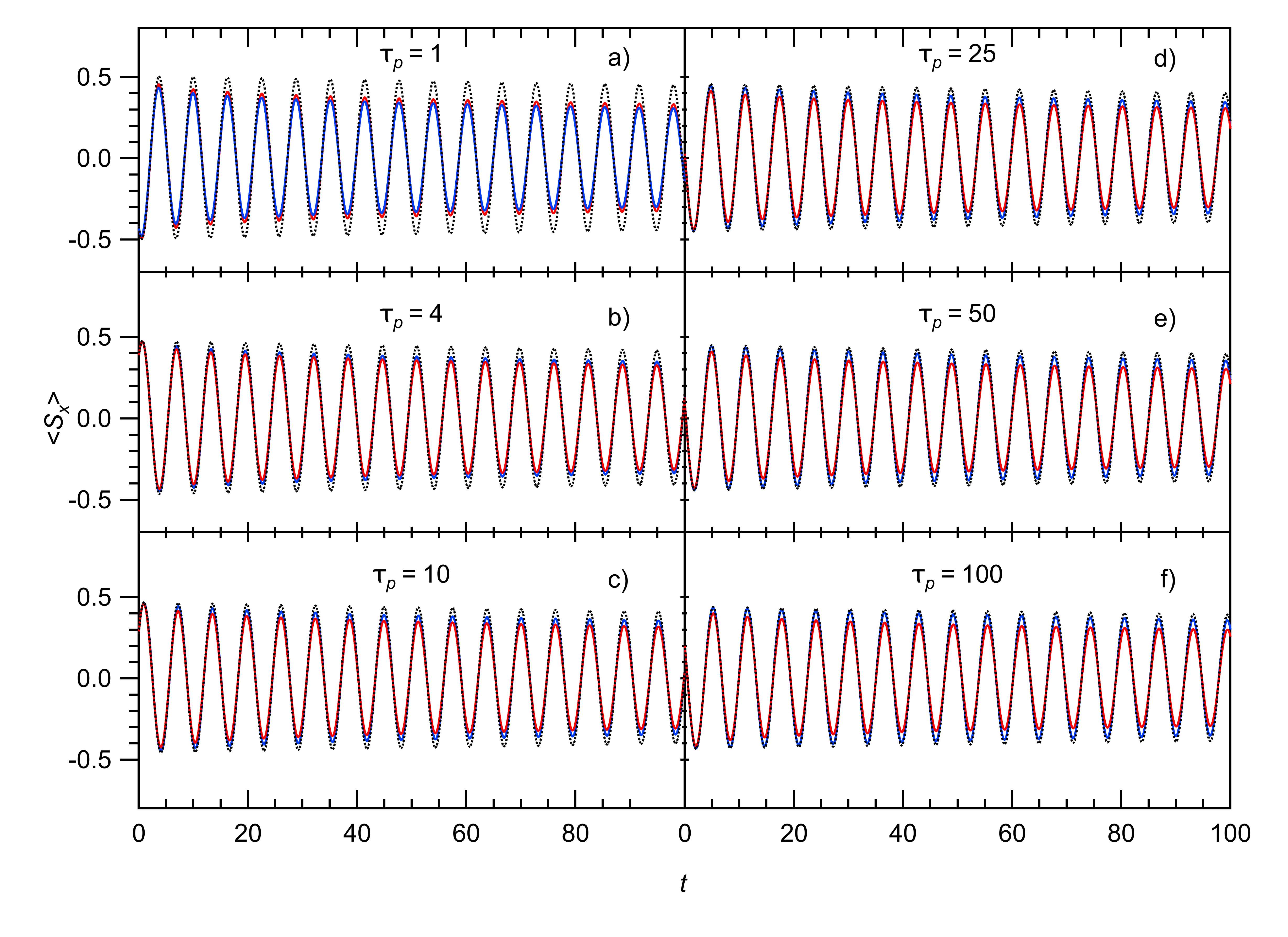}
    \captionsetup{justification=raggedright, singlelinecheck=false}
    \caption{X-component of the magnetization as a function of time for Markovian and non-Markovian dynamics \textit{for different pulse lengths}. The initial state is prepared using the in-gate dissipator, as defined in Eq.~(\ref{Eq: full_dissipator_in_gate}).
    The pulse length, characterized by \(\tau_p\), increases from (a) to (c), while all other parameters remain fixed. Dotted black line: Markovian dynamics with decoherence time The decoherence time $T_2=865$ equal to twice the relaxation time. Red solid line: non-Markovian dynamics with a factorized initial state. Blue solid line: non-Markovian dynamics with a dynamically prepared initial state for finite pulse length. The dephasing dynamics of the dynamically prepared state converge slowly to those of the Markovian case as the pulse length increases, gradually driving the system toward the adiabatic limit.
    Parameters: $\lambda^2 = 0.002$, $\phi = 0$, $\Delta = 1$, $\xi = 4$, $T = 0$, and $\omega_c = 1$.
}
    \label{Fig: Dephasing_Transition}
\end{figure}

Fig.~\ref{Fig: Dephasing_Transition} displays the \( x \)-component of the magnetization as a function of time following the application of pulses of various durations. While the initial perpendicular component of the magnetization is always \( 1/2 \), the value of \( s_x(0) \) varies between panels due to the changing pulse duration \( \tau_p \). For short pulses (\( \tau_p = 1 \)), the post-gate coherence dynamics is non-Markovian, closely resembling that obtained from a factorized initial condition and consistent with the instantaneous gate regime. 
\begin{figure}[htbp]
    \centering
    \includegraphics[width=1\columnwidth]{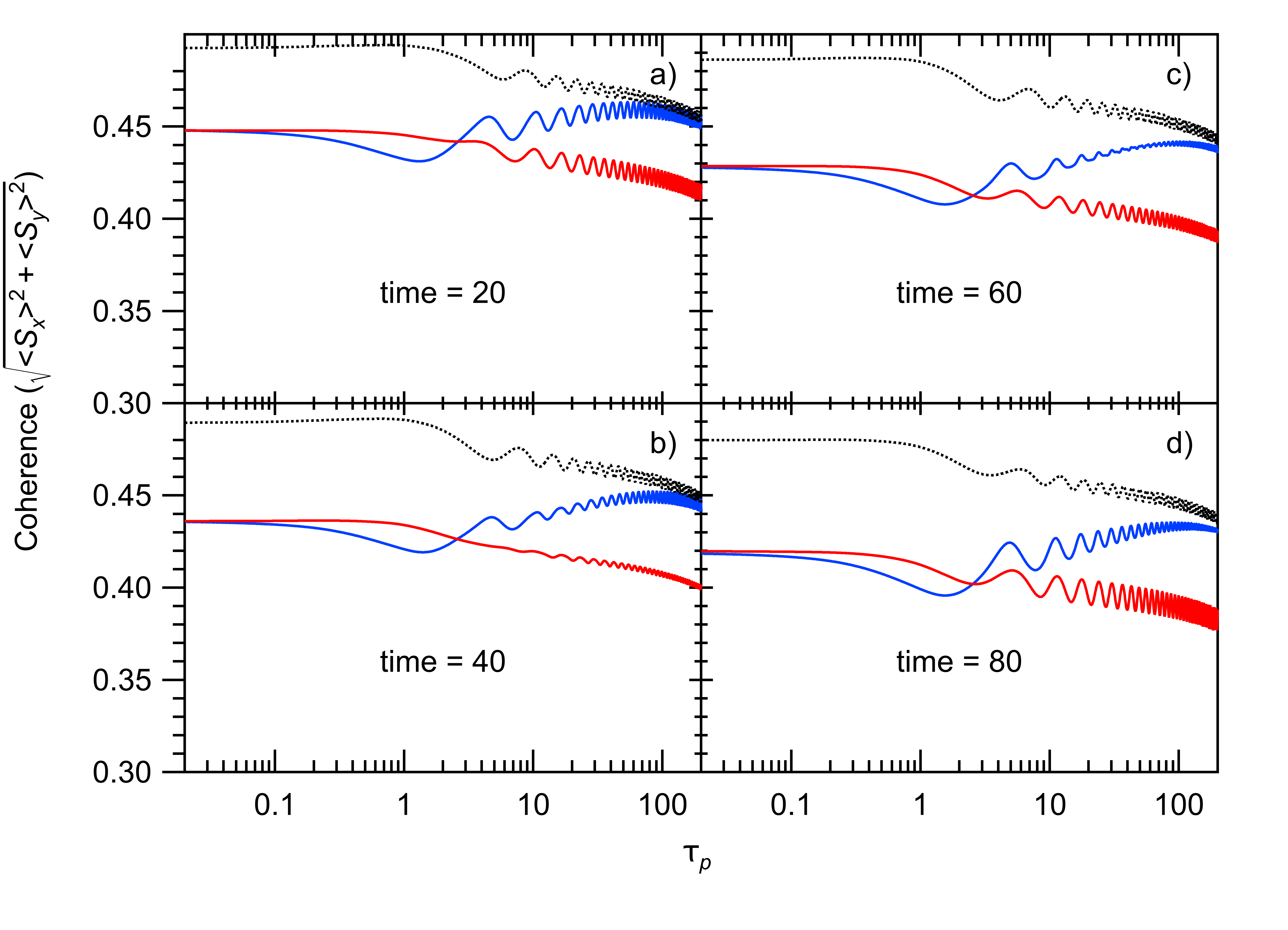}
    \captionsetup{justification=raggedright, singlelinecheck=false}
    \caption{Dephasing dynamics as a function of pulse duration, evaluated at four times relative to the end of the pulse. The red, blue, and black curves correspond to a factorized initial condition, dynamically initialized state, and Markovian dynamics, respectively.
 The data illustrate a crossover from non-Markovian to Markovian coherence dynamics as the pulse duration increases. While the coherence gradually diminishes with time from $20$ to $80$, this temporal decay does not affect the pulse-length–dependent crossover. Parameters: $\lambda^2 = 0.001$, $\phi = 0$, $\Delta = 1$, $\xi = 4$, $T = 0$, and $\omega_c = 1$.
}
    \label{Fig: coherence_analysis}
\end{figure}

As the pulse duration increases, the coherence dynamics gradually transition into the Markovian regime.
Figure~\ref{Fig: coherence_analysis} shows the magnitude of the perpendicular component of the magnetization as a function of pulse length, for a fixed $\pi/2$ rotation. The key feature is the crossover in the coherence dynamics of the dynamically prepared state—from non-Markovian dephasing, associated with a factorized initial condition at short pulse durations, to Markovian dephasing at longer pulses. The characteristic crossover time is approximately $\tau_p = 30$.

As the panels progress from (a) to (b), the post-gate time (relative to $\tau_{p_2}$) increases, demonstrating that the influence of non-Markovian dynamics on the coherence is uniform over this range of times.

\begin{figure}[htbp]
    \centering
    \includegraphics[width=1\columnwidth]{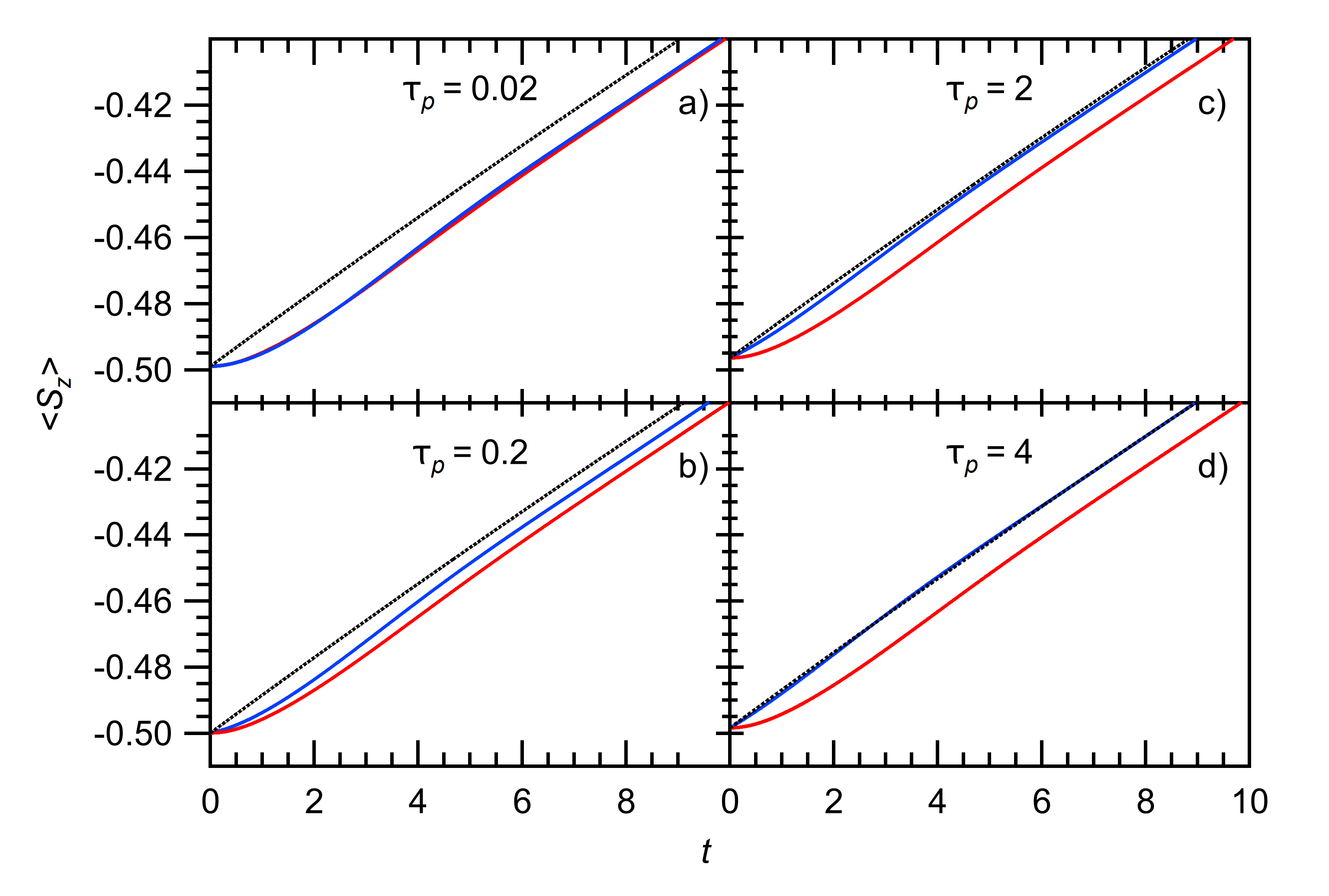}
    \captionsetup{justification=raggedright, singlelinecheck=false}
    \caption{Z-component of the magnetization as a function of time for Markovian and non-Markovian dynamics \textit{for different pulse lengths}. The initial state is prepared using the in-gate dissipator, as defined in Eq.~(\ref{Eq: full_dissipator_in_gate}).
    The pulse length, characterized by \(\tau_p\), increases from (a) to (c), while all other parameters remain fixed. Dotted black line: Markovian dynamics with decoherence rate equal to half the relaxation rate. Red solid line: non-Markovian dynamics with a factorized initial state. Blue solid line: non-Markovian dynamics with a dynamically prepared initial state for finite pulse length. The relaxation dynamics of the dynamically prepared state converge rapidly to those of the Markovian case as the pulse length increases.
    Parameters: $\lambda^2 = 0.01$, $\phi = 0$, $\Delta = 1$, $\xi = 0$, $T = 0$, and $\omega_c = 1$.
}
    \label{Fig: relaxation_transition}
\end{figure}

The relaxation dynamics, \( \langle s_z(t) \rangle \), are shown in Fig.~\ref{Fig: relaxation_transition}, following a rotation of the qubit about the x-axis by \( \pi \) (i.e., a spin flip). For the shortest pulses, the dynamically prepared states exhibit a relaxation delay similar to that observed for a factorized initial condition. As the pulse duration increases, however, the relaxation delay decreases in an oscillatory fashion and eventually reaches the Markovian regime, where no relaxation delay is observed.

\begin{figure}[htbp]
    \centering
    \includegraphics[width=1\columnwidth]{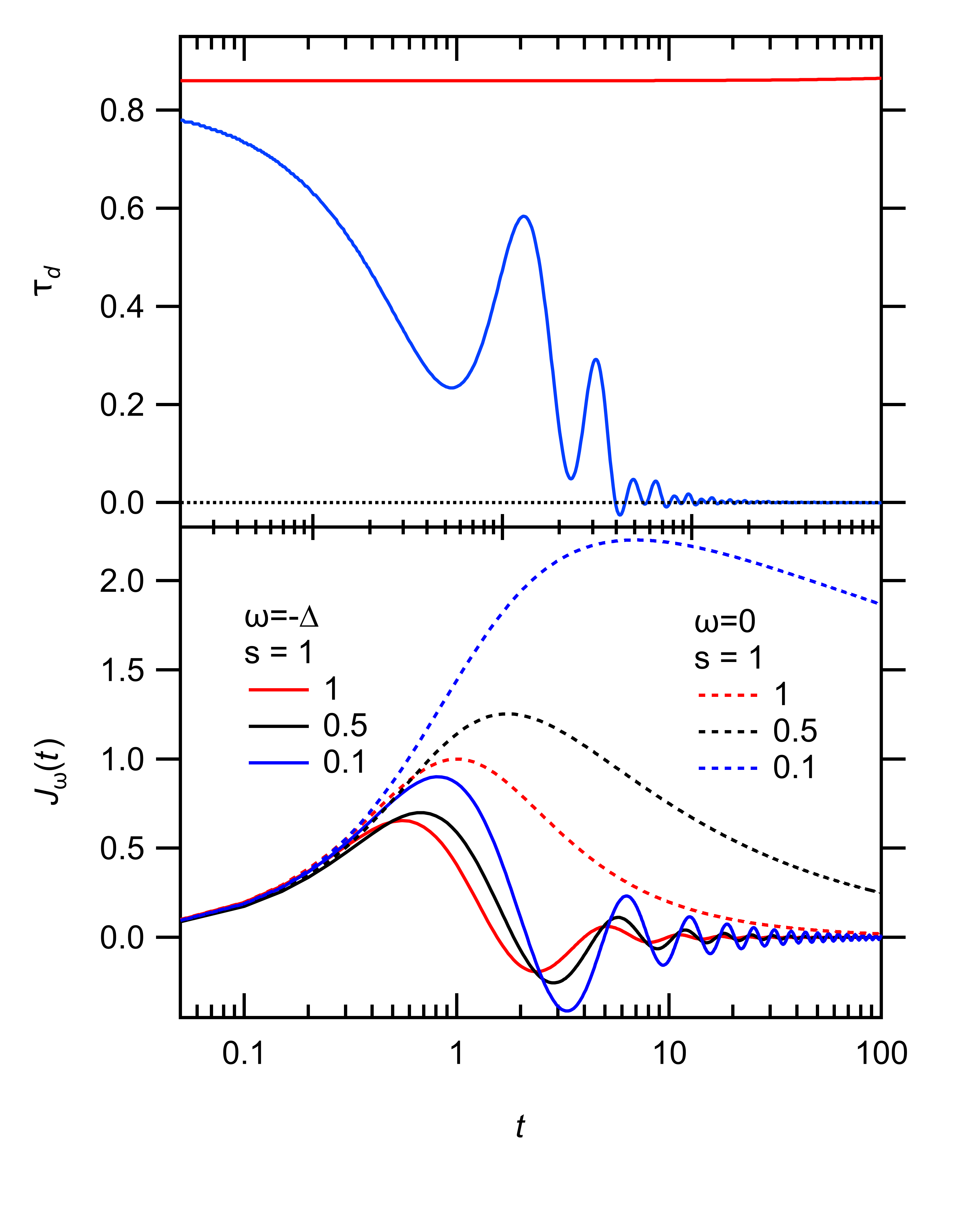}
    \captionsetup{justification=raggedright, singlelinecheck=false}
    \caption{(a): Relaxation delay of the dynamically initialized state as a function of pulse duration. The red and black lines correspond to the factorized initial condition (maximum delay) and Markovian dynamics (no delay), respectively. The data illustrate the crossover from non-Markovian to Markovian population dynamics with increasing pulse duration.  Parameters: \( s=1 \), \( \lambda^2 = 0.001 \), \( \phi = 0 \), \( \Delta = 1 \), \( \xi = 0 \), \( T = 0 \), and \( \omega_c = 1 \). (b) Discrepancy in the time-scales for the time-dependent spectral density for factorized initial conditions between  nonzero and zero frequency. The dephasing rate saturates at much longer time scales than the relaxation (Fermi-Golden rule) rate, partially explaining the non-uniformity in the transition to the Markovian regime as a function of pulse length. Red, black, and blue: $s=1,0.5$ and $0.1$. Full (dashed) line: spectral density at $\omega=-\Delta (0)$
}
    \label{Fig: relaxation_delay}
\end{figure}

The characteristic timescale for this transition is shown in Fig.~\ref{Fig: relaxation_delay}, which displays the offset in \( \langle s_z(t) \rangle \) in the linear regime at large \( t \). The characteristic timescale is \( \tau_p = 1 \), approximately thirty times shorter than that observed for the dephasing dynamics, illustrating the differing sensitivity of relaxation and dephasing channels to the underlying bath correlations.

Thus, the relaxation and dephasing channels exhibit a non-uniform transition to the Markovian regime as a function of pulse duration. For intermediate pulses, the post-gate relaxation dynamics are approximately Markovian, while the dephasing dynamics remain similar to those of a factorized (uncorrelated) initial condition. This behavior implies the necessity of invoking two non-equivalent completely positive maps to describe the dissipative effects of such pulses, exemplifying the results of Paz-Silva \textit{et al.}~\cite{paz2019dynamics}.

The microscopic origin of this dichotomy can be partially attributed to the differing time dependence of the dephasing and relaxation rates, starting from a factorized initial condition.
Figure~\ref{Fig: relaxation_delay}(b) shows the time evolution of the spectral density at frequencies $\omega = -\Delta$ and $\omega = 0$ for an Ohmic bath ($s = 1$) and two sub-Ohmic baths ($s = 0.5$ and $0.1$). (At $\omega = +\Delta$, $J_\omega(t)$ saturates on a similar time scale as at $\omega = -\Delta$, data not shown.) Comparing Figs.~\ref{Fig: relaxation_delay}(a) and (b), we observe that for $s = 1$ there is good quantitative agreement between the reaction delay and the characteristic time scale of the spectral density at nonzero frequency.  This time scale is comparable to $\tau_c$ for each bath—namely, the bath reorganization time associated with transitions between the qubit’s two energy levels. It also defines the pulse duration at which the relaxation dynamics transitions into the Markovian regime.

In contrast, the characteristic time scale over which the dephasing rate in Fig.~\ref{Fig: relaxation_delay}(b) to zero is substantially longer than $\tau_c$, consistent with the observation that longer pulse durations are required to bring post-gate dephasing into the Markovian regime. The quantitative agreement between the characteristic times for dephasing is less precise, however, as the pulse length required to suppress dephasing is {\it significantly longer} than that needed to suppress dephasing from uncorrelated initial conditions. This dichotomy between dephasing and relaxation time scales becomes increasingly pronounced as the bath becomes more sub-Ohmic. This behavior is particularly relevant for quantum computing with superconducting and semiconducting qubits, where sub-Ohmic noise is commonly encountered. These results underscore the importance of carefully optimizing pulse shapes to control dephasing induced by the gate in the presence of preexisting correlations, an issue we will discuss further in Sec.~\ref{Sec:application: optimization}.

\subsection{Gate Fidelity as a Signature of Correlated Dynamics in Open Quantum Systems}

The time-dependent Bloch-Redfield master equation, combined with dynamically prepared states, provides a systematic framework for evaluating gate fidelity. In realistic quantum devices, fidelity is influenced by numerous factors \cite{krantz2019quantum}, including pulse amplitude, waveform shape \cite{ku2017single,kuzmanovic2024high,yi2024robust}, and control protocols \cite{carr1954effects,meiboom1958modified,pokharel2018demonstration}. Our approach begins from the microscopic system Hamiltonian and accounts for both the transfer of bath correlations and intra-gate dissipation, without relying on phenomenological assumptions. In Sec.~\ref{Sec:application: optimization}, we demonstrate that pulse-shape refinement can yield significant improvements in fidelity. This result points toward a broader parameter space that warrants exploration in future research.

As introduced in Secs.~\ref{Sec:dp} and~\ref{Sec:Finite-Duration}, the ideal flip rotation is determined by the pulse intensity \(\epsilon\). We quantify the fidelity using the trace distance between the density matrix \(\rho(\epsilon)\), obtained via the dynamically prepared Bloch-Redfield equation, and a reference pure state \(\rho'(\theta,\phi)\). Here, \(\theta\) and \(\phi\) denote the polar and azimuthal angles of a Bloch vector \(\vec{r} = (r\sin\theta \cos\phi, r\sin\theta \sin\phi, r\cos\theta)\), where \(r < 1\) for mixed states. The pure-state density matrix is given by 
\[
\rho'(\theta, \phi) = \frac{1}{2} \left( \mathbbm{I} + \vec{r} \cdot \vec{\sigma} \right),
\]
at $r=1$.

The trace distance is computed using the singular value decomposition (SVD):
\begin{equation}
    \rho(\epsilon)-\rho'(\theta,\phi) = U \Sigma V^\dagger,
\end{equation}
where \(U\) and \(V\) are unitary matrices and \(\Sigma\) is diagonal with nonnegative entries \(\{\sigma_i\}\). The trace distance is then 
\[
D = \frac{1}{2} \sum_i \sigma_i,
\]
and the fidelity \(F\) is defined as
\begin{equation}
    F(\epsilon, \theta, \phi) = 1 - \frac{1}{2} \sum_i \sigma_i.
\end{equation}

Although this definition of gate fidelity is not conventional, it is adopted here because it captures fidelity loss arising from phase uncertainty—i.e., deviations in the direction of the Bloch vector. Specifically, the Bloch-Redfield master equation yields vanishing population corrections (to order \(\lambda^2\)) at zero temperature, while retaining accurate coherence corrections at the same order~\cite{crowder2024invalidation}. As such, the fidelity measure used here should be interpreted as \textit{phase fidelity}. A complete description of gate fidelity, including population uncertainty due to system-bath interactions, lies beyond the scope of this treatment and would require a fourth-order master equation.

The pulse amplitude \(\epsilon\) is set to produce the desired flip angle—e.g., \(\pi/2\) or \(3\pi/2\)—in the absence of environmental coupling. For a perfect \(\pi/2\) rotation about the \(X\) axis, the target state is \(\rho_t(\theta = \pi/2, \phi = -\pi/2)\), and the fidelity should ideally reach its maximum at \(F(\epsilon, \theta = \pi/2, \phi = -\pi/2)\). However, in the presence of a bath, the final state is generally mixed, and its Bloch vector deviates from the target orientation. The displacement of the fidelity maximum in the \((\theta, \phi)\) parameter space reflects phase latency and drift arising from system-bath interactions during the gate operation.

\begin{figure}[h]
    \centering
    \includegraphics[width=1\columnwidth]{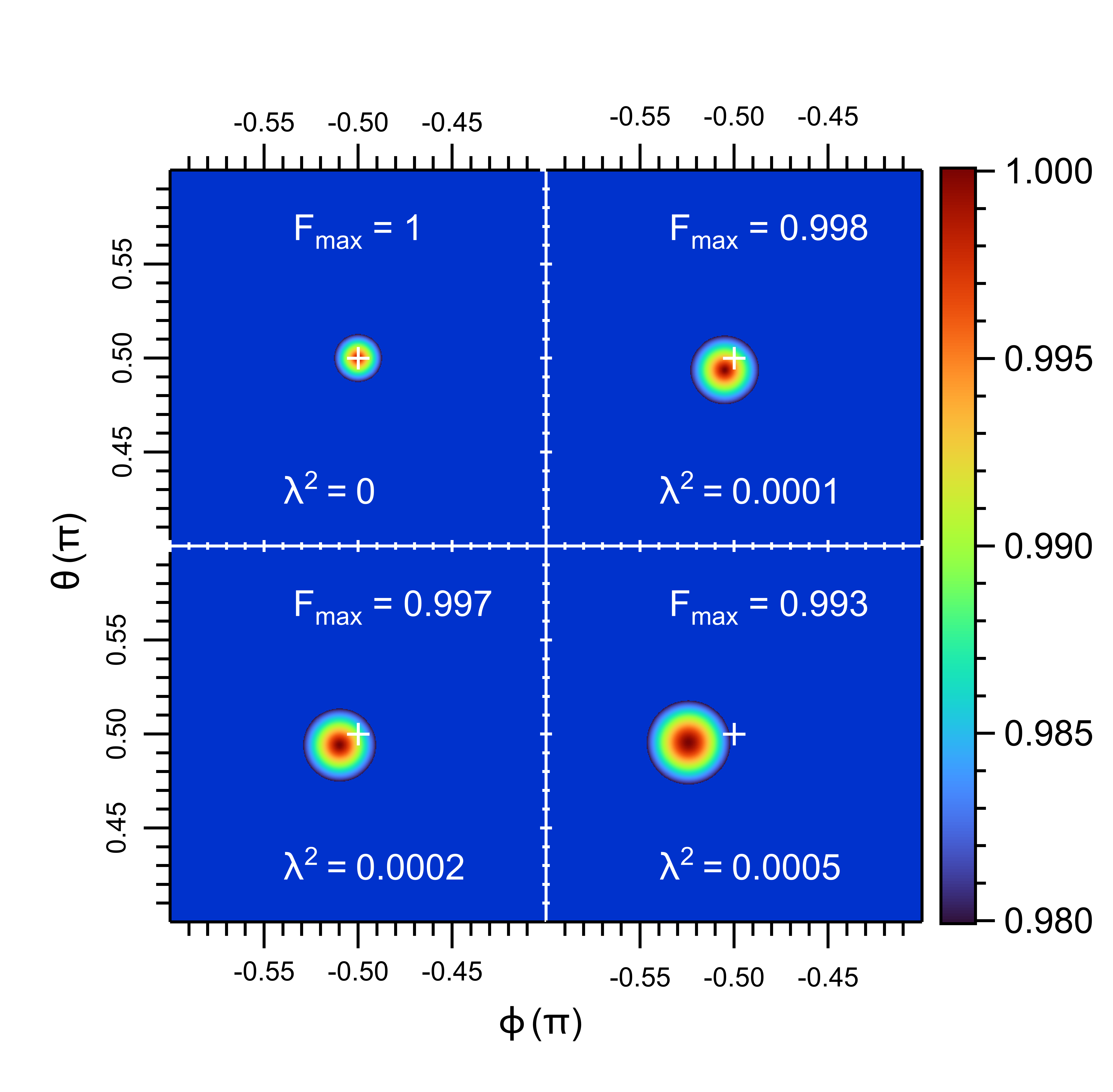}
    \captionsetup{justification=raggedright, singlelinecheck=false}
    \caption{Fidelity ratio (\(F/F_{\text{max}}\)) of a rotation operation about the x-axis as a function of the rotation angle \(\pi/2\). Only fidelity in range \([0.98F_{\text{max}}, F_{\text{max}}]\) is plotted.
    Parameters: $\phi = 0$, $\Delta = 1$, $\tau_p = 200$, $\xi=1$, $T = 0$, and $\omega_c = 1$. 
}
    \label{Fig: Color_Fidelity}
\end{figure}

As shown in Fig.~\ref{Fig: Color_Fidelity}, for a \(\pi/2\) pulse—which drives the qubit from the ground state to a superposition state—the maximum fidelity decreases with increasing \(\lambda^2\). In addition, the angle \(\theta_m\) at which the fidelity is maximized is slightly less than the ideal \(\pi/2\), indicating a relaxation bias toward the ground state. A modest overdrive of the pulse can compensate for this effect, restoring the peak fidelity angle to \(\pi/2\). Similarly, the phase angle \(\phi_m\) deviates from its ideal value, but can likewise be corrected through appropriate change in the target state. 
The shifts in \(\theta_m\) and \(\phi_m\) arise from the anisotropic spreading of the quantum state, induced by its interaction with the environment. 

Even with optimal adjustments to \(\theta\) and \(\phi\), the maximum fidelity remains below unity. This indicates a fundamental phase uncertainty from system-bath interaction—irreducible by target redefinition. In the analysis below, we redefine gate fidelity as \(F_\text{max}\), which quantifies loss of quantum phase information rather than mere phase shift.

\begin{figure}[h]
    \centering
    \includegraphics[width=1\columnwidth]{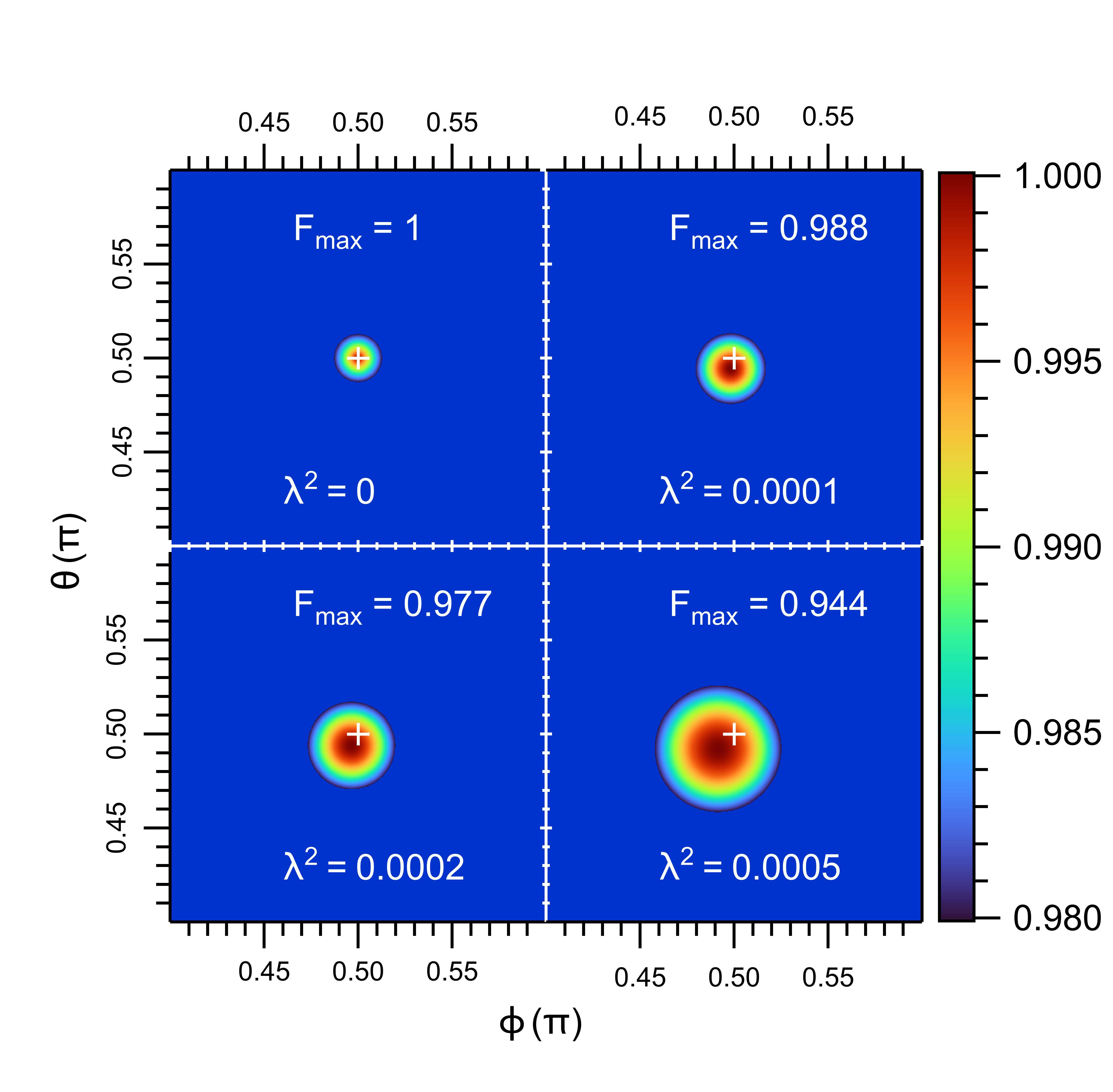}
    \captionsetup{justification=raggedright, singlelinecheck=false}
    \caption{Fidelity ratio (\(F/F_{\text{max}}\)) of a rotation operation about the x-axis as a function of the rotation angle \(3\pi/2\). Only fidelity in range \([0.98F_{\text{max}}, F_{\text{max}}]\) is plotted.
    Parameters: $\phi = 0$, $\Delta = 1$, $\tau_p = 200$, $\xi=1$, $T = 0$, and $\omega_c = 1$. 
}
    \label{Fig: Color_Fidelity_3pi2}
\end{figure}

\begin{figure}[h]
    \centering
\begin{subfigure}[t]{1\columnwidth}
    \centering
    \includegraphics[width=\linewidth]{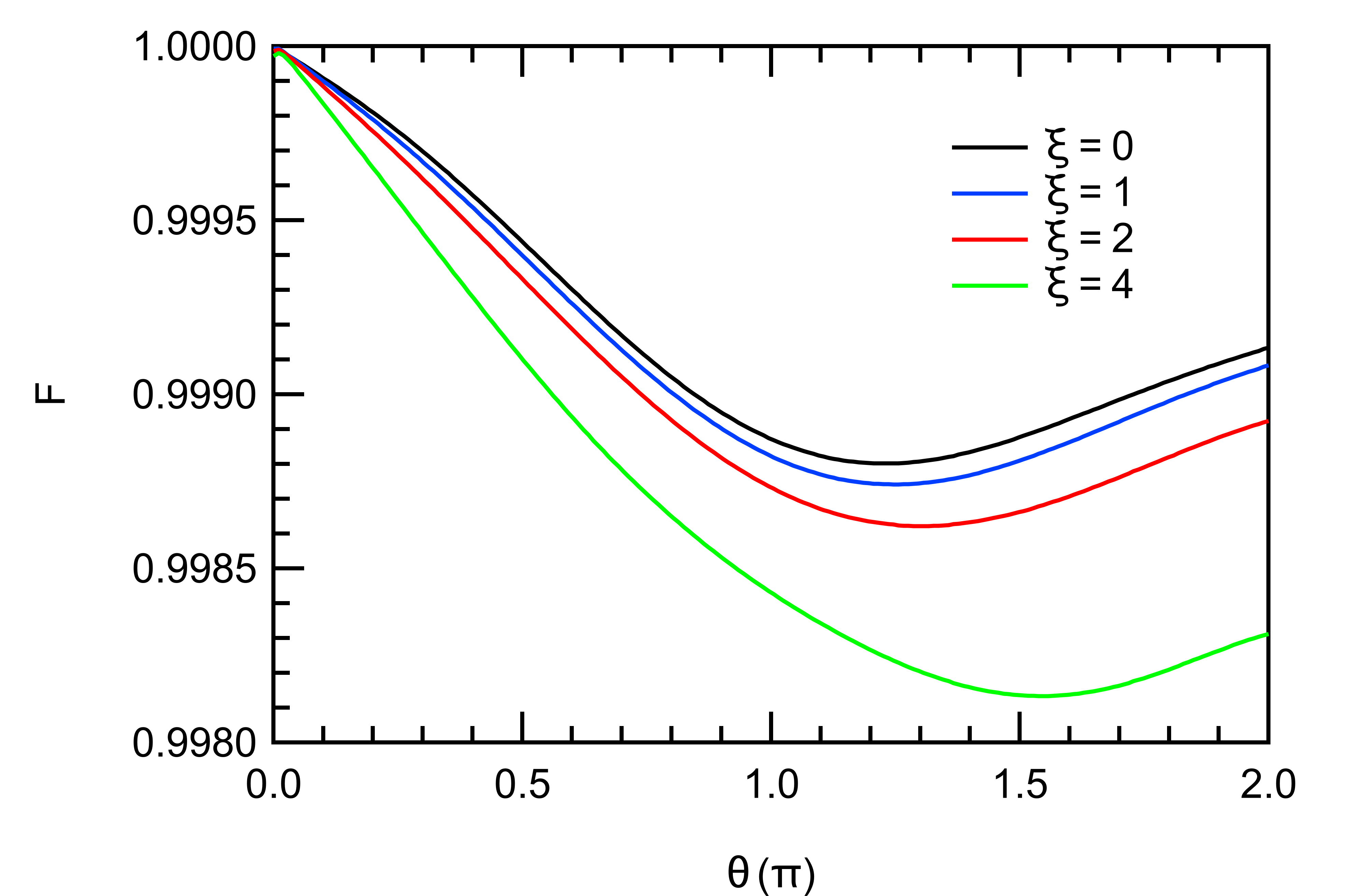}
    \caption{}
\end{subfigure}
    \hfill
\begin{subfigure}[t]{1\columnwidth}
    \centering
        \includegraphics[width=\linewidth]{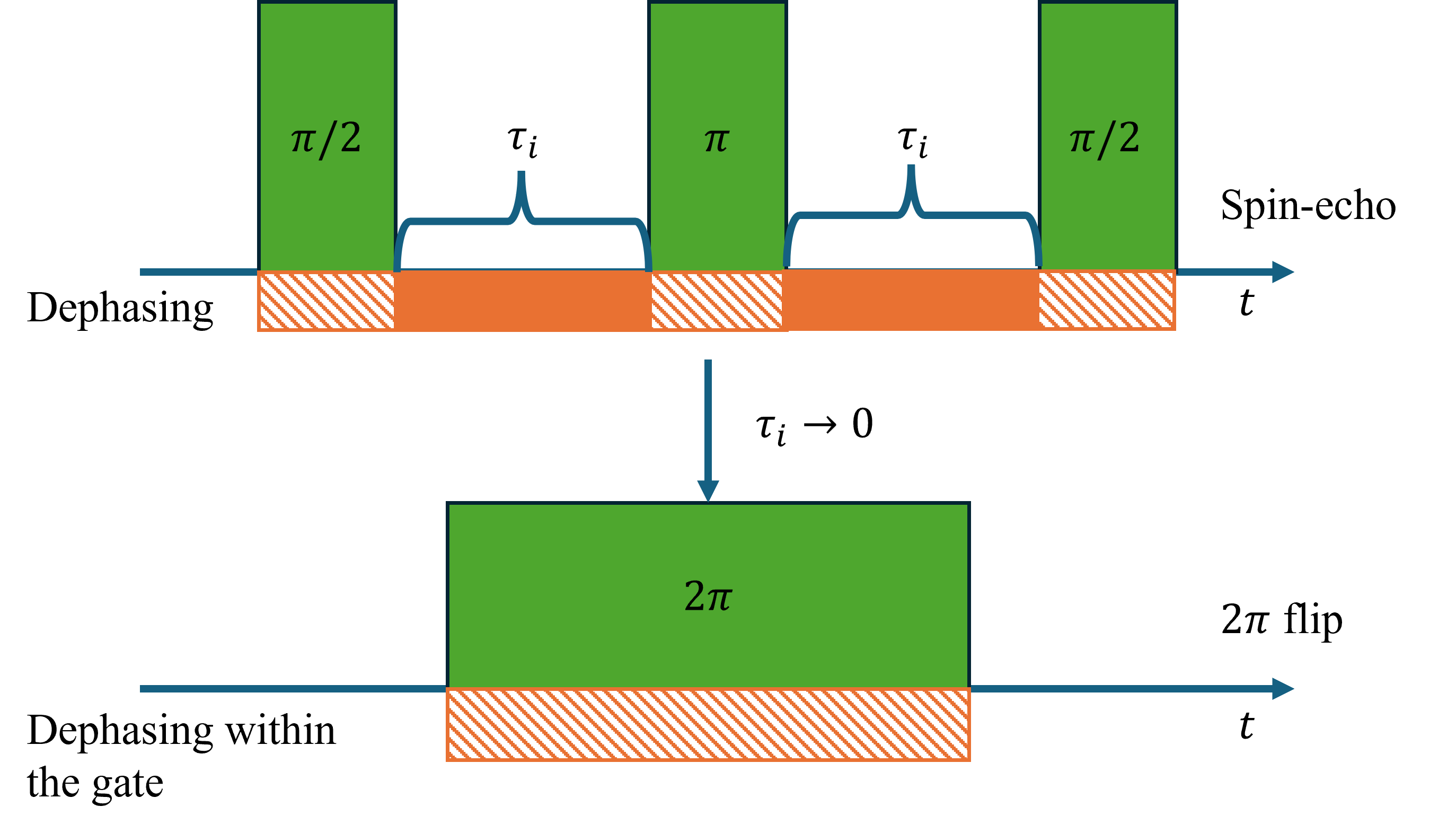}
    \caption{}
\end{subfigure}
\captionsetup{justification=raggedright, singlelinecheck=false}
\caption{(a) Fidelity of a rotation operation about the x-axis as a function of the rotation angle \(\theta\). The final state is obtained using the in-gate dissipator, as defined in Eq.~(\ref{Eq: full_dissipator_in_gate}), evaluated at the end of the pulse. Fidelity is computed via the trace distance between the ideally rotated state and the actual state evolved under in-gate dissipation.
Parameters: $\lambda^2 = 10^{-5}$, $\phi = 0$, $\Delta = 1$, $s = 1$, $\tau_p = 200$, $T = 0$, and $\omega_c = 1$. Black solid line: $\xi=0$.
Blue solid line: $\xi=1$. Red solid line: $\xi=2$. Green solid line: $\xi=4$. (b) An analogy can be drawn with spin echo. Regarding dephasing during gate operation, the \(2\pi\) rotation may be interpreted as a "continuous" spin-echo.
}
    \label{Fig: Fidelity}
\end{figure}

\begin{figure}[h]
    \centering
    \includegraphics[width=1\columnwidth]{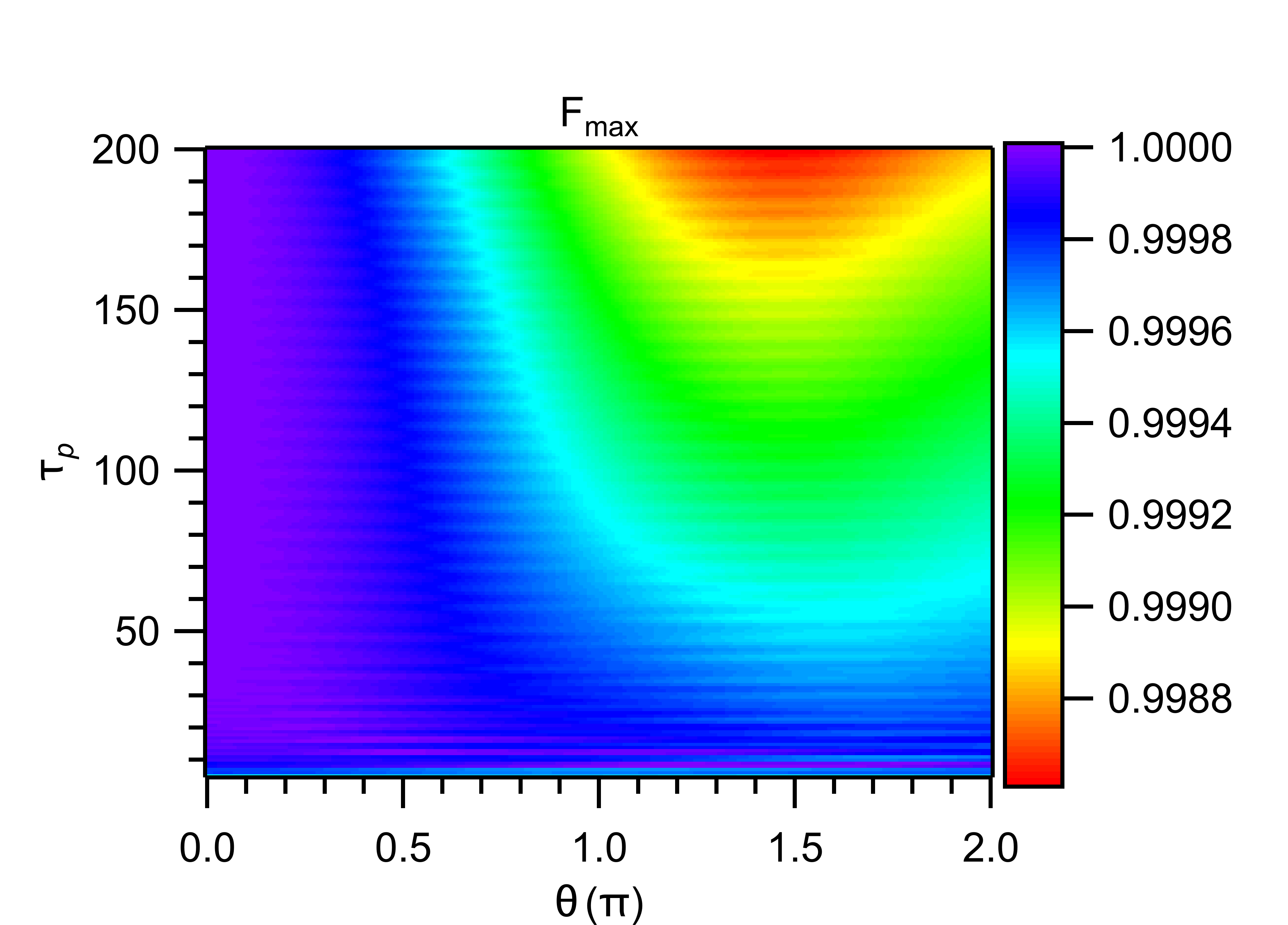}
    \captionsetup{justification=raggedright, singlelinecheck=false}
    \caption{Max fidelity of a rotation operation about the x-axis as a function of the rotation angle \(\theta\). The final state is obtained using the in-gate dissipator, as defined in Eq.~(\ref{Eq: full_dissipator_in_gate}), evaluated at the end of the pulse. Fidelity is computed via the trace distance between the ideally rotated state and the actual state evolved under in-gate dissipation.
    Parameters: $\lambda^2 = 10^{-5}$, $\phi = 0$, $\Delta = 1$, $s = 1$, $\xi=2$, $\tau_p = 200$, $T = 0$, and $\omega_c = 1$.
}
    \label{Fig: Fidelity_tptheta}
\end{figure}

The reduction in \(F_\text{max}\) from one is not irreversible, suggesting that fidelity loss is at least partly due to inhomogeneous broadening during the gate. As shown in Fig.~\ref{Fig: Fidelity}, fidelity recovers near a \(2\pi\) rotation—resembling a continuous spin-echo effect. Furthermore, in Fig.~\ref{Fig: Fidelity_tptheta}, plotting \(F_\text{max}\) versus pulse duration \(\tau_p\) and rotation angle \(\theta\) reveals a clear revival at long pulse durations, indicating a refocusing process.

Fig.~\ref{Fig: Fidelity_tptheta} reveals a qualitative difference between short and long pulse behavior. For short pulses, as shown in Fig.~\ref{Fig: relaxation_delay}(b), the system exhibits relaxation delay during the gate, leading to complex and strongly non-Markovian dynamics. In contrast, longer pulses allow refocusing of \(F_\text{max}\).

Fidelity recovery at long pulses and large angles suggests the possibility of enhancement via dynamical pulse engineering. Practical implementation of this strategy is discussed in the next section.

\newpage
\section{Applications\label{Sec:Applications}}

\subsection{Optimizing Pulse Shapes for High-Fidelity Gate Operations} \label{Sec:application: optimization}
In the preceding discussion, the external drive was modeled as a square-shaped pulse. This choice induces a well-defined gate operation frequency and allows the relevant integrals to be expressed analytically in terms of time-dependent spectral densities with frequency shifts.

For more general pulse shapes, the dissipator in Eq.~(\ref{Eq: dissipator in gate}) can still be evaluated directly by substituting the bath correlation function \(C(t - \tau)\) into the integral expression:
\begin{equation}
\begin{aligned}
    \Lambda_{\text{gate}}^S(t) &= U_r(t - \tau_{p_1}) \left[\Lambda^{\text{SM}} - \Lambda_{\text{static}}(t - \tau_{p_1})\right] U_r^\dagger(t - \tau_{p_1}) \\
    &\quad + \int_{\tau_{p_1}}^{t} d\tau\, C(t - \tau)\, U_S(t, \tau)\, A\, U_S(\tau, t).
\end{aligned}
\label{Eq:OptimizeGate}
\end{equation}

This formulation provides a natural and systematic route to optimizing pulse shapes based on the spectral response and back-action from the bath. By tuning the pulse parameters, one can identify configurations that minimize the influence of \( \Lambda_{\text{gate}} \), thereby improving the overall gate fidelity.

\begin{figure}[h]
    \centering
    \includegraphics[width=\columnwidth,clip,trim=180pt 150pt 100pt 120pt]{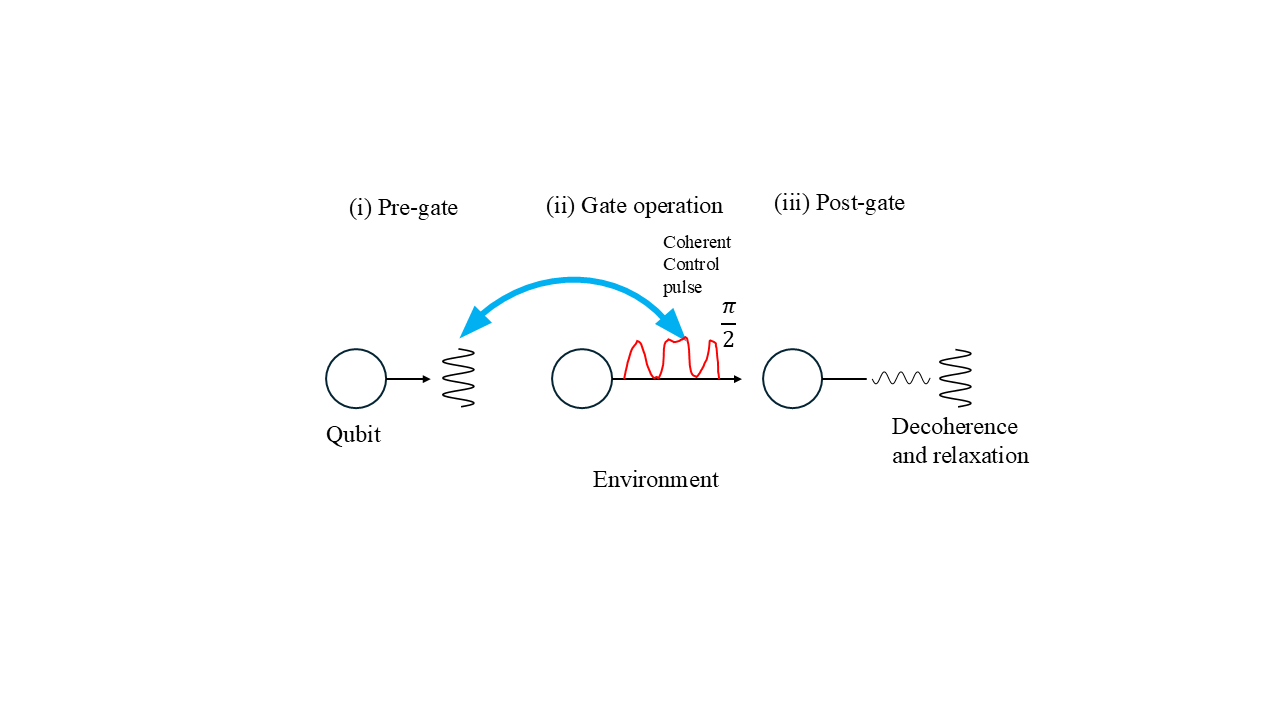}
    \captionsetup{justification=raggedright, singlelinecheck=false}
    \caption{
    Schematic of a qubit’s evolution during the gate operation stage. Due to coupling with a bath possessing memory, dissipation has two contributions: one arising from the transfer of preexisting correlations (blue arrowed line), and another from dissipation occurring within the gate itself. Optimizing the pulse shape (red line) requires considering the interplay between these two components.
    }
    \label{Fig:FidelityNM}
\end{figure}

Figure~\ref{Fig:FidelityNM} illustrates an adaptation of Fig.~\ref{Fig:PrePostGate} for the context of gate-fidelity optimization. In this setting, it is essential to balance the contribution of preexisting system–environment correlations—the first term in Eq.~(\ref{Eq:OptimizeGate})—with the dissipation that takes place during the finite-duration gate pulse, represented by the second term.

The interaction picture quantum state at the end of the pulse satisfies:
\begin{align}
\begin{split}
\varrho(\tau_{p_2}) &= \varrho(\tau_{p_1}) + \int_{\tau_{p_1}}^{\tau_{p_2}} dt\, \mathcal{D}_{\text{gate}}^I(t) \varrho(t) \\
&\approx \varrho(\tau_{p_1}) + \int_{\tau_{p_1}}^{\tau_{p_2}} dt\, \mathcal{D}_{\text{gate}}^I(t) \varrho(\tau_{p_1}),
\end{split}
\end{align}
where the approximation in the second line assumes that the gate duration is much shorter than the relaxation timescale, allowing us to neglect changes in the state within the interaction picture during the gate.
Here, \( \mathcal{D}_{\text{gate}}^I(t) \) is the dissipative generator in the interaction picture, obtained by first computing the Schrödinger-picture dissipator using Eq.~(\ref{Eq:DBloch-Redfield}) with \( \Lambda_{\text{gate}}^S(t) \), and then transforming it to the interaction picture via Eq.~(\ref{Eq:X_IS}).

The optimal gate fidelity is obtained by minimizing the norm of the dissipative contribution,
\begin{equation}
\label{Eq:minD}
   \mathbf{D} = \left\Vert \int_{\tau_{p_1}}^{\tau_{p_2}} dt\, \mathcal{D}_{\text{gate}}^I(t) \right\Vert.
\end{equation}
Notably, the optimal parameters in this setting are independent of the coupling strength \( \lambda^2 \), and are determined entirely by the spectral properties of the environment and the structure of the system operators \( A \) and \( H_S(t) \). However, the gate fidelity itself is reduced from unity by an amount proportional to \( \lambda^2 \).

Here we provide a simple model, optimizing the pulses' robustness for dephasing. We first expressed the amplitude of pulse shape as:
\begin{equation}
    \epsilon(t) = \omega_p + \sum_{n=1}^7 a_n \frac{n\pi}{\tau_p} \cos(n \pi \frac{t}{\tau_p})
\end{equation}
As an example, we first fix the coupling operator to be \(A = \sigma_z\) for simplicity. Under this choice, we evaluate the integral and obtain the gate-induced dissipator in the form:
\begin{equation}
    \Lambda_{\text{gate}}(t) = c_x(t)\, \sigma_x + c_y(t)\, \sigma_y + c_z(t)\, \sigma_z,
\end{equation}
where the coefficients \(c_x(t)\), \(c_y(t)\), and \(c_z(t)\) capture the time-dependent dissipative contributions along each Pauli direction. We then perform a parameter optimization in \textsc{Matlab} to identify the optimal set of pulse parameters that minimize
$\mathbf{D}$. Parameter optimization was carried out using \textsc{Matlab}'s \texttt{fmincon} with the interior–point algorithm, parallel evaluation enabled, a maximum of 300 iterations, and iterative progress output (\texttt{Display = iter}).
For reproducibility, all \textsc{Matlab} optimization settings and parameter choices are documented in the publicly available code linked at the end of this paper.  

\begin{figure}[htbp]
    \centering
    \includegraphics[width=1\columnwidth]{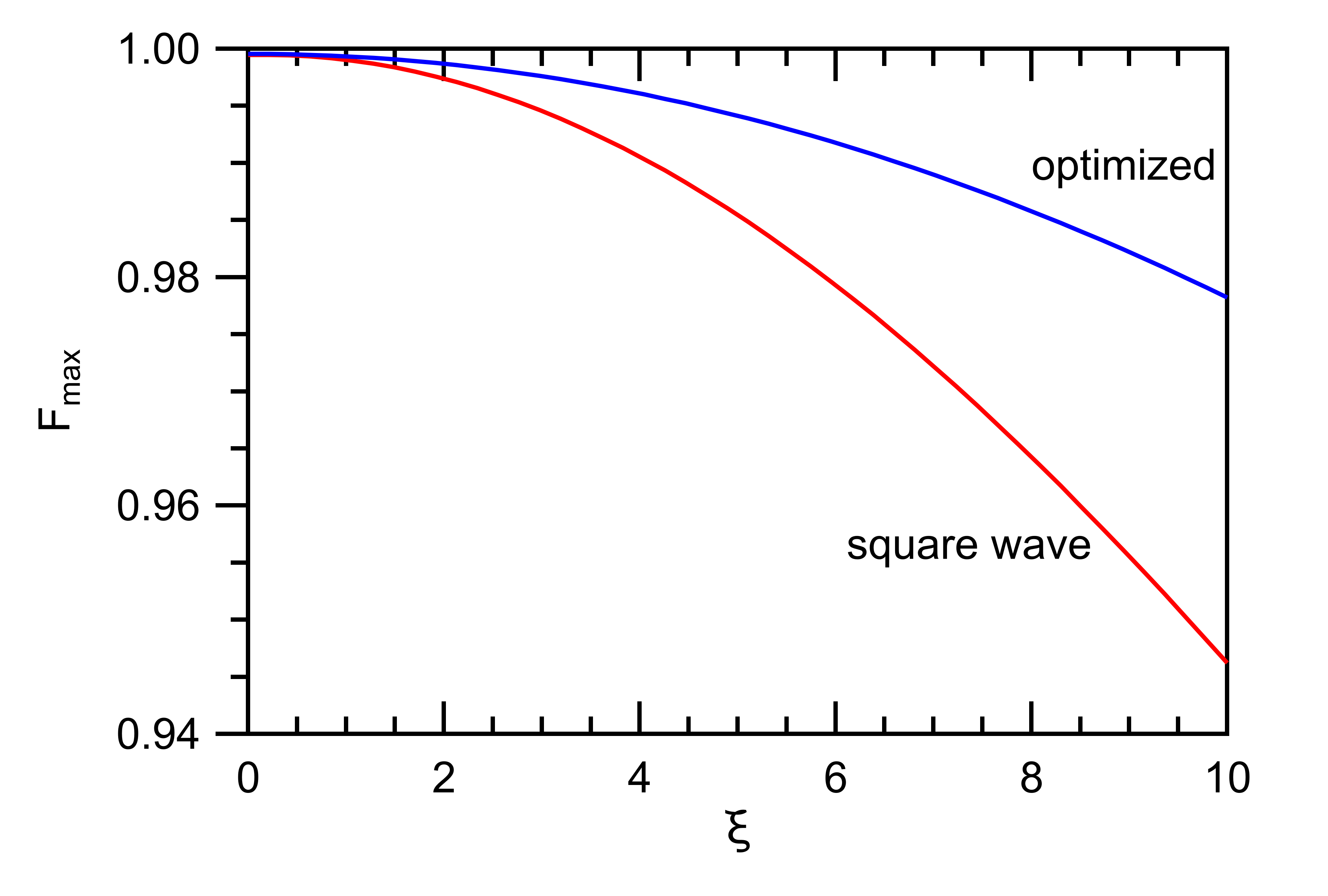}
    \captionsetup{justification=raggedright, singlelinecheck=false}
    \caption{Fidelity of a  \(\pi/2\) rotation operation about the x-axis. Blue solid line: Fidelity of optimized pulse shape.
    Parameters: $\lambda^2=10^{-5}$, $\phi = 0$, $\Delta = 1$, $\tau_p = 200$, $T = 0$, $s = 1/2$, and $\omega_c = 1$. 
    }
    \label{Fig: Optimized_Fidelity}
\end{figure}

We first consider the sub-Ohmic case with spectral exponent \(s = 1/2\), motivated by the ubiquity of \(1/f^x\) noise in realistic environments and its strong connection to dephasing dynamics~\cite{paladino20141}. In this regime, we find that the optimized pulse parameters \(a_n = \{-1.06,\, 0.44,\, -0.12,\,0.11,\, -0.24,\, -0.30,\, 0.38\}\) significantly enhance gate fidelity, as illustrated in Fig.~\ref{Fig: Optimized_Fidelity}. 

By contrast, for an Ohmic environment (\( s = 1 \)), the same optimization procedure yields much smaller improvements in fidelity, indicating that the effectiveness of pulse shaping is strongly influenced by the spectral properties of the environment. This type of analysis can be extended to a broad range of coupling operators, system Hamiltonians, spectral densities, and pulse shapes, with the potential to inform the development of protocols and algorithms for characterizing the system--bath Hamiltonian. A comprehensive treatment of such extensions, however, lies beyond the scope of this paper.

In order to utilize this approach in realistic quantum device settings, additional constraints should be incorporated to ensure experimental feasibility and compatibility with specific system requirements. For instance, bounding the peak amplitude of the control field \(\epsilon(t)\) can help suppress leakage into non-computational space, which may arise when the drive becomes excessively strong~\cite{motzoi2009simple, gambetta2011analytic, chen2016measuring}.
\subsection{Coherence Dynamics in FMO Complexes\label{Sec:FMO}}

The Fenna–Matthews–Olson (FMO) complex is a prototypical pigment-protein structure mediating exciton transport in green sulfur bacteria~\cite{matthews1979structure,blankenship2002molecular}. 
Two-dimensional electronic spectroscopy (2DES) experiments have revealed long-lived coherence signals in FMO, both at cryogenic (77~K)~\cite{engel2007evidence} and room temperature~\cite{panitchayangkoon2010}. These observations have raised the possibility that quantum coherence plays a functional role in biological energy transfer~\cite{scholes2017using, lambert2013quantum}.

In 2DES, the system is excited by a sequence of femtosecond laser pulses. Fourier analysis converts time delays into the frequency domain, allowing access to the time evolution of the density matrix and identification of energy-transfer pathways~\cite{jonas2003two}. The characteristic excitonic splittings ($\sim 200~\mathrm{cm}^{-1}$) are small relative to the inverse pulse bandwidth ($\sim1.67\times10^{5}~\mathrm{cm}^{-1}$)~\cite{nalbach2011exciton}. Therefore, the instantaneous gate approximation (see Sec.~\ref{Sec:InstaGate}) applies.

The seminal experiment by Engel \textit{et al.} reported an anticorrelation between peak amplitude and spectral width in the 2DES signal, suggesting the presence of excitonic coherence~\cite{engel2007evidence}.
However, standard open quantum system models, based on static noise and factorized initial conditions, predict rapid decoherence at high temperature~\cite{breuer2002theory,ishizaki2009unified}. This led to alternative interpretations based on vibronic coherence, where nuclear vibrations mimic coherent features in the spectrum~\cite{chin2013role,christensson2012origin}. In such cases, the signal is classical in origin.

These models, however, neglect the correlations between the system and the dispersive environment that are transferred by the short pulse through the unitary dressing described in Eq.~\ref{Eq:UnitaryDressing}, and illustrated schematically in Fig.~\ref{Fig:PrePostGate}. These correlation transfers can induce long-lived, stable coherence in the post-gate regime, as shown in Sec.~\ref{Section: Coherence Dynamics and Recovery}.

Instead, such models typically assume factorized initial conditions, Markovian dynamics, or employ secular approximations—all of which lead to rapid suppression of coherence, especially at elevated temperatures or in the presence of static noise. As a result, the coherence generation and non-secular population–coherence transfer mechanisms revealed by the instantaneous gate framework presented in this work are not captured by conventional approaches~\cite{tanimura2006stochastic,nalbach2011exciton,chen2015efficient}.

In our explanation, ultrafast pulses in strongly non-Markovian environments induce non-secular population-to-coherence transfers, which persist even under time-averaging in the interaction picture (see Sec.~\ref{Section: Coherence Dynamics and Recovery}). When the pulse duration is shorter than the bath correlation time, the pre-gate component of the dissipator becomes conjugated by the instantaneous unitary. This transformation shifts the effective dephasing action from the coherence–coherence block to the population–coherence block of the dissipative generator.

In contrast to static models where coherence decays rapidly, the dynamics presented here exhibit a long-lived non-secular transfer process that mitigates dephasing of the static post-gate component and enables partial coherence refocusing. This mechanism persists until the initially correlated system–bath state has fully relaxed across the bath degrees of freedom. In such slow (i.e., strongly correlated) environments, coherence lifetimes can be substantially extended~\cite{rebentrost2009environment,plenio2008dephasing}.

\begin{figure}[htbp]
    \centering
    \includegraphics[width=1\columnwidth]{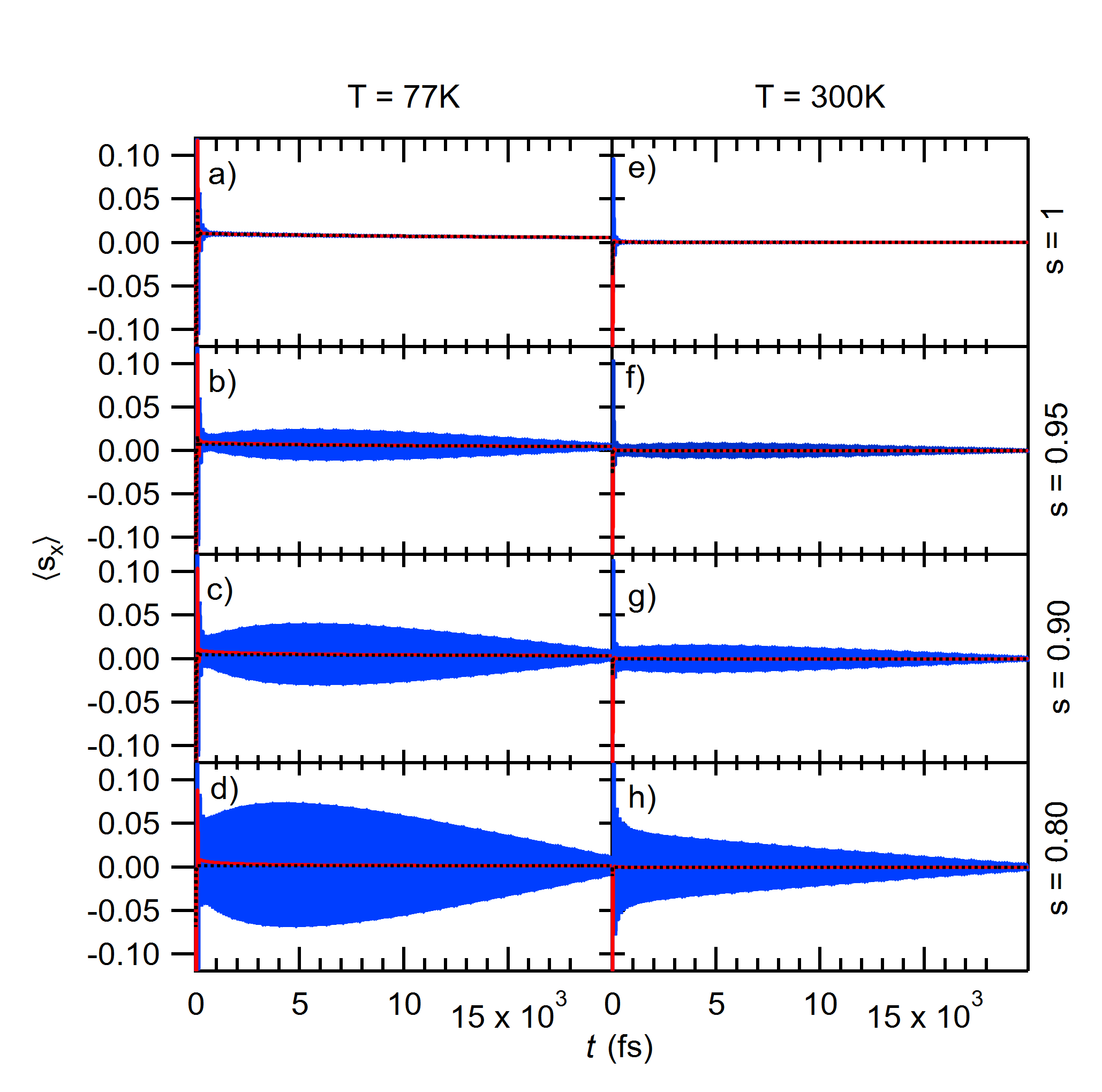}
    \captionsetup{justification=raggedright, singlelinecheck=false}
    \caption{
        X-component of the magnetization as a function of time for factorized (red) and dynamically prepared (blue) initial conditions at 77~K and 300~K, under the instantaneous gate approximation. The bath dispersion parameter $s$ varies as indicated along the rightmost axis. Coherence recovery is suppressed with increasing temperature, but remains possible in the sub-Ohmic regime with dynamical state preparation (blue). No coherence recovery occurs in the static model (red).
    }
    \label{fig:FMO}
\end{figure}

In the remainder of this section, we map the excitonic dynamics of FMO onto the spin-boson model and estimate the relevant parameters. We then show that under realistic conditions, coherence recovery and rephasing persist. This provides a consistent theoretical explanation for the long-lived excitonic coherence observed in FMO, based on non-Markovian dynamics and consistent with the findings of Engel \textit{et al.}~\cite{engel2007evidence} rather than vibronic effects.

\subsubsection{Model}

The well-established Frenkel-exciton model is a standard framework for describing exciton transport in pigment-protein complexes, where individual pigments are typically coupled to distinct baths. For a two-site (dimer) system, the Frenkel Hamiltonian can be mapped onto the spin-boson model~\cite{gilmore2006criteria,lampert2025sixth}. In this mapping, the qubit’s diagonal matrix elements, $\pm\Delta/2$, correspond to the energy splitting $\Delta$ between the exciton sites, and the system-bath coupling is initially purely diagonal, characteristic of pure dephasing.

However, the presence of a small dipole-dipole interaction between the sites, with coupling strength $U \ll \Delta$, leads to coherent superpositions of site states in the excitonic eigenbasis. In this basis, the energy splitting remains approximately $\Delta$, but the system-bath coupling acquires a weak off-diagonal component. Specifically, the ratio of off-diagonal to diagonal coupling elements becomes $A_{12}/A_{11} \approx 2U/\Delta$, as shown in Eq.~\ref{Eq:A0}.

The resulting effective model is a strongly biased spin-boson system, where $\Delta$ defines the energy asymmetry between the sites, and the dimensionless bias parameter $\xi = \Delta/(2U) \gg 1$. Despite the basis transformation, the overall system-bath coupling strength remains comparable to that of the original exciton-bath interaction~\cite{lampert2025sixth}.

For the spin-boson model, we adopt parameters typical in the field, following Ref.~\cite{nalbach2011exciton}: the reorganization energy is $\lambda_r = -S(0) = 2\lambda^2\omega_c = 50\,\mathrm{cm}^{-1}$, with $\Delta = 250\,\mathrm{cm}^{-1}$, $U = 10\,\mathrm{cm}^{-1}$, $\omega_c = 100\,\mathrm{cm}^{-1}$, and $\lambda^2 = 0.25$. We compute the perpendicular component of the magnetization as a function of time, assuming an instantaneous $\pi/2$ pulse as described in Sec.~\ref{Section: Coherence Dynamics and Recovery}.

The results at 77 and 300~K are shown in Fig.~\ref{fig:FMO}~(a--d) and (e--h), respectively. Under these conditions, an Ohmic bath does not support coherence recovery. However, introducing weak sub-Ohmicity—by reducing the spectral exponent \(s\)—substantially enhances the relaxation time associated with bath-induced dephasing. As a result, nonsecular coherence pumping via population–coherence transfer enables long-lived excitonic coherences, which persist until the bath fully relaxes, including its slow zero-frequency spectral components.

This suggests that the coherence recovery observed in light-harvesting complexes results from the interplay between ultrafast optical pulses and the transfer of preexisting system--bath correlations, followed by nonsecular evolution after the pulse. These correlations decay on long timescales, allowing coherent oscillations to persist well beyond what would be expected in a fully relaxed environment. Thus, long-lived coherences may be ubiquitous in such systems when probed with fast pulses. However, in natural energy transfer during photosynthesis, where the system remains in a steady state, this mechanism is absent. This implies that the observed coherences are unlikely to play a functional role in energy efficiency.

\section{Conclusion}

We have developed a unified framework for modeling quantum gate operations in open systems under continuous environmental coupling. By extending the Bloch--Redfield master equation to include time-dependent Hamiltonians and dynamically prepared initial states, we captured the subtle interplay between control dynamics and system--bath correlations across gate operations.

Our analysis reveals that such gates do not produce fully factorized or fully equilibrated post-gate states. Instead, a hybrid dynamical regime emerges: coherences decay as if initialized from an uncorrelated state, while populations reflect residual correlations. This mismatch drives both dephasing and relaxation to their fastest permissible rates—posing a fundamental limitation for quantum information processing.

Despite this, the same correlations that contribute to decoherence can also be harnessed. We demonstrated how pulse shaping can enhance fidelity by coordinating intra-gate dissipation with transferred bath correlations. In sub-Ohmic environments, such control yields substantial improvements. We further uncovered spin-echo–like revivals near $2\pi$-rotations, as well as echo-like rephasing under ultrafast gating—offering intrinsic mechanisms for partial error cancellation even without explicit error correction.

Finally, we showed that our formalism applies beyond engineered quantum devices. Long-lived excitonic coherences observed in light-harvesting complexes such as FMO arise from the same mechanisms: slowly relaxing bath correlations conjugated by unitary dynamics induced by fast pulses. This suggests a broader unifying principle in the behavior of coherently driven open systems across physical domains.

The methods presented here are extensible to multi-qubit architectures, spin baths, and more structured environments. They provide a foundation for understanding the fundamental limits and hidden opportunities in the control of quantum coherence in noisy, real-world systems.

Looking ahead, this framework raises important questions about the role of system–environment correlations in quantum control. Extending the analysis to multi-qubit systems and entangling gates may reveal how such correlations influence collective decoherence and scalable gate design. The spin-echo–like features and coherence revivals observed here suggest that mechanisms akin to spin-echo and dynamical decoupling may arise intrinsically from the structure of non-Markovian dynamics. These predictions are amenable to verification in current experimental platforms, including superconducting qubits and ultrafast spectroscopies. More broadly, our results point to the possibility that environmental memory—when properly understood and incorporated—could be harnessed as a functional resource in coherent quantum technologies.

\section*{ACKNOWLEDGMENTS}
We thank Elyana Crowder, Nicholas Ezzell and Daniel Lidar for fruitful discussions. Support by the school of physics in Georgia Institute of Technology is gratefully acknowledged.
\section*{DATA AVAILABILITY}
The data that support the findings of this article are openly available~\cite{Chen2025DataCode}.
\newpage
\appendix

\section{Explicit dissipative generator for Instantaneous Gates} 
In this section, we present the explicit form of the dissipative generator for an instantaneous $\pi/2$ $X$-gate in the Schrödinger picture. This formulation elucidates the resulting relaxation and dephasing dynamics, including non-secular population-to-coherence transfers that emerge during the operation and persist even after averaging over system oscillations in the interaction picture---unlike in static models.

The explicit form of the dissipative generator is obtained by evaluating the dissipator \(\Lambda^S_{dp}(t)\) in Eq.~\eqref{Eq:finalLambdaDP}, following the procedure outlined in Sec.~\ref{Sec:staticRegime}. The resulting generator takes the form:
\begin{equation}
\label{Eq: dDBloch-RedfieldFull7}
\begin{alignedat}{1}
&\mathcal{D}_{dp}'(t)
= \mathcal{D}_{BR}'(t)
+ \delta \mathcal{D}_{1+}'(t)
+ \delta \mathcal{D}_{1-}'(t)
+ \mathcal{D}_{2}'(t)
\\[-2pt]
&\quad
+ \underbrace{\delta \mathcal{D}_{3-}'(t)
+ \delta \mathcal{D}_{3+}'(t)
+ \mathcal{D}_{4}'(t)}_{{\text{rapidly oscillating terms}}}.
\end{alignedat}
\end{equation}

Here, we define \(\delta\mathcal{D}(t)\) as the deviation of the time dependent dissipative generator from its asymptotic value:
\begin{equation}
    \delta\mathcal{D}(t) = \mathcal{D} - \mathcal{D}(t),
\end{equation}
where \(\mathcal{D}\) denotes the asymptotic value of the generator. We hereby give the explicit values of the last 6 terms in Eq.~\eqref{Eq: dDBloch-RedfieldFull7}:

\begin{equation}
\mathcal{D}_{1\pm}'(t) = \frac{1}{4}
\begin{pmatrix}
0 & 0 & 0 & 0 \\
\pm\xi  J_{\pm\Delta}(t) & 0 & 0 & -\xi J_{\pm\Delta}(t)\\
-\xi  S_{\pm\Delta}(t) & \pm S_{\pm\Delta}(t) &   J_{\pm\Delta}(t)& \pm\xi  S_{\pm\Delta}(t) \\
\mp J_{\pm\Delta}(t) & 0 & 0 &   J_{\pm\Delta}(t)
\end{pmatrix},
\label{d1+}
\end{equation}

\begin{equation}
\mathcal{D}_{2}'(t) = 
\begin{pmatrix}
0 & 0 & 0 & 0 \\
0 & - \xi^2 J_0(t) & 0 & 0 \\
-\xi S_0(t) & 0 &  -  \xi^2 J_0(t) & 0 \\
0 &  \xi J_0(t) & 0 & 0
\end{pmatrix},
\label{d2}
\end{equation}

\begin{equation}
\resizebox{\columnwidth}{!}{$
 \mathcal{D}_{3+}'(t) = \frac{1}{2}
\begin{pmatrix}
0 & 0 & 0 & 0 \\
\xi(  J_{\Delta}(t)M+  S_{\Delta}(t)N)  & \xi(  S_{\Delta}(t)G-  J_{\Delta}(t)H) & 0 & \xi(  J_{\Delta}(t)M+  S_{\Delta}(t)N) \\
-  J_{\Delta}(t)(G+\xi N)-  S_{\Delta}(t)(H-\xi M) &   S_{\Delta}(t)M-  J_{\Delta}(t)N &  -  J_{\Delta}(t)(M +\xi H)-  S_{\Delta}(t)(N-\xi G) & -\xi(  J_{\Delta}(t)N-  S_{\Delta}(t)M) \\
-  J_{\Delta}(t)M-  S_{\Delta}(t)N & -  S_{\Delta}(t)G+  J_{\Delta}(t)H & 0 & -  J_{\Delta}(t)M-  S_{\Delta}(t)N
\end{pmatrix},
$}\label{Eq:d3+}
\end{equation}

\begin{equation}
\resizebox{\columnwidth}{!}{$
   \mathcal{D}_{3-}'(t) = \frac{1}{2}
\begin{pmatrix}
0 & 0 & 0 & 0 \\
\xi(-   J_{-\Delta}(t)M+   S_{-\Delta}(t)N)  & -\xi(   S_{-\Delta}(t)G-   J_{-\Delta}(t)H) & 0 & \xi(   J_{-\Delta}(t)M-   S_{-\Delta}(t)N) \\
   J_{-\Delta}(t)(G+\xi N)-   S_{-\Delta}(t)(H-\xi M) & -   S_{-\Delta}(t)M-   J_{-\Delta}(t)N &  -   J_{-\Delta}(t)(M +\xi H)-   S_{-\Delta}(t)(N-\xi G) & -\xi   J_{-\Delta}(t)N-   S_{-\Delta}(t)M \\
   J_{-\Delta}(t)M-   S_{-\Delta}(t)N &    S_{-\Delta}(t)G-   J_{-\Delta}(t)H & 0 & -   J_{-\Delta}(t)M+   S_{-\Delta}(t)N
\end{pmatrix},
$}\label{Eq:d3-}
\end{equation}

\begin{equation}
\mathcal{D}_{4}'(t) = 
\begin{pmatrix}
0 & 0 & 0 & 0 \\
\xi^2 S_0(t)G & 0 & 0 & \xi^2 J_0(t)H \\
\xi^2 S_0(t)H &  \xi J_0(t)G &  -  \xi J_0(t)G &  \xi^2 J_0(t)G \\
\xi S_0(t)G & 0 & 0 & - \xi J_0(t)H
\end{pmatrix},
\label{Eq:d4}
\end{equation}
where
\begin{equation}
\begin{aligned}
& M = \frac{1}{2}\cos(2\Delta t),\, N= \frac{1}{2}\sin(2\Delta t),\\&
  G = \cos(\Delta t),\, H =\sin(\Delta t).
\end{aligned}
\label{Eq:d4}
\end{equation}
It is important to note that, although the last three terms are rapidly oscillating, they cannot be neglected in the Schrödinger picture. The justification for this is discussed in Appendix~\ref{appendix:cg}.

\section{Coarse-Graining in Interaction Picture}\label{appendix:cg}
In this section, we elaborate on the coarse-graining procedure referenced in Section~\ref{Section: Coherence Dynamics and Recovery}. The motivation for performing coarse-graining in the interaction picture stems from the form of the master equation. In the Schrödinger picture, the equation is given by:
\begin{equation}
    \frac{d\rho}{dt} = -i[H_0, \rho] + \lambda^2 \mathcal{D}^S \rho,
\end{equation}
whereas in the interaction picture it becomes:
\begin{equation}
    \frac{d\varrho}{dt} = \lambda^2 \mathcal{D}^I \varrho.
\end{equation}
In this latter form, the density matrix evolves slowly in time, making it more suitable for coarse-graining procedures. Specifically, the interaction picture allows for the elimination of rapidly oscillating components by averaging over short time scales, thereby isolating the dominant, slowly varying dynamics.

Recalling discussion in section~\ref{Section: Coherence Dynamics and Recovery}, we separate the interaction picture dissipator into two parts:
\begin{equation}
    \Lambda_{dp}^I(t) = \underbrace{U_c\left[\Lambda^{IM} - \Lambda_{\text{static}}^I(t)\right] U_c^\dagger }_{\text{pre-gate}}+ \underbrace{\Lambda_{\text{static}}^I(t)}_{\text{post-gate}}.
    \label{Eq:APPBDP}
\end{equation}

In section~\ref{Section: Coherence Dynamics and Recovery} we have that the pre-gate part is dressed by the unitary rotation: 
\begin{equation}
\label{Eq:FullNonsecular-appendix}
\resizebox{\columnwidth}{!}{$
\displaystyle
\mathcal{D}_{\text{pre-gate}}^{I,\text{cg}}(t)=\frac{1}{2}
\begin{pmatrix}
0 & 0 & 0 & 0 \\
2\xi^2 \delta S_0(t) + \frac{1}{2}(\delta S_{-\Delta}(t)+ \delta S_\Delta(t)) & 0 & \frac{1}{2}(\delta S_{-\Delta}(t)+ \delta S_\Delta(t)) & 0 \\
\frac{1}{2}(\delta J_{-\Delta}(t)- \delta J_\Delta(t)) & 0 &  -\frac{1}{2}(\delta J_{-\Delta}(t)+\delta J_\Delta(t)) & -2\xi^2\delta J_0(t) \\
-\frac{1}{2}(\delta J_{-\Delta}(t)- \delta J_{\Delta}(t)) & 0 & 0 & -\frac{1}{2}(\delta J_{-\Delta}(t)+\delta J_{\Delta}(t))
\end{pmatrix}
$}
\end{equation}

whereas the post-gate part remains identical with factorized initial condition:

\begin{equation}
\label{Eq:standardDSP_appendix}
\resizebox{\columnwidth}{!}{$\displaystyle
\mathcal{D}_{\text{static}}^{I,\text{cg}}(t)=\frac{1}{2}
\begin{pmatrix}
0 & 0 & 0 & 0 \\
0 & -2\xi^2 J_0(t) - \frac{1}{2}\left(J_{\Delta}(t) + J_{-\Delta}(t)\right) & \frac{1}{2}\left(S_{\Delta}(t) - S_{-\Delta}(t)\right) & 0 \\
0 & \frac{1}{2}\left(S_{-\Delta}(t) - S_{\Delta}(t)\right) & -2\xi^2 J_0(t) - \frac{1}{2}\left(J_{\Delta}(t) + J_{-\Delta}(t)\right) & 0 \\
- J_{-\Delta}(t) + J_{\Delta}(t) & 0 & 0 & - J_{-\Delta}(t) - J_{\Delta}(t)
\end{pmatrix}
$}
\end{equation}

\begin{table}[htbp]
\centering
\setlength{\tabcolsep}{3.5pt}
\renewcommand{\arraystretch}{1.18}

\begin{subtable}[t]{\columnwidth}
\centering
\begin{tabular}{|c|P{.42\columnwidth}|P{.42\columnwidth}|}
\hline
 & \textbf{Pre-gate} & \textbf{Post-gate} \\
\hline
$\mathcal{D}_{33}$ &
\resizebox{\linewidth}{!}{$\displaystyle
\begin{aligned}
&-\tfrac{1}{4}\!\bigl(1+\cos(2\Delta t)+2\xi\sin(\Delta t)\bigr)\,\delta J_{-\Delta}(t)\\
&-\tfrac{1}{4}\!\bigl(1+\cos(2\Delta t)+2\xi\sin(\Delta t)\bigr)\,\delta J_{\Delta}(t)\\
&+\tfrac{1}{2}\cos(\Delta t)\bigl(-\xi+\sin(\Delta t)\bigr)\bigl(\delta S_{-\Delta}(t)-\delta S_{\Delta}(t)\bigr)
\end{aligned}$} &
\resizebox{\linewidth}{!}{$\displaystyle
\begin{aligned}
&-\xi^2 J_0(t)\;-\;\tfrac{1}{2}\cos(\Delta t)\Bigl(
\cos(\Delta t)\bigl(J_{-\Delta}(t)+J_{\Delta}(t)\bigr)\\
&\qquad\qquad\qquad\qquad\qquad\qquad
+\;\sin(\Delta t)\bigl(-S_{-\Delta}(t)+S_{\Delta}(t)\bigr)\Bigr)
\end{aligned}$}
\\
\hline
$\mathcal{D}_{34}$ &
\(-\xi^2 \delta J_0(t)\) &
\resizebox{\linewidth}{!}{$\displaystyle
\frac{\xi}{2}\!\left(
\begin{aligned}[t]
&\sin(\Delta t)\bigl(J_{-\Delta}(t)+J_{\Delta}(t)\bigr)\\
&+\;\cos(\Delta t)\bigl(S_{-\Delta}(t)-S_{\Delta}(t)\bigr)
\end{aligned}\right)$}
\\
\hline
\end{tabular}
\caption{Pre-gate and Post-gate components \textit{before} coarse-graining in the interaction picture.}
\end{subtable}

\vspace{0.8em}

\begin{subtable}[t]{\columnwidth}
\centering
\begin{tabular}{|c|P{.42\columnwidth}|P{.42\columnwidth}|}
\hline
 & \textbf{Pre-gate} & \textbf{Post-gate} \\
\hline
$\mathcal{D}_{33}$ &
\(-\tfrac{1}{4}\bigl(\delta J_{-\Delta}(t)+\delta J_{\Delta}(t)\bigr)\) &
\(-\xi^2 J_0(t)-\tfrac{1}{4}\bigl(J_{\Delta}(t)+J_{-\Delta}(t)\bigr)\)
\\
\hline
$\mathcal{D}_{34}$ &
\(-\xi^2 \delta J_0(t)\) &
\(0\)
\\
\hline
\end{tabular}
\caption{Pre-gate and Post-gate components \textit{after} coarse-graining in the interaction picture.}
\end{subtable}

\caption{Comparison of \(\mathcal{D}_{33}\) and \(\mathcal{D}_{34}\) before and after coarse-graining.}
\label{Table: CG comparision}
\end{table}

The terms associated with dephasing dynamics and non-secular population-to-coherence transfer are denoted by $\mathcal{D}_{33}$ and $\mathcal{D}_{34}$, respectively. Table~\ref{Table: CG comparision} presents a comparison of these two terms before and after coarse-graining, for both the pre-gate and post-gate segments. In the strong-dephasing regime, the dominant dissipative contributions are $-\xi^2 \delta J_0(t)$ in $\mathcal{D}_{34}$ and $-\xi^2 J_0(t)$ in $\mathcal{D}_{33}$ (as well as in $\mathcal{D}_{22}$), for the pre-gate and post-gate components, respectively.

As a result, the net dissipator given by Eq.~\ref{Eq:APPBDP} drives the coherences to values on the order of one, during the time scale in which the pre-gate component is significant. This time scale is much longer than $\tau_c$, owing to the slow decay of the spectral density at zero frequency (see Eq.~\ref{Eq:OhmicSDas}).

\begin{figure}[htbp]
    \centering
    \includegraphics[width=1\columnwidth]{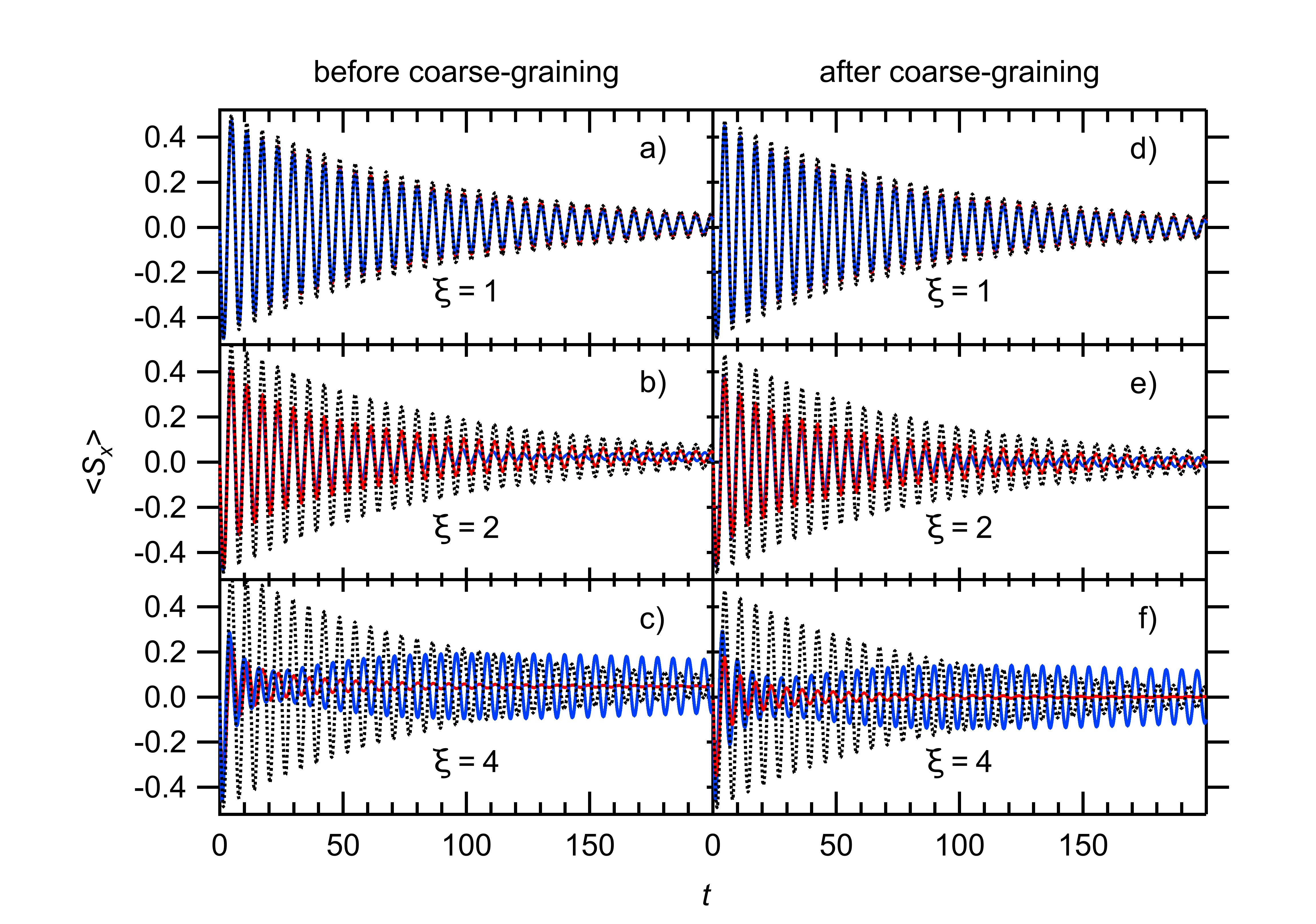}
    \captionsetup{justification=raggedright, singlelinecheck=false}
    \caption{X-component of the magnetization as a function of time for Markovian and non-Markovian dynamics before and after coarse-graining, which is performed in the interaction picture. Dotted black line: Markovian dynamics with decoherence rate equal to half the relaxation rate. Red solid line: non-Markovian dynamics with a factorized initial state. Blue solid line: non-Markovian dynamics with a dynamically prepared initial state under the instantaneous gate approximation. 
    Parameters: $\lambda^2 = 0.02$, $\phi = 0$, $\Delta = 1$, $T = 0$, and $\omega_c = 1$.
}
    \label{Fig: SecularApproximation}
\end{figure}

The coarse-graining procedure eliminates the rapidly oscillating terms in the dissipative generator while retaining the dominant effects induced by the instantaneous gate operation. As shown in Fig.~\ref{Fig: SecularApproximation}, the resulting dynamics obtained in the interaction picture with coarse-graining exhibit excellent agreement with those computed directly by the Bloch-Redfield equation, thereby validating the accuracy of this approximation.

\section{Positivity Violations in the Time-dependent Bloch--Redfield Master Equation}

Because the time-dependent Bloch--Redfield equation is derived perturbatively, it does not, in general, guarantee complete positivity of the reduced system dynamics~\cite{fleming2011accuracy}. Consistent with the established accuracy profile of the Bloch--Redfield master equation, the steady-state \emph{coherences} are reproduced with quadratic accuracy in the system--bath coupling, $\mathcal{O}(\lambda^{2})$~\cite{fleming2011accuracy,thingna2012generalized,hartmann2020accuracy,lee2022perturbative,crowder2024invalidation,tupkary2022fundamental}, in agreement with the mean-force Gibbs state. By contrast, the \emph{populations} can deviate already at the same order: while they are typically $\mathcal{O}(\lambda^{4})$ in the asymptotic states at zero temperature, they should in fact be $\mathcal{O}(\lambda^{2})$. This mismatch in precision induces non-positivity.

To assess positivity in practice, we \emph{audit} all data sets by monitoring the minimum eigenvalue of the density operator,
\begin{equation}
\epsilon_{\min}(t) = \min\{\epsilon_i(t)\},
\label{eq:A1}
\tag{A1}
\end{equation}
where $\epsilon_i(t)$ denote the instantaneous eigenvalues of $\rho(t)$. 

\begin{figure}[htbp]
    \centering
    \includegraphics[width=1\columnwidth]{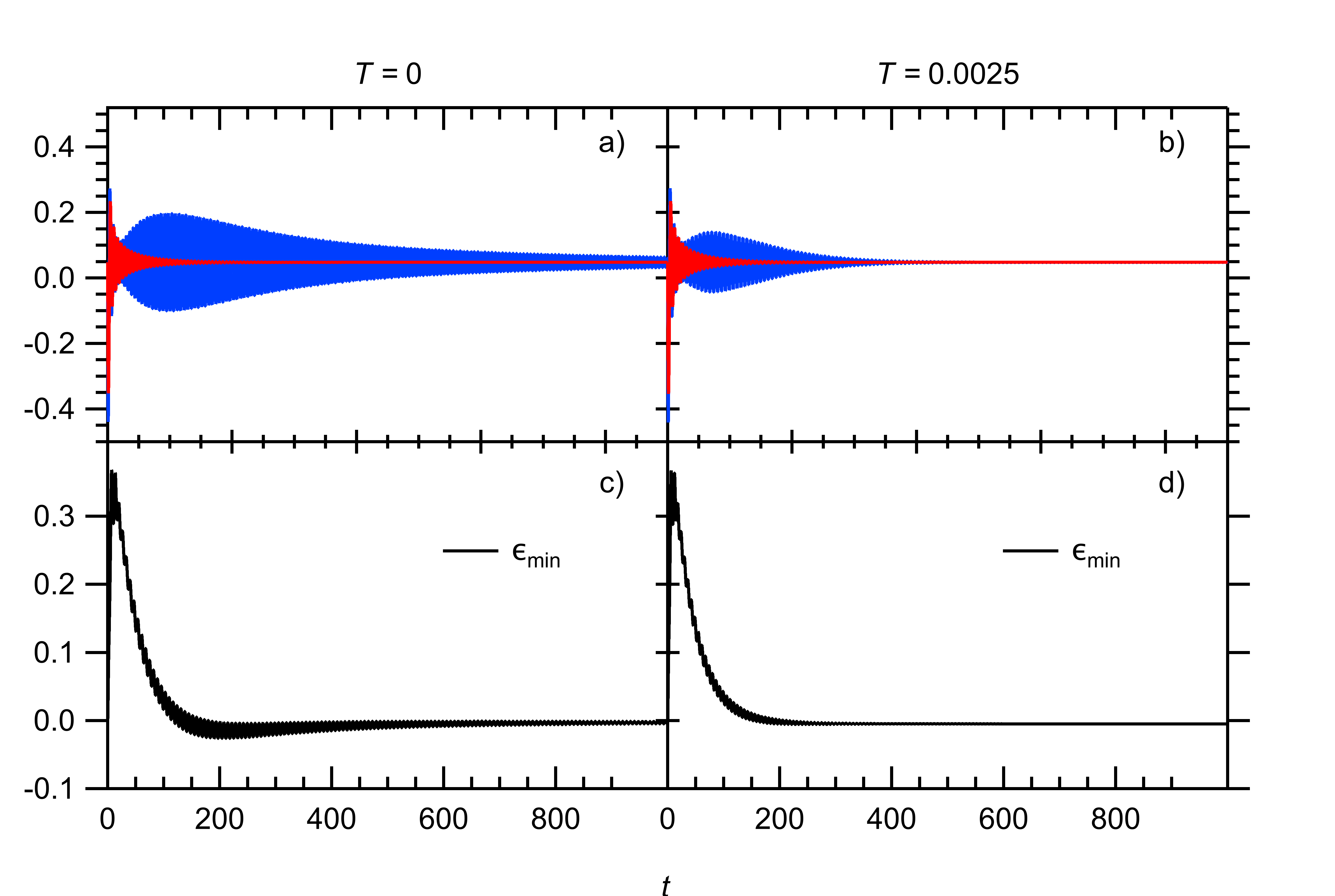}
    \captionsetup{justification=raggedright, singlelinecheck=false}
   \caption{(a,b) X-component of the magnetization as the function of time. The initial magnetization is along the \(y\)-axis. (c,d) The minimum eigenvalue of the density matrix versus time. Red: non-Markovian dynamics with a factorized initial state. Blue: non-Markovian dynamics with the instantaneous gate approximation. Panels (a,c) correspond to zero temperature, while (b,d) correspond to finite temperature \(T=0.0025\). Parameters: \(\lambda^2 = 0.02\), \(\phi = 0\), \(\Delta = 1\), \(\xi = 4\), \(\omega_c = 1\).}

    \label{Fig: PossitivityCheck}
\end{figure}

As shown in Fig.~\ref{Fig: PossitivityCheck}, at $T=0$ we observe a slight transient loss of positivity, manifested as a small dip below zero in $\epsilon_{\min}(t)$ around $t=200$. Even in the most extreme conditions (large coupling amplitude $\lambda^2$ and strong bath anisotropy $\xi$), the deviation remains bounded, with $\epsilon_{\min}(t) > -0.02$ . The residual asymptotic non-positivity is within the established accuracy profile of the
Bloch–Redfield master equation. At a slightly higher temperature, $T=0.0025$, this transient non-positivity disappears. At all higher temperatures relevant to the main text, no violation of positivity is observed.

\newpage

\bibliography{quantum-template2}

\newpage

\end{document}